# *High obliquity, high angular momentum Earth as Moon's origin revisited by Advanced Kinematic Model of Earth-Moon System[1].*

Corresponding Author: Bijay Kumar Sharma[2]

**ABSTRACT**:

Matija Cuk et.al (2016) have proposed a new model for the birth and tidal evolution of our natural satellite Moon, born from lunar accretion of impact generated terrestrial debris in the equatorial plane of high obliquity, high angular momentum Earth. This paper examines their findings critically in the light of advanced kinematic model (AKM) which includes Earth's obliquity($\phi$), Moon's orbital plane inclination ($\alpha$) , Moon's obliquity ($\beta$) and lunar's orbit eccentricity (e). For the real Earth-Moon (E-M) system, the history of evolution of $\phi$, $\alpha$, $\beta$, e and (length of month)/(length of day) or LOM/LOD is traced from $45R_E$ to $60.33R_E$ where $R_E$ is Earth Radius. It is shown that AKM's valid range of application is from $45R_E$ to $60.33R_E$ . The evolution of $\alpha$, $\beta$, e is in correspondence with the simulation results of Matija Cuk et.al (2016) but evolution of Earth's obliquity has a break at $45R_E$ . According to AKM, earlier than $45R_E$ Earth should achieve 0° obliquity in order to achieve the modern value of 23.44° obliquity. Cuk et al (2016) donot explain how this can be achieved. AKM stands vindicated because using protocol exchange algorithm http://doi.org/10.1038/protex.2019.017 (described in S6 SOM), AKM has successfully given precise theoretical formalism of Observed LOD curve for the last 1.2Gy time span opening the way for early warning and forecasting methods for Earth-quake and sudden volcanic eruptions(See S7 of SOM).

## 1. Keplerian Era

The  Kepler's Third Law for a given Planet-Sun configuration is:

$$a^3\Omega^2 = G(M+m) \qquad\qquad 1$$

Eq. (1) does not specify if the given orbital configuration is stable. Newton derived this law assuming that centripetal force (GMm/a²) = centrifugal force(mv$_{tang}$²/a) where a = semi-major axis of Earth-Moon orbital configuration, M = mass of the Earth and m = mass of  our Moon. By implication it was assumed that all configurations predicted by (1) are stable. By the end of 19th century George Howard Darwin put a question mark on this stability by publishing two papers on E-M system (Darwin, 1879, 1880).

---

[1] Presented at CELMEC-VII, 3rd to 8th September 2017, Viterbo, Rome.
[2] Emeritus Fellow, National Institute of Technology, Patna 800005; email: bksharma@nitp.ac.in,
Phone:+919334202848.

In 18th Century, German Philosopher Kant had suggested the theory of retardation of Earth's spin based on the ancient records of Solar Eclipses (Stephenson &Houldon, 1986; Stephenson 2003). Similar kind of studies had been carried out by Kevin Pang at Jet propulsion Laboratory at Pasadena (Morrison1978; Jong & Soldt 1989). He happened to step upon certain ancient records regarding Solar Eclipses. A total Solar Eclipse had been observed in the town of Anyang, in Eastern China, on June 5, 1302 B.C. during the reign of Wu Ding. Had Earth maintained the present rate of spin, the Eclipse should have been observed in middle of Europe. This implies that in 1302 B.C. i.e. 3,291 years ago Earth's spin period was shorter by 0.047 seconds. This leads to a slowdown rate of 1.428 seconds per 100,000 years.

In 1879 George Howard Darwin carried out a complete theoretical analysis of Earth-Moon System and put forward a sound hypothesis for explaining the slowdown of Earth's spin on its axis. This marked the end of Keplerian Era. Gravitationally bound bodies were necessarily tidally interacting and tidal interaction led to tidal dissipation with inherent instability and hence a post-Keplerian physics was required to deal with gravitationally bound binary pairs. Tidally dissipative system because of loss of energy cannot be stable. The system will evolve to a minimum energy state which is a stable configuration by necessity.

## 2. The beginning of Evolutionist view of Universe.

By mid 20th century it was increasingly felt that celestial bodies pair behave as electrons orbiting the nucleus in individual atoms. Within an atom electrons had radiation-less stable permissible orbits propounded by Niels Bohr in 1913:

$$Angular\ Momentum\ of\ electron = I \times \omega = m \times r^2 \times \frac{v_{Tang}}{r} = m \times v_{Tang} \times r$$

$$= n \times \frac{h}{2\pi} \qquad 1$$

(1) Eq. (1) Simplifies to de Broglie standing wave condition:

$$2\pi r = n \times \frac{h}{m \times v_{Tang}} = n \times \frac{h}{p} = n \times \lambda_{de-Broglie} \qquad\qquad 2$$

(2) Eq.(2) Simply states that electrons are permitted to stay in radiation-less stable orbits where electrons behaving as matter wave forms a Standing Wave and is inhibited from making synchrotron radiation and ensured stable orbits. Any other orbit would collapse and electron would be launched on a death spiral towards its respective nucleus.

(3) In exactly the same manner celestial body binaries are born at $a_{G1}$ (inner Clarke's Orbit) which is a Keplerian Orbit , an equilibrium orbit where centripetal force is exactly balanced by centrifugal force but it is an energy maxima (see S5 in SOM) hence the secondary tumbles short or tumbles long of $a_{G1}$. In 2002 at World Space Cogress, Houston, Author proposed Kinematic Model(KM) [see S4 in SOM]. According to KM celestial body pairs have two triple synchrony orbits ($a_{G1}$ and $a_{G2}$) where they are conservative systems and no dissipation of energy is involved (Sharma, Ishwar & Rangesh,2009; Sharma, 2011; Krasinsky, 2002). Here triple synchrony orbits implies:

$$\omega(spin\ angular\ velocity\ of\ the\ primary) = \Omega(orbital\ angular\ velocity)$$
$$= \Omega'(spin\ angular\ velocity\ of\ the\ secondary \hspace{2cm} 3$$

The orbits of triple synchrony are defined as geo-synchronous orbits in E-M system and Clarke's orbits in context of planet-satellite pairs, star-planet pairs, star-star pairs, neutron star-neutron star(NS) pairs and NS and BH (black hole) pairs.

i.   Planet-satellite pairs, star-planet pairs and star pairs are non-relativistic systems. They, within months/years, lock-in at outer Clarke's orbit unless impacted by a third celestial body. Non-relativistic systems are stable at outer triple synchrony orbits.

**ii.**  NS pairs, NS and BH pairs or BH pairs are relativistic systems. Relativistic systems are radiating gravitational waves and they are being driven towards coalescence hence they are always experiencing in-spiral until the final ring-down and merger and always negatively off-setted with respect to outer Clarke's Orbit and never locked-in at outer Clarke's orbit. The magnitude of off=set is decided by the relativistic strength of NS pair/NS-BH pair/BH pair which in turn is decided by the rate of apsidal precession (long axis of the elliptical orbit of the planet turning in the same direction as the planet's orbital motion).

## 3. Two competing schools of thought on tidally interacting binaries.

The whole scientific community was using the elasto-viscous model for analyzing the tidally interacting binaries but this was based on the knowledge of Love number and Q factor of the celestial bodies which in turn depended on the knowledge of density, rigidity, viscosity and rate of periodic forcing. These parameters are known with large uncertainties for different Planets and their Satellites and hence their Tidal Evolutionary History will be arrived at with equal uncertainty in Seismic Model based analysis (see S3 in SOM). The Author developed a Kinematic Model (KM) of E-M system to study its tidal evolutionary history from its birth to the final lock-in orbit or to its final doom (see S4_KM section in SOM). The KM required only the globe-orbit parameters and the age of the system. Since the system parameters were known with high degree of confidence level hence the results arrived at were reliable and accurate. In spite of this improvement KM failed to resolve the conundrum in E-M puzzle. This had baffled the whole scientific community.

## 4. The conundrum of E-M puzzle.

The Apollo mission had confirmed the age of E=M system with high degree of certainty as 4.467Gy (Stevenson 2008, Toubol et.al.2007). With this age of E-M system the present rate of recession of Moon should have been 2.4cm/y but Lunar Laser Experiment operational since 20[th] July 1969 was giving a recession rate of 3.82±0.07cm/y which indicated anomalously high dissipation rate in Earth's oceans and continents .If the present rate of recession was assumed and extrapolated into the past it led to the birth of Moon at 2.8Gy. In addition the KM with age as 4.467Gy was not giving a matching theoretical formalism of observed LOD curve based on Coral fossils (Well 1963) and ancient tidalites (Charles P.Sonnett 1997)(see S4.3 in SOM). The reconstruction of the history of Lunar recession from existing data of tidal rhythmites (Coughenour et.al. 2009) and that of length of Earth day (LOD) record in coral fossils ( Wells 1963) indicate that Earth-Moon is not just a Clockwork or orbitally fixed pair of monoliths as viewed by Newton in Mathematica Principia but instead Earth-Moon system is a tidally evolving system where Moon is tidally receding from Earth since its birth. The findings of Stephenson (1997) and Lambeck (1980) firmly established that Earth-Moon system is a non-linearly tidally evolving system with a complex history of interrupted tidal evolution due to intermittent gravitational resonances with lunar $M_2$ tides and solar $S_2$ tides . Kant (1754) hypothesis included the lunar tidal perturbation as well as solar tidal perturbation on tidal evolution of E-M system. G.H. Darwin (1889,1890) clearly established that Moon's tidal brake and Solar tidal brake have slowed down the Earth's spin from 5 hrs and orbital radius of 18,000 Km (just beyond Roche's Limit) (Ida, Canup & Stewart 1997) to the present Earth's spin rate of 24 h and orbital radius of 384,400 Km and in the process the Earth's angular excess spin energy released has led to tidal heating of Earth and spin angular momentum of Earth spin has been transferred to E-M orbital angular momentum and Lunar tidal recession. The results seem to indicate a non-linear variable rate of tidal dissipation throughout E-M system's history. Oceans may enter and exit $M_2$ and $S_2$ tide resonance in geologically short time intervals (Kagan 1997). It was this tension between observation and theory which compelled the Author to remove the first degree approximations from E-M model and propose a more comprehensive and detailed model of E-M system which the Author is referring to as Advanced Kinematic Mode;(AKM) and as will be shown in this paper AKM has dramatically improved the model to real world correlation.(see S6 in SOM).

5. **The origin of Moon from high obliquity, high angular momentum (AM) Earth and impact generated circum-terrestrial debris disk.**

Matija Cuk et.al (2016) have proposed a new model for the birth and tidal evolution of our natural satellite Moon in which lunar tidal dissipation due to lunar obliquity tides during Cassini State transition plays an important role in stabilizing and allowing E-M system to arrive at climatically favorable E-M configuration with a low Earth's obliquity ($\phi$ =23.44°). High angular momentum and high obliquity Earth provides a more robust mechanism to remove excess AM and provides Earth's mantle like isotopic composition properties of Moon. Their proposal is as follows:

Stage 1: A high energy collision between Earth and Theia (Mars-like impactor) impact generated debris (which is iron depleted and hence Moon has unusually small metallic core < 3% of total mass of Moon) forms a circum-terrestrial accretion disc coplanar with the equatorial plane of highly oblique Earth ($\Phi > 70°$) (Canup and Asphaug 2001). The impact resulted in well mixed vaporized and equilibrated molten material from which both Earth's crust and mantle and Moon formed. This resulted in identical isotopic signatures of Earth and Moon (Burkhardt 2015, Young 2016) . Impact had peeled off the mantle of Earth. Old canonical theory assumes the initial terrestrial day to be 5 hr but Cuk et.al. have done their simulation using 2hr terrestrial day. Extraordinary impact left rapidly spinning highly oblate Earth with a tilt angle of $\phi > 70°$ with respect to the ecliptic.

Stage 2: Laplace plane transition at lunar orbit at 'a' (semi-major axis of lunar orbit) = 17R$_E$ (Nicholson et.al. 2008, Tremaine et.al. 2009). This abrupt transition from geo-centric Laplace plane to heliocentric Laplace plane , due to solar secular perturbation in highly oblique Earth's environment, excites sizeable lunar eccentricity, high lunar orbit inclination ($\alpha = 30°$) and draining of AM from lunar orbit to Earth's helio-centric orbit. Simultaneously Earth's obliquity falls from 70° to 30°. The lunar eccentricity causes large stretching and squeezing of Moon leading to internal tidal flexing within Moon. This causes strong eccentricity- damping satellite tides. These eccentricity damping Moon's tides balance the Earth's tides and stall the tidal evolution for a prolonged period (Atobe and Ida 2007). During this stalled period, Moon's orbital plane inclination increases to $\alpha = 30°$.

 Hydrostatic equilibrium shape of Moon at 'a' = 15 to 17R$_E$ with Moon's orbit eccentricity at 0.2 got frozen because of rigid lithosphere  and that fossil oblateness is retained till the modern times (Keane and Matsuyama 2014)

Stage 3: Earth becomes rigid enough to maintain C (principal moment of inertia of Earth around spin axis) = $8.019 \times 10^{37}$ Kg-m$^2$ constant from 'a' = 25R$_E$ to the present day at 'a' = 60.33R$_E$ .  At 'a' = 30R$_E$ to 40R$_E$, lunar spin axis underwent Cassini State Transition (Ward 1975, Chyba 1989). Moon's obliquity increases from $\beta = 10°$ to 50°. This generates strong and forced lunar obliquity tides which help suppress the lunar orbital inclination from $\alpha = 30°$ to 10°. Simulation study show that from 'a' = 29.7 to 35R$_E$, Moon is in non-synchronous state and beyond 35R$_E$ to the present 60.33R$_E$ , Moon is locked in a synchronous orbit with its face always showing towards Earth. Moon is tidally locked with Earth.

At 33R$_E$ , Cassini State transition occurs while transiting from Cassini state 1 to Cassini state 2. Moon's Obliquity ($\beta$) is as high as 70°.Once Moon settles in Cassini State 2, Moon sedately spirals out from Earth. The inclination angle is dampened from 30° to 15° due to Moon's obliquity tides (tidal flexing within the interior of our Moon) .

Stage 4: from 'a' = 30$R_E$ to 'a' = 60.33$R_E$ (in modern times), lunar obliquity tides bring down $\Phi$ =30° to $\Phi$=23.5° and $\alpha$ = 15° to $\alpha$ = 5° at the same time ensuring the current value of AM.(Rubicam 2016)

Here there is a conundrum. As Moon's orbital plane inclination drops from 15° to 5°, Earth's obliquity must rise from 0° to 23.5°. This requires that at Cassini State Transition Earth's obliquity $\Phi$ must be 0.

Rubicam (1993) has discussed this problem. At present, the Earth's mean obliquity is slowly increasing as a result of tidal interactions with the Moon. The lunar inclination is decreasing at the same time, so the angular momentum of the Earth–Moon system is conserved.

The conservation of total angular momentum is given as follows (Rubicam 1993):

$$Sin\left[\frac{\Delta\varphi}{2}\right] = -\frac{J_{orbit}}{J_{spin\_Earth}} \times Sin\left[\frac{\Delta\alpha}{2}\right] \qquad\qquad 1$$

This implies that if lunar orbital plane inclination angle decreases by 5° and $J_{orbit}$/ $J_{spin\_Earth}$ =10 then Earth's obliquity must increase by 60.6°. This precisely is predicted by Advanced Kinematic Model as seen by close examination of Figure 1..

6. **The total resultant angular momentum vector of Earth-Moon system.**
    6.1. **Calculation of Total Angular Momentum(AM) of Earth-Moon System as the vector sum of constituent AMs**

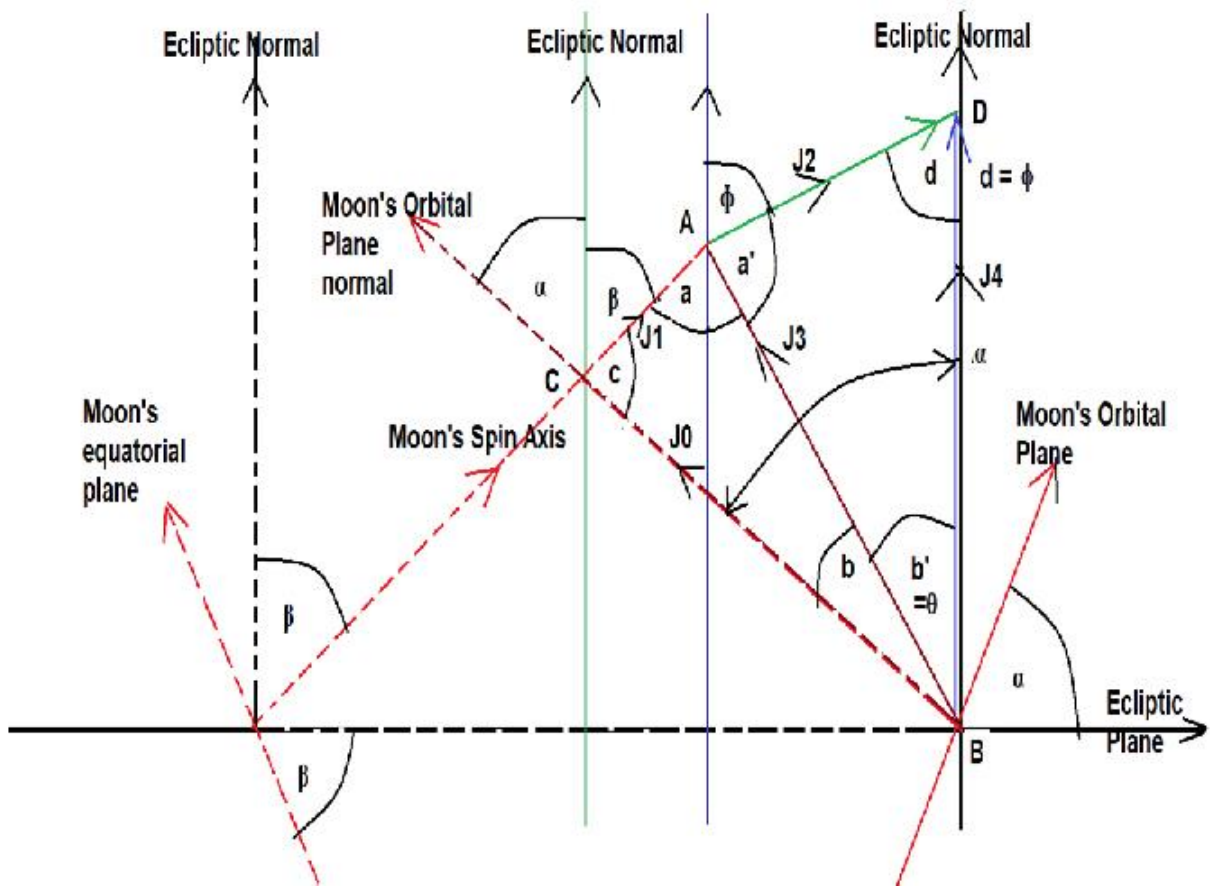

**Figure 1. Spin-Orbital configuration of Earth-Moon System.**

In Figure 1,
J0 = orbital angular momentum (AM) of Earth-Moon (E-M) system.
J1= spin angular momentum of Moon.
J2=spin angular momentum of Earth.
J3 = vector sum of J0 and J1.
J4= Total AM of E-M system.

J0, J1 and ecliptic normal are coplanar according to Cassini Law 3.
Presently E-M system is in Cassini State II hence J0 vector and J1 are on the two sides of
Ecliptic Normal as shown in the Figure 1.
J3, J2 ,J4 and Ecliptic normal are coplanar.
But the plane containing J0,J1 and Ecliptic Normal and the plane containing J2, J3, J4
and Ecliptic normal are two separate planes hence J0 and J1 are shown by dotted lines.

Definitions of Earth's Obliquity(ϕ), Moon's orbital plane inclination (α) and Moon's
Obliquity (β):

Axial tilt of Earth's spin axis with respect to (w.r.t.) Ecliptic Normal = $\phi$ = 23.44° = 0.4091051767 radians;

Axial tilt of Moon's spin axis w.r.t. Ecliptic Normal = $\beta$ = 1.54° = 0.02687807 radians;

Angle between Moon's equatorial plane and ecliptic plane = $\beta$;

Total axial tilt of Moon's spin axis w.r.t. E-M orbital AM vector = $\alpha+\beta$ = 6.68° =0.11658 radians.

All these are observational Astronomy data in the current era.

6.2. **The total resultant angular momentum vector of Earth-Moon system.**

According to Cassini Law, Moon's spin axis Normal to the equatorial plane of Moon, Moon's orbital plane Normal and Ecliptic plane Normal are co-planar hence these three NORMALS can be drawn on the same page but Earth's spin axis are not co-planar hence Earth's spin will be kept out while determining the resultant angular momentum J3 = vector sum of J0 and J1 of E-M system.

**Cassini's laws**, three underlined empirical rules that accurately describe the rotation of the Moon, formulated in 1693 by Gian Domenico Cassini. They are: (1) the Moon rotates uniformly about its own axis once in the same time that it takes to revolve around the Earth; (2) the Moon's equator is tilted at a constant angle ($\beta$ = about 1.54° of arc) to the ecliptic, the plane of Earth's orbit around the Sun; and (3) the ascending node of the lunar orbit (*i.e.,* the point where the lunar orbit passes from south to north on the ecliptic) always coincides with the descending node of the lunar equator (*i.e.,* the point where the lunar equator passes from north to south on the ecliptic). As a consequence of the third law, the north pole of the Moon as projected on the sky (point *z*), the north pole of the ecliptic (point *Z*), and the north pole of the lunar orbit (point *P*, inclined at an angle of about = $\alpha$ = 5.14° to the ecliptic) all lie close to one another on a great circle.

Total Angular Momentum Vector of E-M system is determined in two parts.

*In first part*: $J_0$ the orbital angular momentum and $J_1$ Moon's spin angular momentum and the Ecliptic Normal are taken coplanar and co-processing and $J_0$ and $J_1$ are placed on the two sides of the Ecliptic Normal since for lunar semi-major axis 'a' > 33$R_E$ E-M system has settled down in Cassini State II. And AKM is valid within the range 45$R_E$ < a < 60.33$R_E$ hence E-M system being in Cassini State II is a valid assumption.

*In second part:* $J_2$ the spin angular momentum of oblique Earth, the ecliptic plane normal and $J_3$ the vector sum of $J_0$ and $J_1$ will be treated as co-planar and the vector sum :
$J_4 = J_2 + J_3$ will be determined.

6.3. *Determination of $J_3$ vector = $J_0$ vector + $J_1$ vector*

Here J0 (orbital angular momentum of E-M system and J1(spin angular momentum of Moon) and Ecliptic normal are coplanar hence the vector triangle ABC can be drawn on one plane.

Sum of the interior angles = 180° = $\pi$ =3.14 radians

Hence $\Delta$ABC the sum of the interior angles : a + b + c = 3.14

But c = π – (α+β) = 3.14 - 0.116588 = 3.025012654 radians

Working out the vector sum of constituent angular momentum vectors , we arrive at the following results:

$$J_0 = \overrightarrow{BC} = \overrightarrow{J_{ORB}} \ (Earth - Moon \ System) \ is \ normal \ to \ the \ Moon's \ orbital \ plane \ and$$

$$Moon's \ orbital \ plane \ is \ inclined \ to \ ecliptic \ plane \ by \ \alpha$$
$$= 5.14° \ and \ sidereal \ orbital \ period = 27.3217d$$

$$J_{ORB} = \boldsymbol{J_0} = \frac{m}{1 + \frac{m}{M}} \times a^2 \times \frac{2\pi}{T_{ORB}} \sqrt{1 - e^2}$$

$$= 1.07066 \times 10^{40} Kg - m^2 \times \frac{2.6617 \times 10^{-6} rad}{s}$$
$$= 2.84978 \times 10^{34} \frac{Kg - m^2}{s} \qquad\qquad 2$$

Here 'a' (semi-major axis of Moon's orbit) = $3.844 \times 10^8$m; mass of Earth M= $5.9723 \times 10^{24}$ Kg and mass of our Moon m = $0.07346 \times 10^{24}$ Kg, m/(1+m/M) = reduced mass of Moon = $7.25674 \times 10^{22}$ Kg, $T_{ORB}$ orbital period of Moon around Earth(sidereal period) = 27.3217d and e is eccentricity = 0.0549.

*Moon's spin angular momentum is in the direction of Moon'spin axis normal to Moon'sequaorial pl*

As seen in Figure 1 , Moon's spin axis is tilted w.r.t. Ecliptic normal by 1.54° and tilted w.r.t. orbital normal by 6.68º to the right of the orbital normal because presently we are in Cassini State II

$$J_M = J_1 = Moon'spin \ angular \ momentum = \overrightarrow{CA} = I \times \Omega$$
$$= 2.32541 \times 10^{29} \frac{Kg - m^2}{s} \qquad\qquad 3$$

Where

$$R_{Moon} = 1737.4Km; I = 0.394 \times m \times R_{Moon}^2 = 8.73669 \times 10^{34} Kg - m^2, \Omega$$
$$= 2.6617 \times 10^{-6} radians/s$$

Since Moon is in synchronous orbit:

$$Moon's \ spin \ period = Moon's \ orbital \ period = 27.3217d \qquad\qquad 4$$

As seen in Figure 3:

$$\overline{J_3}(\overrightarrow{BA}) = \overline{J_0} + \overline{J_1}$$

$$J_3 = vector\ sum\ of\ \overline{J_0}\ and\ \overline{J_1}$$

From ΔABC in Figure 3 we obtain:

$$J_3^2 = J_0^2 + J_1^2 - 2J_0J_1 Cos\{\pi - (\alpha + \beta)\}$$

$$where\ \alpha\ =\ 5.14°\ =\ 0.08970992355250854\ \text{radians}\ ;\ \beta = 1.54°$$
$$= 0.02687785118484197 radians$$

$$and\ (\pi - \alpha - \beta) = \widehat{c} = 3.025\ radians \qquad\qquad 5$$

(5) can be simplified to:

$$J_3^2 = J_0^2 + J_1^2 + 2J_0J_1 Cos\{(\alpha + \beta)\} = J_0^2 + J_1^2 + 2J_0J_1\{Cos\alpha Cos\beta - Sin\alpha Sin\beta\} \qquad 6$$

Here we define the following Trignometric Identities:

$$Sin[\alpha] = A; Sin[\beta] = D; Sin[\varphi] = B \qquad\qquad 7$$

Substituting these identities in (6) we get:

$$J_3^2 = J_0^2 + J_1^2 + 2J_0J_1 Cos\{(\alpha + \beta)\}$$
$$= J_0^2 + J_1^2 + 2J_0J_1\left\{\sqrt{1-A^2}\sqrt{1-D^2} - A \times D\right\} \qquad\qquad 8$$

For modern times values of inclination and lunar obliquity we obtain:

$$\sqrt{1-A^2}\sqrt{1-D^2} - A \times D = 0.993211 \qquad\qquad 9$$

Sunstituimg (9) and the magnitudes of J0 and J1 we obtain:

$$J_3^2 = 8.12138 \times 10^{68}\ (\frac{Kg-m^2}{s})^2$$

$$J_3 = 2.8498 \times 10^{34}\frac{Kg-m^2}{s} \qquad\qquad 6$$

Inspecting Figure 3, we see that J₃ makes an angle θ with respect to the normal of the ecliptic and J₃ lies left to the normal.

By Sin Law:

$$\frac{J_0}{Sin[a]} = \frac{J_1}{Sin[b]} = \frac{J_3}{Sin[c]} \qquad\qquad 7$$

From (7) the three angles are:

$$angle\ c = (3.14 - 0.116588) = 3.02501265\ radians,$$

$$angle\ b = 9.49129 \times 10^{-7} radians$$

$$and\ angle\ a = \ 0.116579\ radians\ and\ a + b + c = \pi \qquad\qquad 8$$

The angle of inclination of $J_3$ w.r.t. ecliptic normal and left to normal = θ =α –b = 0.08970905087 radians = 5.13995⁰ ~ α ;

### 6.4. Determination of $J_4$ vector = $J_2$ vector + $J_2$ vector=Total AM of E-M system.

For calculating the total resultant angular momentum $J_4$ we have to consider  ΔABD in Figure 1..

Applying Cos Law to ABD we get:

$$J_4^2 = \ J_2^2 + J_3^2 - 2J_2J_3 Cos\{a'\}$$

$$where\ \theta = 0.08970905087\ radians, a' = \pi - \emptyset - \theta \sim \pi - \emptyset - \alpha = 2.64277\ rad$$

For the period from a = 45R$_E$ to a = 60.336R$_E$ Moon's spin AM is several order of magnitudes smaller than orbital AM hence it is valid to assume that angle a'~ (π-ϕ-α) hence total AM J4 is given as follows:

$$J_4^2 = \ J_2^2 + J_3^2 - 2J_2J_3 Cos\{\pi - \emptyset - \alpha\} \qquad\qquad 9$$

(9) simplifies to the following:

$$J_4^2 = \ J_2^2 + J_3^2 + 2J_2J_3 Cos\{\emptyset + \alpha\} = \ J_2^2 + J_3^2 + 2J_2J_3\{Cos\emptyset Cos\alpha - Sin\emptyset Sin\alpha\} \qquad 10$$

Substituting the Trignometric identities in (10) we obtain:

$$J_4^2 = \ J_2^2 + J_3^2 + 2J_2J_3 Cos\{\emptyset + \alpha\} = \ J_2^2 + J_3^2 + 2J_2J_3\left\{\sqrt{1 - A^2}\sqrt{1 - B^2} - A.B\right\} \qquad 11$$

Taking the modern values of terrestrial Obliquity and lunar orbital inclination:

$$\sqrt{1 - A^2}\sqrt{1 - B^2} - A.B = 0.87815 \qquad\qquad 12$$

From (6):

$$J_3 = 2.8498 \times 10^{34} \frac{Kg - m^2}{s} \qquad\qquad 6$$

Earth's spin axis obliquity with respect to (w.r.t.) the ecliptic normal = Φ = 23.44° = 0.40910 radians ;

$$\overrightarrow{J_E} = \lfloor J_E \rfloor\ 23.44°\ \ to\ the\ right\ of\ the\ ecliptic\ normal \qquad\qquad 10$$

$$J_E = J_2 = C \times \frac{2\pi}{T_{spin\_Earth}} = 0.3308 \times M_{Earth} \times R_{Earth}^2 \times \frac{2\pi}{23.9345 \times 3600}$$

$$= 5.84758 \times 10^{33} \frac{Kg - m^2}{s} \qquad 11$$

$$where\ Sidereal\ spin\ period = 23.9345h,$$

$$Volumetric\ mean\ radius\ of\ Earth = 6371.008 Km, \qquad M_E = 5.9723 \times 10^{24} Kg;$$

Substituting the magnitudes of the parameters we get:

$$C = 8.01906 \times 10^{37} Kg - m^2\ and\ \omega = 7.29211 \times 10^{-5} radians\ per\ s.$$

Substituting the numerical values in (9);

$$J_4^2 = J_2^2 + J_3^2 - 2J_2 J_3 Cos\{a'\} = 1.13901 \times 10^{69} \frac{Kg - m^2}{s}^2$$

Therefore:

$$J_4 = 3.37492 \times 10^{34} \frac{Kg - m^2}{s} \qquad\qquad 12$$

To determine the angles b' and d the Sine Law is used namely:

$$\frac{J_4}{Sin[a']} = \frac{J_2}{Sin[b']} = \frac{J_3}{Sin[d]} \qquad\qquad 13$$

From (13) the three angles are:

$$a' = 2.64277 rad = 151.42^0;\ b' = 0.0829842 rad = 4.75^0;\ d = 0.415838\ rad = 23.825^0;$$

The sum of the internal angles comes to be:

$$a' + b' + d = 3.1415922 = 180^0\ as\ it\ should\ be.$$

Therefore total angular momentum of E=M system is:

$$J_{Total} = J_4 = 3.37492 \times 10^{34} \frac{Kg - m^2}{s} \angle(\theta - b')$$

$$= 3.37492 \times 10^{34} \frac{Kg - m^2}{s} \angle\gamma = 0.39^0 \qquad\qquad 14$$

In scalar analysis,

$J_T$ = total angular momentum of E-M system= $3.43584 \times 10^{34}$Kg-m²/s , (eccentricity was zero)

By vector analysis of AM of E-M system very simple picture emerges.

In real world situation, vectorial total angular momentum of E-M system has been constrained to be almost but not exactly normal to ecliptic plane after Laplace Plane Transition and its magnitude has remained constant at $3.3749 \times 10^{34}$ (Kg-m$^2$)/s.

So the vector diagram of Figure 1 is valid in assuming that total AM J4 has remained invariant for last 1.5Gy and has remained coincident with the Ecliptic Normal.

7. *Determination of the initial and final lock-in orbits and orbital period = spin period of Moon = spin period of Earth at initial and final lock-in orbit.*

Rewrite (8)

$$J_3^2 = J_0^2 + J_1^2 + 2J_0J_1Cos\{(\alpha + \beta)\}$$
$$= J_0^2 + J_1^2 + 2J_0J_1\left\{\sqrt{1 - A^2}\sqrt{1 - D^2} - A \times D\right\} \qquad 8$$

Rewrite (11) and substitute (8) in (11):

$$J_4^2 = J_2^2 + J_3^2 + 2J_2J_3Cos\{\emptyset + \alpha\} = J_2^2 + J_3^2 + 2J_2J_3\left\{\sqrt{1 - A^2}\sqrt{1 - B^2} - A.B\right\} \qquad 11$$

$$J_4^2 = J_2^2 + \left[J_0^2 + J_1^2 + 2J_0J_1\left\{\sqrt{1 - A^2}\sqrt{1 - D^2} - A \times D\right\}\right]$$
$$+ 2J_2\left[\sqrt{J_0^2 + J_1^2 + 2J_0J_1\left\{\sqrt{1 - A^2}\sqrt{1 - D^2} - A \times D\right\}}\right]\left\{\sqrt{1 - A^2}\sqrt{1 - B^2}\right.$$
$$\left. - A.B\right\} \qquad 12$$

Let us redefine J0, J1, J2 J3

$$J_O = F^* \times a^2 \times \varOmega \qquad 13$$

$$where \; F^* = \frac{m}{1 + \frac{m}{M}} \times \sqrt{1 - e^2}$$

$$J_1 = I \times \varOmega \qquad 14$$

$$J_2 = C \times \omega \qquad 15$$

Substituting (13), (14) and (15) in (12) we obtain:

$$J_4^2 = (C \times \omega)^2 + \left(m^* \times \sqrt{1 - k^2} \times a^2 \times \varOmega\right)^2 + (I \times \varOmega)^2 +$$

$$2\left(m^* \times \sqrt{1 - k^2} \times a^2 \times \varOmega\right)(I \times \varOmega)\left\{\sqrt{1 - D^2}\sqrt{1 - A^2} - AD\right\} + 2(C \times \omega)$$

$$\times \sqrt{\left(m^* \times \sqrt{1-k^2} \times a^2 \times \Omega\right)^2 + (I \times \Omega)^2 + 2\left(m^* \times \sqrt{1-k^2} \times a^2 \times \Omega\right)(I \times \Omega)\left\{\sqrt{1-D^2}\sqrt{1-A^2} - AD\right\}}$$

$$\times \left\{\sqrt{1-A^2}\sqrt{1-B^2} - A.B\right\} \qquad 16$$

Divide (16) by $(C \times \Omega)^2$ and let $\omega/\Omega = X$ we get:

$$\frac{J_4^2}{(C \times \Omega)^2} = \frac{(C \times \omega)^2}{(C \times \Omega)^2} + \frac{\left(m^* \times \sqrt{1-k^2} \times a^2 \times \Omega\right)^2}{(C \times \Omega)^2} + \frac{(I \times \Omega)^2}{(C \times \Omega)^2} +$$

$$2\left(m^* \times \sqrt{1-k^2} \times a^2 \times \Omega\right)\frac{(I \times \Omega)}{(C \times \Omega)^2}\left\{\sqrt{1-D^2}\sqrt{1-A^2} - AD\right\} + \frac{2(C \times \omega)}{(C \times \Omega)^2}$$

$$\times \sqrt{\left(m^* \times \sqrt{1-k^2} \times a^2 \times \Omega\right)^2 + (I \times \Omega)^2 + 2\left(m^* \times \sqrt{1-k^2} \times a^2 \times \Omega\right)(I \times \Omega)\left\{\sqrt{1-D^2}\sqrt{1-A^2} - AD\right\}}$$

$$\times \left\{\sqrt{1-A^2}\sqrt{1-B^2} - A.B\right\} \qquad 17$$

Simplifying (17)

$$\left(\frac{J_4}{C \times \Omega}\right)^2 = X^2 + \left(\frac{m^* \times \sqrt{1-k^2}}{C}\right)^2 \times (a^2)^2 + \left(\frac{I}{C}\right)^2 +$$

$$2\left(\frac{m^* \times \sqrt{1-k^2}}{C} \times a^2\right)\left(\frac{I}{C}\right)\left\{\sqrt{1-D^2}\sqrt{1-A^2} - AD\right\} + 2 \times X$$

$$\times \sqrt{\left(\frac{m^* \times \sqrt{1-k^2}}{C} \times a^2\right)^2 + \left(\frac{I}{C}\right)^2 + 2\left(\frac{m^* \times \sqrt{1-k^2}}{C} \times a^2\right)\left(\frac{I}{C}\right)\left\{\sqrt{1-D^2}\sqrt{1-A^2} - AD\right\}}$$

$$\times \left\{\sqrt{1-A^2}\sqrt{1-B^2} - A.B\right\} \qquad 18$$

Let

$$\frac{m^*}{C} = F \text{ and } \frac{I}{C} = G \qquad\qquad 19$$

Substitute (19) in (18) we get:

$$\left(\frac{J_4}{C \times \Omega}\right)^2 = X^2 + (F\sqrt{1-k^2})^2 \times (a^2)^2 + G^2 +$$

$$2\left(F\sqrt{1-k^2} \times a^2\right)(G)\left\{\sqrt{1-D^2}\sqrt{1-A^2} - AD\right\} + 2 \times X$$

$$\times \sqrt{\left(F\sqrt{1-k^2}\times a^2\right)^2 + (G)^2 + 2\left(F\sqrt{1-k^2}\times a^2\right)(G)\left\{\sqrt{1-D^2}\sqrt{1-A^2}-AD\right\}}$$
$$\times \left\{\sqrt{1-A^2}\sqrt{1-B^2}-A.B\right\} \qquad 20$$

Substituting Kepler's third law :

$$\frac{1}{\Omega^2} = \frac{a^3}{B^2} \quad in \ (20) \ and$$

$$B = \sqrt{GM+Gm} = \sqrt{0.39860\times10^6 + 0.00490\times10^6} = 2.00873\times10^7\frac{m^{3/2}}{s}$$

We get:

$$(\frac{J_4}{C\times B})^2 \times a^3 = X^2 + (F\sqrt{1-k^2})^2\times(a^2)^2 + G^2 +$$

$$2\left(F\sqrt{1-k^2}\times a^2\right)(G)\left\{\sqrt{1-D^2}\sqrt{1-A^2}-AD\right\} + 2\times X$$

$$\times \sqrt{\left(F\sqrt{1-k^2}\times a^2\right)^2 + (G)^2 + 2\left(F\sqrt{1-k^2}\times a^2\right)(G)\left\{\sqrt{1-D^2}\sqrt{1-A^2}-AD\right\}}$$
$$\times \left\{\sqrt{1-A^2}\sqrt{1-B^2}-A.B\right\} \qquad 21$$

Let

$$\frac{J_4}{C\times B} = N \qquad\qquad 22$$

Substituting (22) in (21) we get:

$$(N)^2 \times a^3 = X^2 + (F\sqrt{1-k^2})^2\times(a^2)^2 + G^2 +$$

$$2\left(F\sqrt{1-k^2}\times a^2\right)(G)\left\{\sqrt{1-D^2}\sqrt{1-A^2}-AD\right\} + 2\times X$$

$$\times \sqrt{\left(F\sqrt{1-k^2}\times a^2\right)^2 + (G)^2 + 2\left(F\sqrt{1-k^2}\times a^2\right)(G)\left\{\sqrt{1-D^2}\sqrt{1-A^2}-AD\right\}}$$
$$\times \left\{\sqrt{1-A^2}\sqrt{1-B^2}-A.B\right\} \qquad 23$$

In ideal case where Lunar Orbital Inclination($\alpha$) , Earth's obliquity($\phi$) and Moon's obliquity( $\beta$) are zero then

Sin[$\alpha$] = A=0 and Cos[$\alpha$] =$\sqrt{(1-A^2)}$=1;

Sin[Φ] = B=0 and Cos[Φ] =√(1-B²)=1;

√(1-A²) √(1-B²)-A.B = 1;

And Sin[β] = D=0 and Cos[β] =√(1-D²)=1

Substituting the above results in (23)

$$(N)^2 \times a^3 = X^2 + (F)^2 \times (a^2)^2 + G^2 +$$

$$2(F \times a^2)(G)\{1\} + 2 \times X$$

$$\times \sqrt{(F \times a^2)^2 + (G)^2 + 2(F \times a^2)(G)\{1\}} \times (1) \qquad 24$$

(24) simplifies to classical KM equation:

$$X = N \times a^{3/2} - F \times a^2 - G \qquad 25$$

(25) is the classical form being used by the Author for Kinematic Modelling with assumptions that Moon's orbital plane Inclination, Earth's Obliquity and Moon's Obliquity are zero degree angle.

F and G have been defined in (19) and N has been defined in (22). Substituting the numerical values of the system parameters we get:

$$B = \sqrt{GM + Gm} = \sqrt{0.39860 \times 10^6 + 0.00490 \times 10^6} = 2.00873 \times 10^7 \frac{m^{3/2}}{s}$$

$$N = \frac{J_4}{B \times C} = 2.09517 \times 10^{-11} \left(\frac{1}{m^{\frac{3}{2}}}\right), where\ J_4 = 3.37492 \times 10^{34} \frac{Kg - m^2}{s}, C$$
$$= 8.01906 \times 10^{37} Kg - m^2,$$

$$G = 0.00108949, \qquad F = \frac{m^*}{C} = 9..04936 \times 10^{-16} \left(\frac{1}{m^2}\right) \qquad 26$$

Our Moon has been born at inner geo-synchronous orbit , a$_{G1}$ . E-M system has tidally evolved from inner geo-synchronous orbit , a$_{G1}$, to the outer geo-synchronous orbit , a$_{G2}$. It is midway in this evolutionary path at a = 384,400Km from the center of the Earth. At both the geo-synchronous orbit E-M system is in triple synchrony state when X =1.

(24) simplifies to (25) because at the two end points E-M system is locked-in triple synchrony with

$$Earth's\ Obliquity\ Angle = \varphi = 0\ radians; Lunar\ Orbital\ Plane\ Inclination = \alpha$$
$$= 0\ radians;$$

$$Moon's\ Obliquity = \beta = 0\ rad. \quad and\ X = \frac{\omega}{\Omega} = \frac{Length\ of\ Sidereal\ Month}{Length\ of\ Sidereal\ Day} = \frac{LOM}{LOD}$$
$$= 1 \qquad\qquad 27$$

Equating (25) to Unity we obtain the two geo-synchronous orbits.

$$a_{G1} = 1.48646 \times 10^7 m; \qquad a_{G2}$$
$$= 5.33505 \times 10^8 m \qquad\qquad 28$$

Using Kepler's third law:

$$\frac{1}{\Omega^2} = \frac{a^3}{B^2}$$

We obtain triple synchrony time period at $a_{G1}$ and $a_{G2}$ as 5 hours and 44.6 solar days respectively.

## 8. Calculation of observed LOM/LOD = ω/Ω = 27.3217

In E-M system LOM (length of month) = sidereal lunar month and LOD (length of day) = the sidereal day.

$$\frac{LOM}{LOD} = \frac{\omega}{\Omega} = 27.3217\ in\ modern\ times \qquad\qquad 29$$

In Section 7, LOM/LOD equation has been formulated in (23) and it is being restated here:

$$(N)^2 \times a^3 = X^2 + (F\sqrt{1-k^2})^2 \times (a^2)^2 + G^2 +$$
$$2\left(F\sqrt{1-k^2} \times a^2\right)(G)\left\{\sqrt{1-D^2}\sqrt{1-A^2} - AD\right\} + 2 \times X$$
$$\times \sqrt{\left(F\sqrt{1-k^2} \times a^2\right)^2 + (G)^2 + 2\left(F\sqrt{1-k^2} \times a^2\right)(G)\left\{\sqrt{1-D^2}\sqrt{1-A^2} - AD\right\}}$$
$$\times \left\{\sqrt{1-A^2}\sqrt{1-B^2} - A.B\right\} \qquad\qquad 23$$

If real world is considered then:

Using current Earth's obliquity (Φ = 23.44°), current Moon's orbital inclination (α = 5.14°) and current Moon's obliquity (β=1.54°) and k= 0.0549

We obtain the following trigonometric identities:

Sin[$\alpha$] = A=0.0895897 and Cos[$\alpha$] =√(1-A$^2$)= 0.995979;

Sin[$\beta$] = D=0.0268768 and Cos[$\beta$] =√(1-D$^2$)= 0.999639;

Sin[$\Phi$] = B=0.397784 and Cos[$\Phi$] =√(1-B$^2$)=0.917479;

$$\left\{ \sqrt{1-D^2}\sqrt{1-A^2} - AD \right\} = 0.993211 \; and \; \left\{ \sqrt{1-A^2}\sqrt{1-B^2} - A.B \right\} = 0.87815$$

Rewriting (23) and substituting the numerical values of the trigonometric identities we get:

$$(N)^2 \times a^3 = X^2 + (F \times 0.99849)^2 \times (a^2)^2 + G^2 +$$

$$2(F \times 0.99849 \times a^2)(G)\{0.993211\} + 2 \times X$$

$$\times \sqrt{(F \times 0.99849 \times a^2)^2 + (G)^2 + 2(F \times 0.99849 \times a^2)(G)\{0.993211\}}$$
$$\times \{0.87815\} \qquad\qquad 30$$

Solving (42) with numerical values of N, F and G and 'a' (the current semi-major axis) substituted we get the following quadratic equations:

$$17826.5 + 234.495X + X^2 - 24979.225 = 0 \qquad\qquad 31$$

The two roots of (31) are: -261.815 and 27.3199.

The negative root is rejected since both the spin of Earth and Moon and orbital motion are retrograde . Hence only 27.3199 is tenable.

We are having a perfect match with observed LOM/LOD

(23) is satisfied for the current epoch ω/Ω (LOM/lOD), α (Inclination angle) , β (lunar obliquity), Φ (terrestrial obliquity) and e (eccentricity).

9. **Evolution of inclination of Lunar orbital plane, eccentricity of Lunar orbit and obliquity of Moon's spin axis based on the information in Cuk et.al.(2016)**

The empirical relation describing the evolution of Moon's orbital plane inclination with respect to the ecliptic is (Appendix A1):.

$$Inclination \; angle \; \alpha = \frac{1.18751 \times 10^{25}}{a^3} - \frac{7.1812 \times 10^{16}}{a^2} + \frac{1.44103 \times 10^8}{a}$$
$$-8.250567342 \times 10^{-3} \qquad\qquad 32$$

The empirical relation describing the evolution of Moon's obliquity angle (β) is given as below (Appendix S2):

$$Moon's\ Obliquity\ angle\ \beta$$
$$= 3.36402 - 1.37638 \times 10^{-8}a + 1.32216 \times 10^{-17}a^2 \qquad 33$$

The empirical relation describing the evolution of Moon's orbit eccentricity is (Appendix A):.

$$e = 0.210252 + 8.38285 \times 10^{-10}a - 3.23212 \times 10^{-18}a^2 \qquad 34$$

### 10. The Determination of the evolutionary history of Earth's Obliquity from Advanced Kinematic Model of tidally interacting E-M system

From a previous personal communication arXiv: http://arXiv.org/abs/0805.0100
LOM/LOD of Earth Moon system is known over the tidal evolutionary history. It is tabulated in Table 1.

In Appendix S1, the evolutionary history expression have been derived for LOM/LOD and Earth's obliquity Φ (radians). They are as follows:

$$\frac{LOM}{LOD} = \frac{\omega}{\Omega} = -12.0501 + 2.6677 \times 10^{-7} \times a - 4.27538 \times 10^{-16} \times a^2 \qquad 35$$

$$\Phi = -0.732299 + 2.97166 \times 10^{-9} \times a \qquad 36$$

**Table 1. LOM/LOD and Earth's Obliquity for past geological epochs.**

| a(×R$_E$) | a(×10$^8$m) | LOM/LOD | Sin[Φ] | Φ (radians) | Φ° |
|---|---|---|---|---|---|
| 30 | 1.9113 | 23.3752 | -0.464076 | unstable | unstable |
| 35 | 2.22985 | 26.1194 | -0.216896 | unstable | unstable |
| 40 | 2.5484 | 28.1147 | 0.02137757 | 0.0213773 | 1.22483 |
| 45 | 2.86695 | 29.2938 | 0.113547 | 0.113792 | 6.51 |
| 50 | 3.1855 | 29.5965 | 0.218451 | 0.220227 | 12.6 |
| 55 | 3.50405 | 28.9877 | 0.309749 | 0.314929 | 18 |
| 60 | 3.8226 | 27.4 | 0.388198 | 0.398676 | 22.84 |
| 60.335897 | 3.844 | 27.32 | 0.397788 | 0.409105 | 23.44 |

Using (23)

$$(N)^2 \times a^3 = X^2 + (F\sqrt{1-k^2})^2 \times (a^2)^2 + G^2 +$$

$$2\left(F\sqrt{1-k^2} \times a^2\right)(G)\left\{\sqrt{1-D^2}\sqrt{1-A^2} - AD\right\} + 2 \times X$$

$$\times \sqrt{\left(F\sqrt{1-k^2}\times a^2\right)^2 + (G)^2 + 2\left(F\sqrt{1-k^2}\times a^2\right)(G)\left\{\sqrt{1-D^2}\sqrt{1-A^2}-AD\right\}}$$
$$\times\left\{\sqrt{1-A^2}\sqrt{1-B^2}-A.B\right\} \qquad 23$$

Obliquity angle is determined.

In (23) all constant and all spatial functions are known except the obliquity angle $\Phi$.

For a given lunar orbit , LOM/LOD is known. Using this information Sin[$\Phi$] is determined and hence $\Phi$ and tabulated in Table 1.

We have six set of data from a =30R$_E$ to the present day semi-major axis.

We clearly see that at Cassini State Transition, Earth's obliquity is indeterminate. From 40R$_E$ to 60.336R$_E$ it is well behaved and obliquity is increasing. It increases from 1.22483° to 23.44°. This means that during angular momentum conservative phase reduction in inclination is accompanied with increase in obliquity by necessity.

Table 2 gives the evolutionary history of  $\omega/\Omega$ (LOM/lOD),$\alpha$ (Inclination angle) , $\beta$ (lunar obliquity), $\Phi$ (terrestrial obliquity) and e (eccentricity)

**Table 2. evolutionary history of  $\omega/\Omega$ (LOM/lOD),$\alpha$ (Inclination angle) , $\beta$ (lunar obliquity), $\Phi$ (terrestrial obliquity) and e (eccentricity)**

| a ($\times$R$_E$) | a ($\times10^8$m) | $\omega/\Omega$ | $\alpha$ radians | $\beta$ | e | $\Phi$(rad) | Sin[$\Phi$] |
|---|---|---|---|---|---|---|---|
| 30 | 1.9113 | 23.3752 | 0.480685 (27.4°) | 1.21635 (69.69°) | 0.2524 | unstable | -0.464076 |
| 35 | 2.22985 | 26.1194 | 0.26478 (15.17°) | 0.952317 (54.56°) | 0.236 | unstable | -0.216896 |
| 40 | 2.5484 | 28.1147 | 0.168969 (9.68°) | 0.71512 (40.97°) | 0.214 | 0.0213773 | 0.0213757 |
| 45 | 2.86695 | 29.2938 | 0.124631 (7.1408°) | 0.504756 (28.92°) | 0.1849 | 0.113792 (6.51°) | 0.113547 |
| 50 | 3.1855 | 29.5965 | 0.103801 (5.04736°) | 0.321225 (18.4°) | 0.1493 | 0.220227 (12.6°) | 0.218451 |
| 55 | 3.50405 | 28.9877 | 0.0941394 (5.39379°) | 0.164527 (9.4267°) | 0.10714 | 0.314929 (18°) | 0.309749 |
| 60 | 3.8226 | 27.4 | 0.0898729 (5.149°) | 0.03466 (1.986°) | 0.0584 | 0.398676 (22.84°) | 0.388198 |
| 60.336 | 3.844 | 27.32 | 0.08971 (5.14°) | 0.0268 (1.54°) | 0.0549 | 0.409105 (23.44°) | 0.397788 |

In Figure 2, the evolution of Earth's obliquity (ɸ) based on AKM data (bold green) and based on Simulation data (dashed green) by Mutja Cuk et.al(2016) is given. We see the discontinuity at $45R_E$ .

.

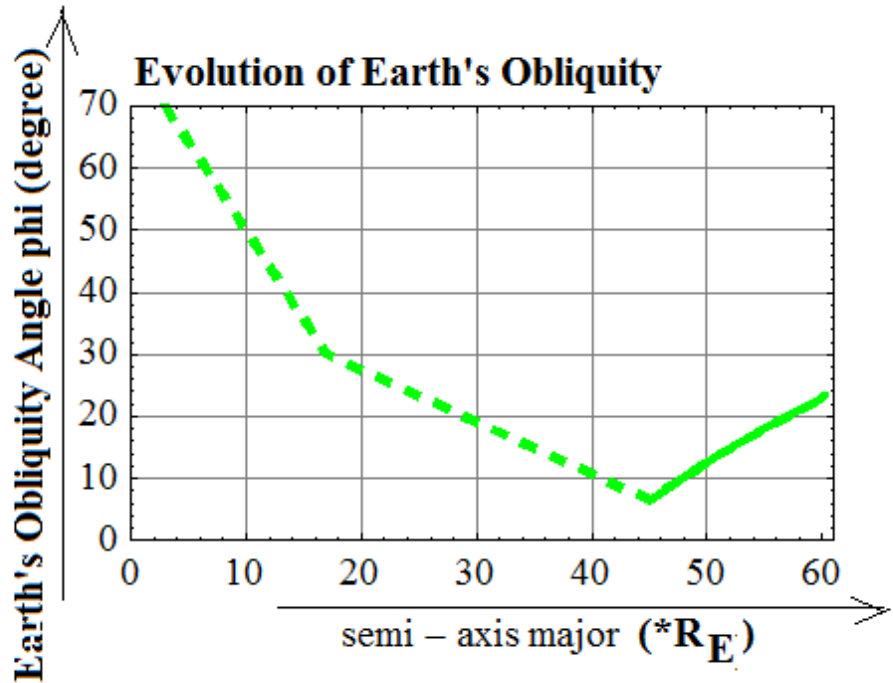

**Figure 2. Earth's Obliquity angle (ɸ°) evolution according to AKM (bold green) and according to Simulation results(dash green) (Cuk et.al.2016)**

In Figure 3, the evolution of Moon's orbital plane inclination (α) based on AKM (bold red) and based on Simulation done (dash red) by Matija Cuk et.al.(2016) is given. Here there is a continuity.

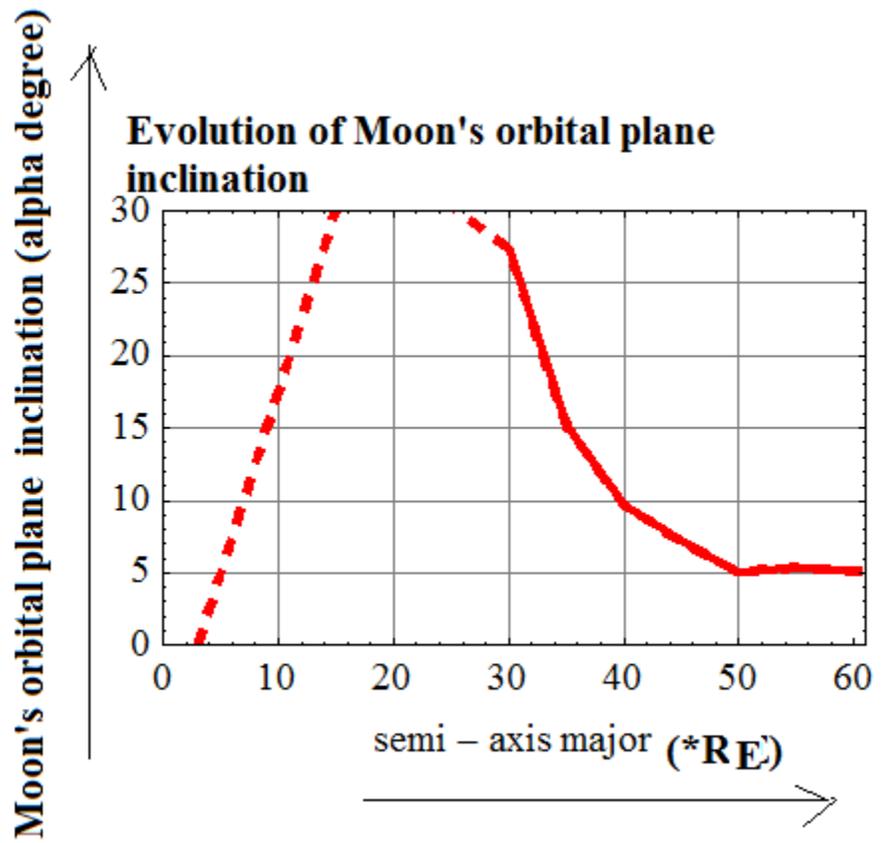

**Figure 3.Moon's orbital plane inclination (α°) based on AKM (bold red) and based on Simulation results (dash red) (Cuk et.al.2016)**

In Figure 4, the evolution of Moon's obliquity (β) based on AKM and based on Simulation done by Matija Cuk et.al.(2016) is given.

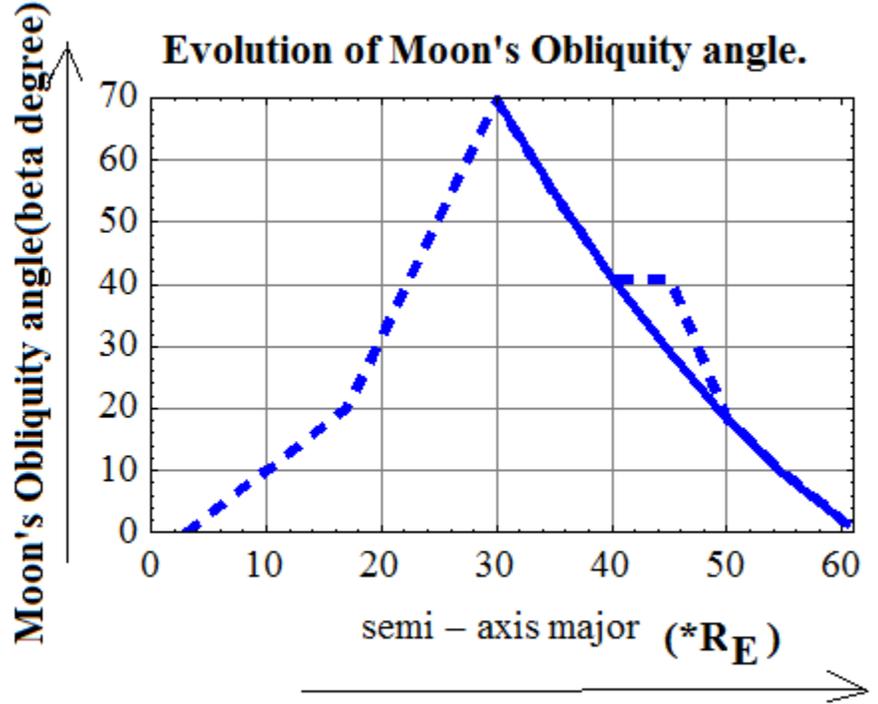

**Figure 4. Moon's Obliquity (β°) based on AKM (bold blue) and based on Simulation results (dash blue) (Cuk et.al.2016)**

**11.** Discussion.

Summing up the findings made by Cuk et.al and by this paper we see the following:

**Table 3. Four stages in tidal evolution of E-M system.**

|  | Post-Impact | Laplace Plane Transition | Cassini State Transition | Cass. State2 | Present |
|---|---|---|---|---|---|
| 'a' | $3R_E$ | $17R_E$ | $33.3R_E$ | $40R_E$ | $60R_E$ |
| Ecc. | Circular orbit | 0.5(excess J drained to heliocentric orbit) | 0.25 | 0.21 | 0.0549 |
| α | 0 | 35° | 28° | 27.54° | 5.14° |
| β | 0 | ? | 69.69° | 40.97° | 1.54° |
| Φ | 70° | 30° | ? | 1.22° | 23.44° |

Inspection of Table 3 leads us to a definite conundrum. After Cassini State 2 is reached, Earth-Moon system enters angular momentum conservative phase. If lunar obliquity tides generated by Cassini state transition help reduce inclination angle from 27.54° to 5.14°. Then by necessity of angular momentum conservation, obliquity must increase. This implies that current obliquity of 23.44° can be achieved only if obliquity angle is zero after Cassini State Transition.

This is obvious by the inspection of Figure 2, Figure 3 and Figure 4. As we see Figure 3 and Figure 4 give a continuity between AKM results and Simulation results in the evolution of inclination and Moon's obliquity data. The two results smoothly merge. But Figure 2 shows a discontinuity near 40R$_E$ for Earth's obliquity. To achieve 23.44° modern value of Earth's obliquity the Earth Moon system must achieve 0° Earth's obliquity just earlier than 40R$_E$ when Moon settles down in Cassini state 2.

Since angular momentum conservation is not required from Laplace Plane Transition to Cassini State Transition it is quite possible that strong obliquity tides are reducing inclination angle as well as Obliquity angle. Then only the climate friendly low obliquity can be achieved.

At this point , Cuk et.al.(2016) are completely quiet . This is a definite conundrum which needs to be addressed before we can assert that

"Our tidal evolutionary model supports high angular momentum, giant impact scenario to explain Moon's isotopic composition and provide a new pathway to reach Earth's climatically favourable low obliquity."

## 12. Conclusion.

This paper brings kinematic model renamed as advanced kinematic model (AKM) of tidally interacting binaries to a new level of maturity whereby it will prove to be more effective in dealing with real life scenario. There is dramatic improvement in the correlation of new model and the real world. In Section 6.4. an assumption has been made that:

$$in\ Figure\ 1, interior\ angle\ \ \widehat{a'} = \pi - \varphi - \widehat{b'} = \pi - \varphi - \alpha \qquad\qquad 37$$

How valid is this assumption over the entire permissible range (40R$_E$ to 60.33R$_E$) of AKM will have to be critically examined in a sequel paper. Cuk Muteja et.al (2016) have proposed that Earth-Moon system while passing through Laplace plane transition and Cassini state transition pass through chaotic and turbulent phase and due to strong obliquity tides in Moon the tidal evolution gets stalled or even reversed for long periods of its existence. E-M system moves in 'Fits' from 3R$_E$ to 17R$_E$ and subsequently to 51.4R$_E$ in 3.267Gy and then it 'Bounds' from 51.4R$_E$ to 60.33R$_E$ in 1.2Gy. At 17R$_E$ Laplace plane transition occurs and at 33R$_E$ Cassini state transition occurs. Cuk Matija have assumed that Moon is born from the Giant impact generated debris disk when Mars sized planetesimal made a glancing angle collision with proto Earth resulting in high obliquity and high Angular Momentum Earth. This resulted in isotopic identity of wide range of materials on Earth and Moon and the subsequent tidal evolution resulted in achieving climatically favorable Earth's obliquity of 23.44° . The application of AKM to this Fits and Bound model of E-M system gives a theoretical LOD curve which has precise match with observed LOD curve over last 1.2Gy as shown in SOM S6 using the Protocol Exchange algorithm http://org.1038/protex.2019.017 . In addition all the observed performance parameters are theoretically justified. The observed parameters are LOD = 24h, LOM/LOD= 27.32 and velocity of recession of Moon as 3.82±0,07cm/y all these are a natural corollary of AKM. Using the Protocol Exchange algorithim in SOM S6 AKM is vindicated on every count. In effect AKM has helped arrive at the correct theoretical formalism of Observed LOD curve. This theoretical formalism will give the datum against which the real time fluctuations in LOD will be compared and the precursors of the impending Earth-quake and sudden volcanic eruptions will be identified and used to give Early Warning and Forecasting for terrestrial disasters triggered by plate tectonic movements(see S7 in SOM).

**Acknowledgement:**

This continued research in E-M system has been made possible under Emeritus Fellowship Scheme sponsored by University Grants Commission, India. The Grant number is EMERITUS/2012-13-GEN-855/. I acknowledge the cooperation extended by the Director, IIT Patna for letting me use the resources at the central Library of the Institute. I acknowledge the cooperation extended by Prof. Pramod Kumar Mishra and Prof Pritam Kumar in EE Department , IIT Patna, for the use of their e-resources. in preparation of this paper. I also the acknowledge the cooperation extended by the Director of NIT, Patna, as well as HOD, Electronics and Communication Department, NIT, Patna, in their continued support of my ongoing Research Programme in both Electronics and in Celestial mechanics.

**Conflict of Interest:**

I have no conflict of interest financial or otherwise whatsoever with anybody.

# *High obliquity, high angular momentum Earth as Moon's origin revisited by Advanced Kinematic Model of Earth-Moon System.*

**Appendix A (all references are in the main text).**

## A1.1. Evolution of inclination of Lunar orbital plane, eccentricity of Lunar orbit and obliquity of Moon's spin axis based on the information in Cuk et.al.(2016)

*A1.1.1. Evolution of Moon's orbital plane inclination angle (α) from 30R$_E$ (Cassini State Transition orbit) to 60R$_E$ (current lunar orbit) based on Cuk et.al.(2016) inTable S2.1. gives the evolution of inclination angle.*

**Table A1.1.. Evolution of inclination angle (α) from 30R$_E$ to 60.336R$_E$ .**

| 'a'(× R$_E$) | 'a'(× $10^8$m) | α (°) | α (radians) |
|---|---|---|---|
| 30 | 1.9113 | 28 | 0.4887 |
| 35 | 2.22985 | 16 | 0.279 |
| 40 | 2.5484 | 9.2 | 0.16057 |
| 45 | 2.86695 | 8 | 0.1396 |
| 50 | 3.1855 | 7 | 0.122 |
| 55 | 3.50405 | 6 | 0.1047 |
| 60 | 3.8226 | | |
| 60.336 | 3.844 | 5.14 | 0.0897 |

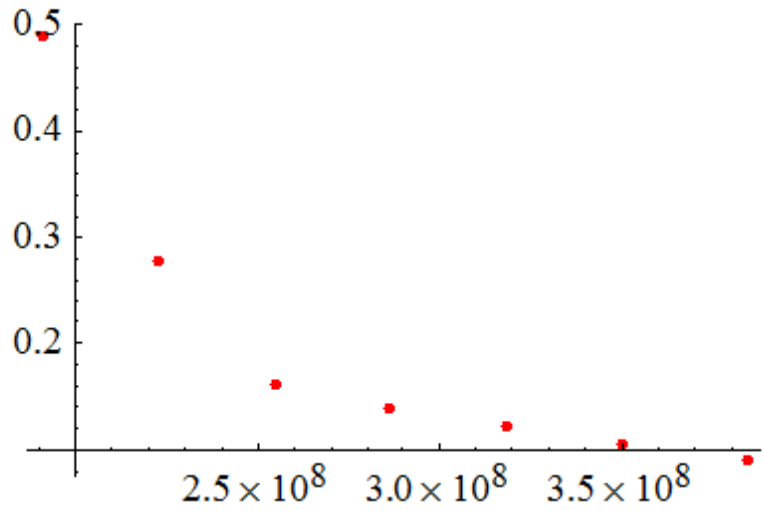

**Figure A1.1. ListPlot of the inclination angles in TableS1.1.**

The approximate FIT to the ListPlot in Figure S1.4.

$$Inclination\ angle\ \alpha$$

$$= \frac{1.18751 \times 10^{25}}{a^3} - \frac{7.1812 \times 10^{16}}{a^2} + \frac{1.44103 \times 10^8}{a} - 8.250567342$$

$$\times 10^{-3} \qquad A1.1.$$

Plot of (A1.1) is as follows:

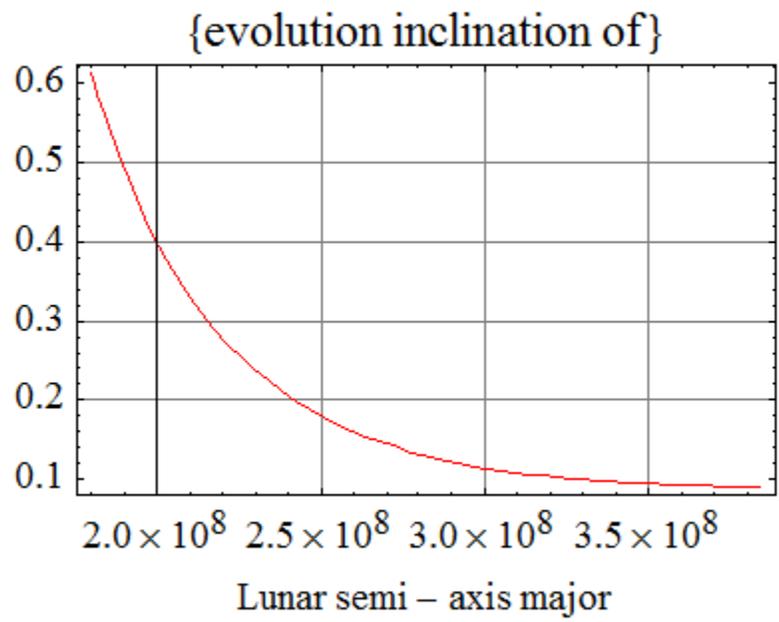

**Figure A1.2. Plot of the FIT function (S1.1).**

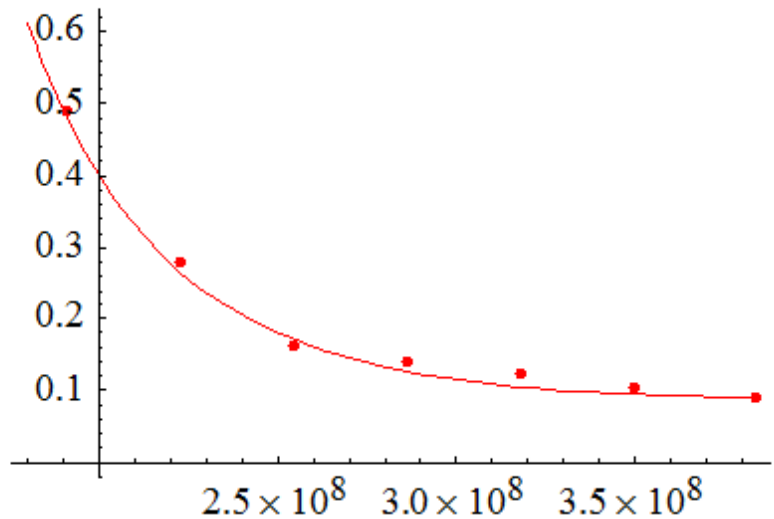

**Fig.A1.3.The superposition of the ListPlot and the FIT Plot.**

Figure A1.3.gives the correspondence between the ListPlot and (S1.1).
The correspondence is good hence (S1.1) gives the evolutionary history of Moon's orbital plane inclination angle in radians.

*A1.2. Evolution of Moon's Obliquity angle which currently is β =1.54°.*

Evolution of Moon's Obliquity angle (β) from 30$R_E$ (Cassini State Transition orbit) to 60$R_E$ (current lunar orbit) based on Cuk et.al.(2016).Table A1.2. gives the evolution of Moon's Obliquity angle.

**Table A1.2. Evolution of Moon's Obliquity angle(β) from 30$R_E$ to 60.336$R_E$ .**

| 'a'(× $R_E$) | 'a'(× $10^8$m) | β(°) | β(radians) |
|---|---|---|---|
| 30 | 1.9113 | 70 | 1.22 |
| 35 | 2.22985 | 55 | 0.96 |
| 40 | 2.5484 | 40 | 0.698 |
| 45 | 2.86695 | 29.27 | 0.5109 |
| 50 | 3.1855 | 19.25 | 0.336 |
| 55 | 3.50405 | 9.27 | 0.1618 |

| 60.336 | 3.844 | 1.54 | 0.0269 |
|--------|-------|------|--------|

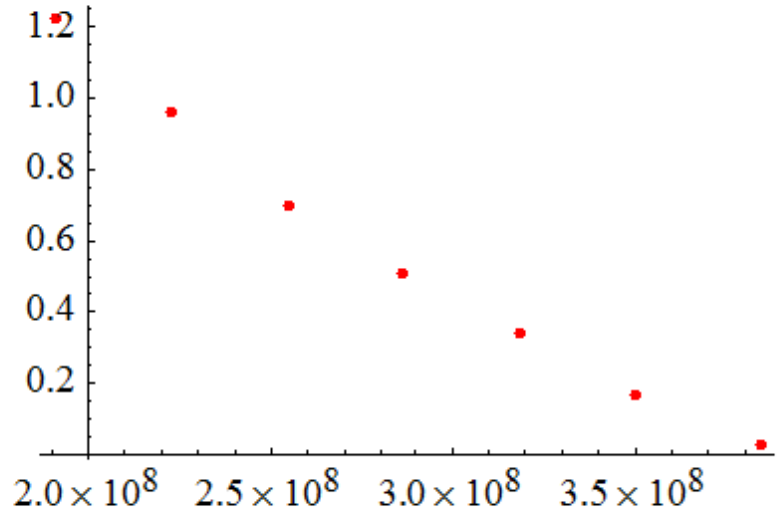

**Figure A1.4. ListPlot of the Moon's Obliquity angles given in Table S1.2..**

The approximate FIT to the ListPlot in Figure S1.4. is:

$$Moon's\ Obliquity\ angle\ \beta$$
$$= 3.36402 - 1.37638 \times 10^{-8}a + 1.32216$$
$$\times 10^{-17}a^2 \qquad (A1.2)$$

Plot of (A1.2) is as follows:

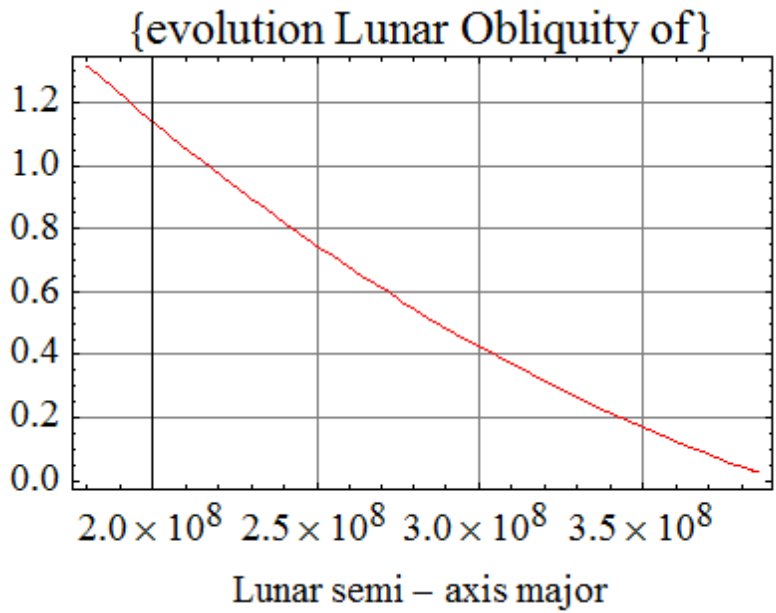

**Figure A1.5. Plot of the FIT function (A1.2).**

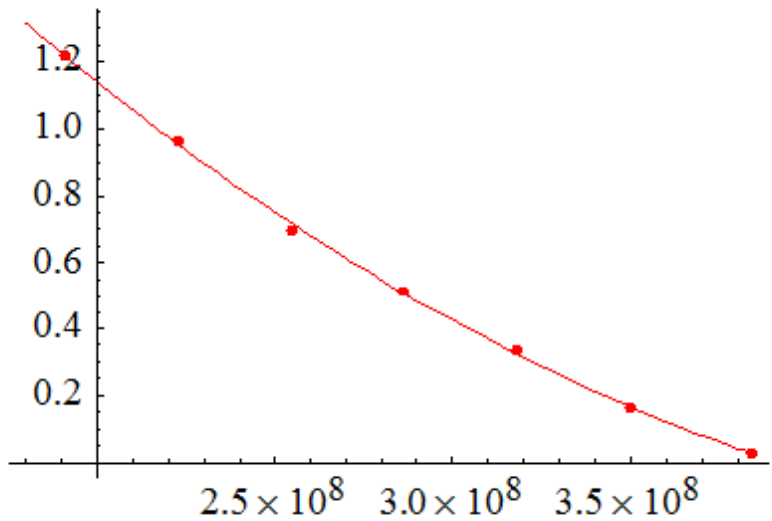

**Figure A1.6.Superposition of ListPlot and Fit Plot of Moon's Obliquity angle.**

Figure A1.6. Superposition of ListPlot and Fit Plot of the Moon's Obliquity Angle curve (A1.2).

The correspondence is good hence (A1.2) gives the evolutionary history of Moon's Obliquity angle in radians.

*A1.3. Evolution of Moon's orbit's eccentricity (e).*

About 80% higher angular momentum(AM) E-M system with highly tilted Earth was born after being impacted by Theia. The Moon accreted from the glancing angle impact generated well mixed Earth's mantle and impactors debris. As fully formed Moon spiraled outward it passed through Laplace Plane Transition ($r_L$) at 17$R_E$. The passage through ($r_L$) in highly oblique Earth's environment excited high eccentricity in Moon's orbit and high inclination of Moon's Orbital Plane. High eccentricity drained the excess AM to heliocentric Earth's orbit and Moon's orbit was circularized through Earth and Moon tidal interaction. Hence highly eccentric orbit excited by Laplace Plane transition was eventually circularized and synchronized. Table (S1.3) gives the evolution of Lunar Orbit's eccentricity (e) from a=30$R_E$ to 60.336$R_E$ .

**Table A1.3.. Evolution of Moon's orbit eccentricity from 30$R_E$ to 60.336$R_E$ .**

| 'a'($\times R_E$) | 'a'($\times 10^8$m) | e |
|---|---|---|
| 30 | 1.9113 | 0.25 |
| 35 | 2.22985 | 0.23 |
| 40 | 2.5484 | 0.21 |
| 45 | 2.86695 | 0.2 |
| 50 | 3.1855 | 0.15 |
| 55 | 3.50405 | 0.1 |
| 60.336 | 3.844 | 0.0549 |

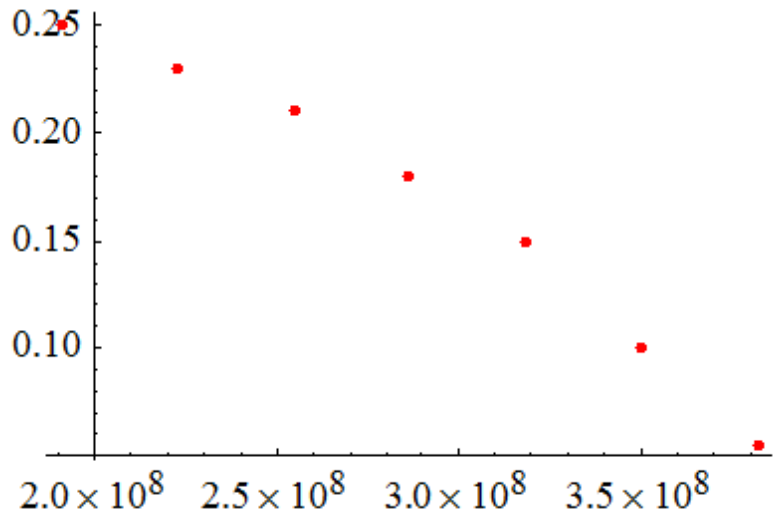

**Figure A1.7. ListPlot of the Moon's Obliquity angles given in Table A1.3..**

The approximate FIT to the ListPlot in Figure S1.7 is:

$$e = 0.210252 + 8.38285 \times 10^{-10} a - 3.23212 \times 10^{-18} a^2 \qquad (A1.3)$$

The Plot of (A1.3) is as follows:

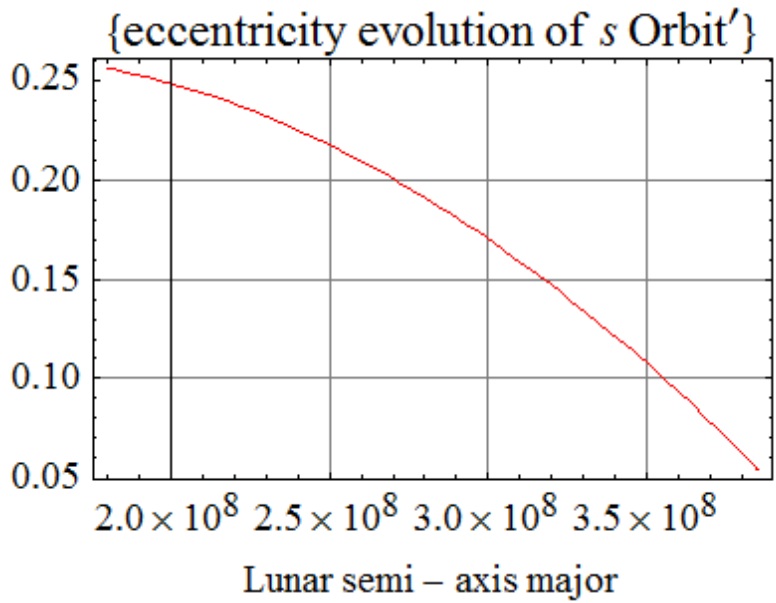

**Figure A1.8. Plot of Fit Function (A1.3).**

Superposition of eccentricity ListPlot and Fit Plot is given in Figure A1.9.

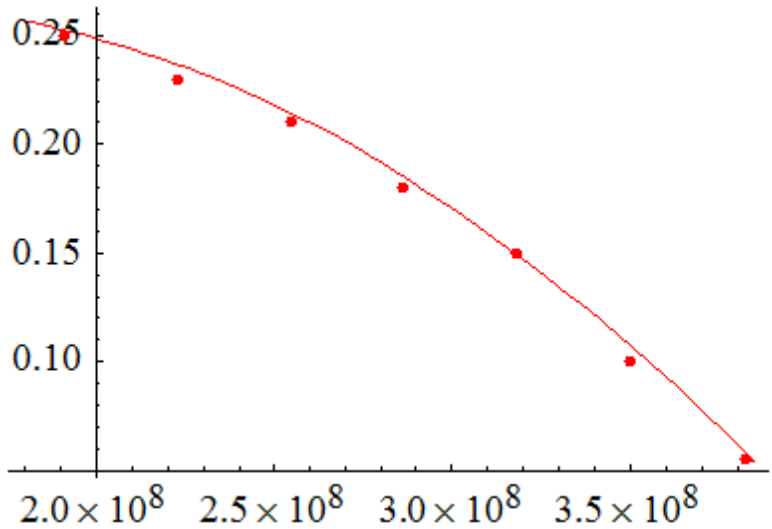

**Figure A1.9. Superposition of ListPlot and Plot of Moon's orbit eccentricity A1.3.**

The correspondence is good hence (S1.3) gives the evolutionary history of Moon's Orbit eccentricity.

**A1.4. The Determination of the evolutionary history of Earth's Obliquity from Advanced Kinematic Model of tidally interacting E-M system**

From a previous personal communication arXiv: http://arXiv.org/abs/0805.0100 LOM/LOD of Earth Moon system is known over the past geologic epochs.

$$\frac{LOM}{LOD} = \frac{1}{C}\left[\frac{J_T \times a^2}{Y \times B} - D - M \times a^2\right]$$

$$where\ D = 8.878598241 \times 10^{34} Kg - m^2,$$

$$M(reduced\ mass\ of\ Moon) = 7.258980539 \times 10^{22} Kg,$$

$$J_T = 3.44048888 \times 10^{34} \frac{Kg - m^2}{s}$$

$$Y = \left(a^{\frac{1}{2}} - A \times a^{\frac{5}{2}} + \frac{A \times a^{\frac{7}{2}}}{Z}\right), A = 9.14257051 \times 10^{-22} m^{-2}$$

$$Z = 5.52887891 \times 10^8 m,$$

$$B = 2.008433303 \times 10^7 \frac{m^{3/2}}{s}, C = 8.02 \times 10^{37} Kg - m^2; \qquad A1.4$$

LOM/LOD for various geological epochs are tabulated in Table A1.4..

**TableA1.4. LOM/LOD and Earth's Obliquity for past geological epochs.**

| a($\times R_E$) | a($\times 10^8$m) | LOM/LOD | Sin[Φ] | Φ (radians) | Φ° |
|---|---|---|---|---|---|
| 30 | 1.9 | 23.3752 | -0.464076 | unstable | unstable |
| 35 | 2.23 | 26.1194 | -0.216896 | unstable | unstable |
| 40 | 2.5484 | 28.1147 | 0.0213757 | 0.0213773 | 1.22483 |
| 45 | 2.867 | 29.2938 | 0.113547 | 0.113792 | 6.51 |
| 50 | 3.1855 | 29.5965 | 0.218451 | 0.220227 | 12.6 |
| 55 | 3.5 | 28.9877 | 0.309749 | 0.314929 | 18 |
| 60 | 3.82 | 27.4 | 0.388198 | 0.398676 | 22.84 |
| 60.335897 | 3.844 | 27.32 | 0.397788 | 0.409105 | 23.44 |

As Moon's orbital plane inclination gets damped from 27.54° to 5.14° under the influence of strong lunar obliquity tides, Earth's obliquity increases from 1.22483° to 23.44°.

Rewriting (23) from the main Text:

$$(N)^2 \times a^3 = X^2 + (F\sqrt{1-k^2})^2 \times (a^2)^2 + G^2 +$$

$$2\left(F\sqrt{1-k^2} \times a^2\right)(G)\left\{\sqrt{1-D^2}\sqrt{1-A^2} - AD\right\} + 2 \times X$$

$$\times \sqrt{\left(F\sqrt{1-k^2} \times a^2\right)^2 + (G)^2 + 2\left(F\sqrt{1-k^2} \times a^2\right)(G)\left\{\sqrt{1-D^2}\sqrt{1-A^2} - AD\right\}}$$

$$\times \left\{\sqrt{1-A^2}\sqrt{1-B^2} - A.B\right\} \qquad A1.5$$

In (A1.5) all constant and all spatial functions are known except the obliquity angle Φ.

For a given lunar orbit , X=LOM/LOD is known. Using this information Sin[Φ] is determined and hence Φ and tabulated in Table S1.4

We have six set of data from a =30R$_E$ to the present day semi-major axis.

We clearly see that at Cassini State Transition i.e. at 33R$_E$ , obliquity is indeterminate. From 45R$_E$ to 60.336R$_E$ obliquity is well behaved and it is increasing. It increases from 6.51° to 23.44°. This means that during angular momentum conservative phase i.e. from Cassini State Transition to the present epoch, reduction in Moon's plane inclination is accompanied with increase in obliquity by necessity.

*A1.4.1. Evolutionary spatial functions of terrestrial obliquity(Φ) and LOM/LOD*

Evolutionary spatial functions of inclination angle (α), Moon's obliquity(β) and of eccentricity 'e' have  been determined in CELE-D-17-00144 and given above. They are as follows:

$$Inclination\ angle\ \alpha$$
$$= \frac{1.18751 \times 10^{25}}{a^3} - \frac{7.1812 \times 10^{16}}{a^2} + \frac{1.44103 \times 10^8}{a} - 8.250567342$$
$$\times 10^{-3} \qquad A1.1$$

$$Moon's\ Obliquity\ angle\ \beta$$
$$= 3.36402 - 1.37638 \times 10^{-8}a + 1.32216 \times 10^{-17}a^2 \qquad A1.2$$

$$e = 0.210252 + 8.38285 \times 10^{-10}a - 3.23212$$
$$\times 10^{-18}a^2 \qquad A1.3$$

The LOM/LOD and Earth's obliquity angles are tabulated in Table A1.4..

*A1.4.1.1. Evolutionary function of LOM/LOD.*

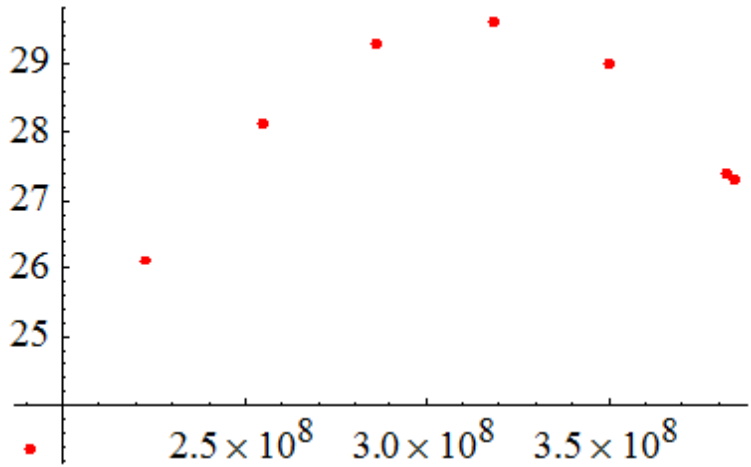

**Figure A1.10. ListPlot of LOM/LOD in different geologic epochs as given in Table A1.4. [Courtesy: Author]**

The approximate FIT function to the ListPlot of LOM/LOD in Tablee S1.14. is:

$$\frac{LOM}{LOD} = \frac{\omega}{\Omega} = -12.0501 + 2.6677 \times 10^{-7} \times a - 4.27538 \times 10^{-16} \times a^2 \qquad A1.6$$

The plot of (A1.6) is as follows:

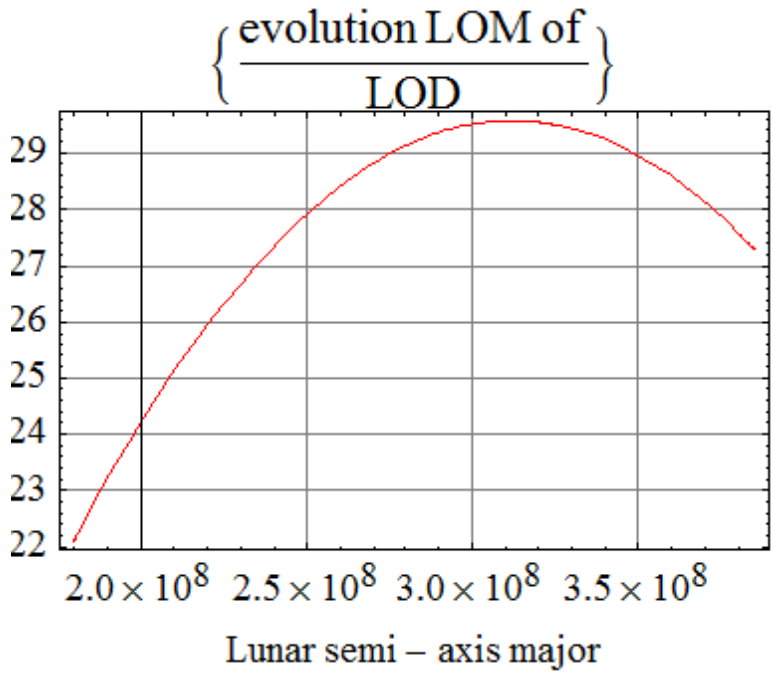

**Figure A1.11. Plot of FIT function given by (A1.6). [Courtesy: Author]**

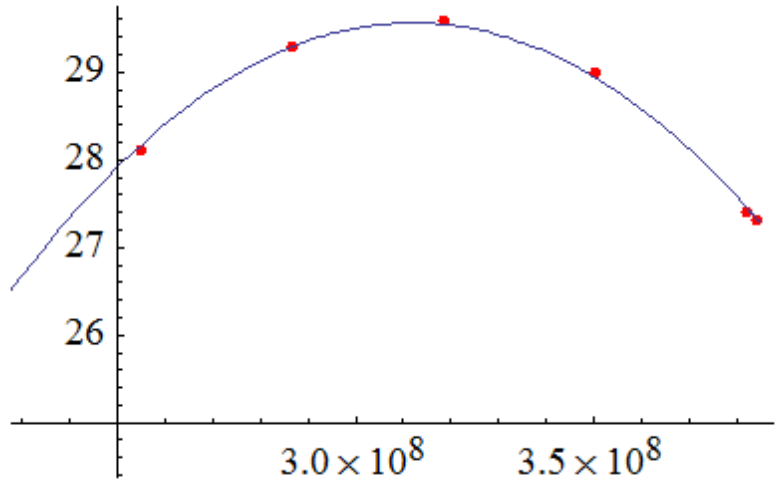

**Figure A1.12. Superposition of LOM/LOD ListPlot and FIT Plot. [Courtesy: Author]**

Superposition of LOM/LOD ListPlot and Fit Plot is given in Figure A1.12.

The correspondence between LISTPLOT and FIT PLOT is good hence (A1.6) gives the evolutionary history of LOM/LOD.

*A1.4.1.2. Evolutionary function of Earth's obliquity.*

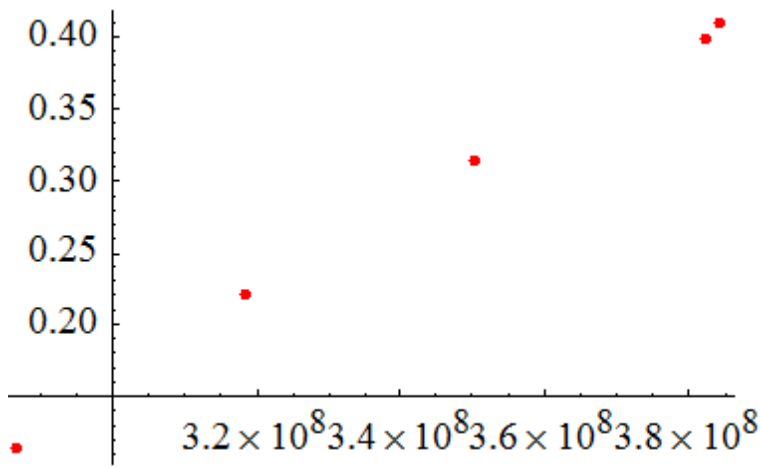

**Figure A1.13. List Plot of Earth's obliquity (Φ) angle over different geological epochs given in Table A1.4. [Courtesy: Author]**

The approximate FIT function to the ListPlot of Earth's obliquity in Figure S1.16 is:

$$\phi = -0.732299 + 2.97166 \times 10^{-9} \times a \qquad A1.7.$$

The Plot of (A1.7) is as follows:

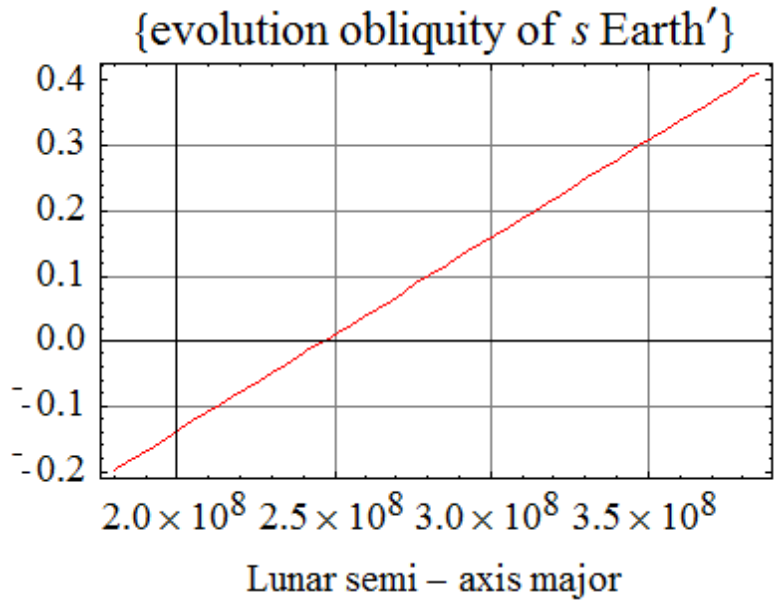

**Figure A1.14. Plot of FIT function given by (A1.7). [Courtesy: Author]**
Superposition of ListPlot and Fit function is given in Figure A1.14.

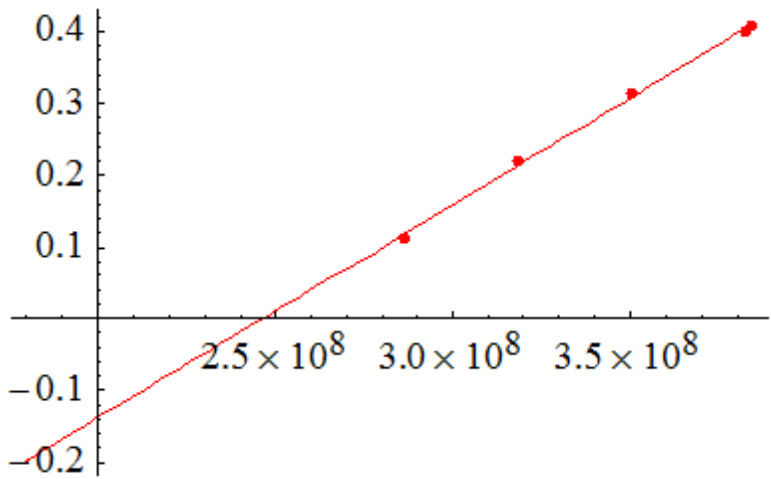

**Figure A1.15. Superposition of ListPlot of Earth's obliquity (Φ) and Fit Plot. [Courtesy: Author]**

The correspondence between LISTPLOT and FIT PLOT is good hence (S1.7) gives an accurate evolutionary history of Earth's obliquity .

We have altogether 5 spatial function (A1.1), (A1.2), (A1.3), (A1.6) and (A1.7) describing the evolution of inclination angle (α), Moon's obliquity (β), eccentricity(e) of lunar orbit, LOM/LOD and Earth's obliquity (Φ) respectively through different geologic epochs. These are tabulated in Table S1.5.

**Table A1.5. evolutionary history of ω/Ω (LOM/lOD),α (Inclination angle) , β (lunar obliquity), e (eccentricity) and Φ (terrestrial obliquity).**

| a (×R_E) | a (×10⁸m) | ω/Ω | α radians | β | e | Φ(rad) | Sin[Φ] |
|---|---|---|---|---|---|---|---|
| 30 | 1.9113 | 23.3752 | 0.480685 (27.4°) | 1.21635 (69.69°) | 0.2524 | unstable | -0.464076 |
| 35 | 2.22985 | 26.1194 | 0.26478 (15.17°) | 0.952317 (54.56°) | 0.236 | unstable | -0.216896 |
| 40 | 2.5484 | 28.1147 | 0.168969 (9.68°) | 0.71512 (40.97°) | 0.214 | 0.0213773 | 0.0123757 |
| 45 | 2.86695 | 29.2938 | 0.124631 (7.1408°) | 0.504756 (28.92°) | 0.1849 | 0.113792 (6.51°) | 0.113547 |
| 50 | 3.1855 | 29.5965 | 0.103801 (5.04736°) | 0.321225 (18.4°) | 0.1493 | 0.220227 (12.6°) | 0.218451 |
| 55 | 3.50405 | 28.9877 | 0.0941394 (5.39379°) | 0.164527 (9.4267°) | 0.10714 | 0.314929 (18°) | 0.309749 |
| 60 | 3.8226 | 27.4 | 0.0898729 (5.149°) | 0.03466 (1.986°) | 0.0584 | 0.398676 (22.84°) | 0.388198 |
| 60.336 | 3.844 | 27.32 | 0.08971 (5.14°) | 0.0268 (1.54°) | 0.0549 | 0.409105 (23.44°) | 0.397788 |

LLR measurement of 3.7cm/y was resulting in too short an age of Moon (~ 3Gy) which was contrary to the observed age of the rocks brought from Moon during Apollo Missions from 1969 to 1972 (curation/Lunar-NASA). These missions brought 382Kg of lunar rock, core samples, pebbles, sand and dust from the Moon surface. It is estimated that Moon's crust formed 4.4by ago. A team of scientist have studied Apollo 14 zircon fragments. They put the age of Moon at 4.51by (Barbanie et.al.2017). Matija Cuk,; Douglas P. Hamilton,; Simon J. Lock,; Sarah T. Stewart (2016) finally have resolved this conundrum. According to this research, from 3R_E to 45R_E , Moon does not have a smooth monotonic and spiral expansion. Infact it is bumpy. It is chaotic, gets stuck in resonances and comes out of the resonances and gets stalled and resumes its tidal evolution. In fact Moon takes 3.267Gy to spirally expand from 3R_E to 45R_E in fits and stalled manner. From 45R_E to 60.336R_E, Moon smoothly coasts in 1.2Gy. This accelerated spiral expansion in the on-going phase results in present day velocity of recession of 3.7cm/y. As we see this consistency with LLR results resolves a long standing problem of mismatch between observed LOD curve and theoretical LOD curve. In this new E-M model, a precise match is obtained between the theory and observation..

The series of papers in CELMEC VII and the main text have set the stage for Advanced KM to be established as a well tested tool for further applications in Space Dynamics. I also envisage the application of this model in earth-quake predictions (see S7 in SOM).

**A2.The algorithm for calculating the Transit Time from an earlier orbit to a later one.**

$$V(a) = \frac{2K}{m^*B} \times \frac{\sqrt{a}}{a^Q} [X - 1] \times 31.5569088 \times \frac{10^6 m}{y} \qquad A2.1$$

The value of the constants in Eq.(A2.1) are as follows:

$$K(structure\ constant) = 8.33269 \times 10^{42} N - m^Q, Exponent\ Q = 3.22684,$$

$$m^* = reduced\ mass\ of\ Moon = 7.256742697 \times 10^{22} Kg.$$

$$B = \sqrt{G(M + m)} = 2.008774813 \times 10^{42} \frac{m^{3/2}}{s}, a = lunar\ semi-major\ axis.$$

The given value of K(structure constant) and Q(exponent of the structure factor) ensure the modern day recession velocity of Moon as 3.82±0.07cm/y as measured in the current Lunar Laser Ranging Experiments (Dickey et.al.1994).

Eq.(23) in the main text is a quadratic equation in X = LOM/LOD. Eq.(23) is solved and two roots are obtained. One is negative and the other is positive. The positive root is retained. The value of lunar orbital inclination angle ($\alpha$ in radians) given in Eq.(31), the value of Moon's obliquity angle ($\beta$ in radians) given in Eq.(32), the value of Moon's orbit eccentricity described in Eq.(33) ,and function of terrestrial obliquity angle ($\phi$ in radians) are substituted in the expression of D and Z . This form of X is substituted in (S2.1).

(S2.1) is used for calculating the transit time from any earlier orbit to the present orbit $3.844 \times 10^8$m.

The transit time is given by the following time integral:

$$Transit\ time\ from\ a_1\ to\ a_2 = \int_{a_1}^{a_2} \frac{1}{V(a)} da; \qquad A2.2$$

***Supplementary-on-line material of High obliquity, high angular momentum Earth as Moon's origin revisited by Advanced Kinematic Model of Earth-Moon System[3].***

Corresponding Author: Bijay Kumar Sharma[4]

**Content of SOM:**

S1. **Back ground of this paper.**

S2. **Keplerian Era.**

S3. **The beginning of Evolutionist view of Universe.**

[4] Emeritus Fellow, National Institute of Technology, Patna 800005; email: bksharma@nitp.ac.in, Phone:+919334202848.

**S3.1. Spin down of the Primary in super-synchronous orbits:**

**S3.2. Spin up of the Primary in sub-synchronous orbits:**

**S3.3. Lock-in at geo-synchronous orbits (a$_{G1}$ and a$_{G2}$ ) :**

**S4. Kinematic Model of E-M System.**

**S4.1. What is Gravitational Sling Shot effect ?**

**S4.2. Deduction of outward radial velocity of Moon.**

**S4.2.1. Deductions of the kinematic parameter of E-M system.**

**S4.2.2. Deduction of LOM/LOD or ω/Ω equation.**

**S4.2.3. The Tidal Torque formalism and Radial Velocity formalism.**

**S4.3. The theoretical formalism of lengthening of day curve poorly validated by observed LOD curve.**

**S5. Appendix (A) and Appendix (B) of S3 and S4.**

**S5.1. Kinematic Model yields two geo-synchronous orbits of E-M system validated by Total Energy Analysis.**

**S5.1.1. To determine the energy extremum points from Total Energy Profile of E-M binary system.**

**S6. Theoretical Formalism of observed LOD within AKM framework.**

**S7.Gutenberg-Richter Law of Chaos.**

**S1. Back ground of this paper.**

On 20[th] July 1994, the author received a NASA Press release that through Lunar Laser Ranging experiment it had been ascertained that our Moon had receded by 1 m in last twenty five years from 20[th] July 1969 to 20[th] July 1994 which gives the recession rate of 3.82±0.07cm/y (Dickey et.al. 1994). This gave the author the second boundary condition for the second order ordinary linear differential equation which he had set for E-M tidally interacting system. The first boundary condition had been provided by George Howard Darwin 100 years earlier. He had calculated that Moon will finally lock-in with Earth at the outer Geo-Synchronous orbit in a triple synchrony state where Earth's spin = Moon's spin = Orbital period = 47days (Darwin 1889,1890). With these two boundary conditions the author was able to solve the differential equation and obtain a time integral which gave the age of Moon as 2.8 Gy and if the Author took the age of Moon as 4.467Gy (as it should be taken) then the lunar recession rate of 2.4cm/y is achieved. The rate of recession as measured by Lunar Laser Ranging experiment is anomalously high and leads to an Age of Moon (2.8Gy) which is completely contradicted by the dated lunar Basalts along with paleontological and sedimentological data. It was only after Matija Cuk et al (2016) published their paper that the Author was able to see that Earth-Moon tidal interaction has not been steady and monotonic. After Laplace Plane transition the tidal interaction was stalled and even reversed and Moon has evolved in Fits and Bound. Once this is taken care of the zig-saw puzzle falls in its place and theoretical formalism of observed LOD provided by John West Wells (1963,1966), Kaula and Harris(1975) and Charles P.Sonett(1998) is exactly matches the observed LOD curve.  It is correct that the present rate of tidal dissipation is anomalously high and this is because Moon is in accelerated tidally expanding spiral path. It is in an accelerated spiraling path because it lost time in halted tidal evolution at Laplace Plane Transition and at Cassini State transition. The first paper by the Author on Earth-Moon System was presented at 82nd Indian Science Congress (Sharma,1995). This work was further expanded to arrive at the theoretical formalism of lengthening of day curve and matching it with the observed LOD curve obtained by John West Wells (1963,1966), Kaula and Harris(1975) and Charles P.Sonnett (1998). This was published in arXiv as a personal communication of the Author : http://arXiv.org/abs/0805.0100  . In 2002 the author had discovered the dynamics of Earth-Moon tidally interacting pairs and the result was reported in World Space Congress held in Houston, Colorado, USA (Sharma and Ishwar,2002). By this time the author realized that George Howard Darwin talked about the outer geosynchronous  orbit only whereas in fact there was an inner geosynchronous orbit at 15,000Km and Roche's Limit (Ida, Canup & Stewart 1997) of 18,000Km fell beyond 15,000Km and hence when Moon was fully formed it was by necessity in a super-synchronous orbit and by gravitational sling shot it was catapulted on an expanding spiral orbital path which we witness today by Lunar Laser Ranging (LLR) and we are recording a  recession rate of 3.82±0.07cm/y by our Moon (Dickey et.al. 1994). In 2004 the author found that Earth-Moon results could be generalized to any tidally interacting pairs and planet-satellite dynamics was extended to Sun-planets system and presented 3 papers at 35th COSPAR Scientific Assembly held in 2004 in Paris. In this Assembly the author presented a NEW PERSPECTIVE on solar and exo-solar systems (Sharma & Ishwar 2004A. Sharma & Ishwar 2004B, Sharma, Ishwar, Rangesh 2009). The author extended this to several exo-solar systems and presented it as The Architectural Design Rules of Solar System at CELMEC V in 2009 in Italy (Sharma,2011). In 2012  the  author presented the Paper No. B0.3-0011-12 at 39th Scientific Assembly-2012 (Sharma,2012)  The correspondence between Newtonian formalism of synchronous orbit and kinematic formalism of Clarke's orbit or synchronous orbit (Clarke's Orbit is geo-synchronous orbit in E-M system) was found and graphically illustrated for vanishingly small mass ratios.  On 20th June 2016 the discovery of an infant planet has been reported (David et.al.,2016). The central tenant of the kinematic model is that planets are always born at inner Clarke's Orbit and from there they either get trapped in a collapsing death spiral as K2-33b is trapped or they get launched on an expanding spiral orbit as our Moon is. K2-33b gave us a rare opportunity to look at the birth orbit of the planets and it was exactly as predicted by kinematic model. The manuscript with the title **"Birth Orbit of K2-33b revealed by kinematic model of tidally interacting binaries" is under preparation**. Mars moons Phobos & Deimos have also been born in the inner Clarke's orbit of M-P-D system. This has been discussed in my manuscript under peer review. Author's predictions made in 35[th] Scientific Assembly (2004) are being confirmed by observational astronomy. All the planets - Giant first and terrestrial planets subsequently- are born at inner Clarke's Orbits as testified by IR imaging of the annular dark rings in circumstellar disc of many young stars (David et.al.2016) and also testified by meteoritic paleomagnetism measurements (Huapei et.al. 2017).

## S2. Keplerian Era

The Kepler's Third Law for a given Planet-Sun configuration is:

$$a^3 \Omega^2 = G(M + m) \qquad\qquad S2.1$$

Eq. (1) does not specify if the given orbital configuration is stable. Newton derived this law assuming that centripetal force $(GMm/a^2)$ = centrifugal force$(mv_{tang}^2/a)$ where a = semi-major axis of Earth-Moon orbital configuration, M = mass of the Earth and m = mass of our Moon. By implication it was assumed that all configurations predicted by (1) are stable. By the end of 19[th] century George Howard Darwin put a question mark on this stability by publishing two papers on E-M system (Darwin, 1879, 1880).

In 18[th] Century, German Philosopher Kant had suggested the theory of retardation of Earth's spin based on the ancient records of Solar Eclipses (Stephenson &Houldon, 1986; Stephenson 2003). Similar kind of studies had been carried out by Kevin Pang at Jet propulsion Laboratory at Pasadena (Morrison1978; Jong & Soldt 1989). He happened to step upon certain ancient records regarding Solar Eclipses. A total Solar Eclipse had been observed in the town of Anyang, in Eastern China, on June 5, 1302 B.C. during the reign of Wu Ding. Had Earth maintained the present rate of spin, the Eclipse should have been observed in middle of Europe. This implies that in 1302 B.C. i.e. 3,291 years ago Earth's spin period was shorter by 0.047 seconds. This leads to a slowdown rate of 1.428 seconds per 100,000 years.

In 1879 George Howard Darwin carried out a complete theoretical analysis of Earth-Moon System and put forward a sound hypothesis for explaining the slow down of Earth's spin on its axis. This marked the end of Keplerian Era. Gravitationally bound bodies were necessarily tidally interacting and tidal interaction led to tidal dissipation with inherent instability and hence a post-Keplerian physics was required to deal with gravitationally bound binary pairs. Tidally dissipative system because of loss of energy cannot be stable. The system will evolve to a minimum energy state which is a stable configuration by necessity.

**S3. The beginning of Evolutionist view of Universe.**

By mid 20[th] century it was increasingly felt that just as electrons had radiation-less stable permissible orbits in exactly the same way celestial body pairs have two triple synchrony orbits ($a_{G1}$ and $a_{G2}$) where they are conservative systems and no dissipation of energy is involved (Sharma, Ishwar & Rangesh,2009; Sharma, 2011; Krasinsky, 2002). Here triple synchrony orbits implies:

$$\omega(spin\ angular\ velocity\ of\ the\ primary) = \Omega(orbital\ angular\ velocity)$$
$$= \Omega'(spin\ angular\ velocity\ of\ the\ secondary \qquad\qquad S3.1.$$

The orbits of triple synchrony means geo-synchronous orbits in E-M system and Clarke's orbits in context of planet-satellite pairs, star-planet pairs, star-star pairs, neutron star-neutron star(NS) pairs and NS and BH (black hole) pairs. Here planet-satellite pairs, star-planet pairs and star pairs are non-relativistic systems. NS pairs, NS and BH pairs or BH pairs are relativistic systems. Relativistic systems are radiating gravitational waves and they are being driven towards coalescence hence they are always unstable. But non-relativistic systems are stable at outer triple synchrony orbits.

From George Howard Darwin's time it is recognized that planets raise body tides in their natural satellites and natural satellites raise body tides in their host planets. It is also recognized that planets and satellites are anelastic bodies (elastoviscous bodies). Hence tidal deformation (tidal stretching and squeezing) leads to dissipation of energy called tidal dissipation. This tidal dissipation causes mis-alignment of the tidal bulge and the radius vector of the secondary and the primary. Because of this mis-alignment an accelerating/or braking Tidal Torque is exerted on the primary body. By assuming different Love Numbers ($k_j$) and different Q parameter, different rate of tidal dissipation can be incorporated in the tidal interaction.

In Love Number, the subscript j is the harmonic degree, and kj is a proportionality constant, or Love number (Love,1911; Munk & MacDonald 1960). The Love numbers depend on internal structure of the deforming body, and reflect a competition between elastic and gravitational influences. If the elastic rigidity is sufficient, the body will deform very little, and the Love numbers will be near zero. If the gravitational effect dominates, the response will be purely hydrostatic. For a purely elastic body, the induced potential will be exactly aligned with the imposed potential, and there will be no torque, no dissipation, and no influence on the orbit. If there is dissipation, as would occur in a viscous or visco-elastic body, then the deformation (the tidal bulge) will lag behind the imposed potential in case of Sub-Synchronous Systems such as Mars-Phobos system and will lead the imposed potential in case of Super Synchronous Systems such as Earth-Moon. The rate of energy dissipation is proportional to the product of the stress times the strain rate, and will depend on the density, rigidity, viscosity, and rate of periodic forcing. From the tidal bulge lag angle , γ degree, the tidal torque as well as the Quality Factor can be determined. Quality Factor is the reciprocal of Tangent of tidal bulge lag angle which is taken as Q = 85.58 for Mars and Tidal Torque is proportional to the product of Tidal Amplitude and Sin (γ degree). Tidal interaction occurs if the tidal bulge of the primary has an angle (leading or lagging) with the radius vector of the secondary component.

### 3.1. **Spin down of the Primary in super-synchronous orbits:**

Tidal interaction inevitably leads to tidal drag (or secular deceleration) or spin down of the primary component if the satellite is long of synchronous orbit. Here the tidal bulge is leading the satellite's radius vector as in Earth-Moon (E-M) system. This is referred to as super-synchronous orbit as shown in Figure S3.1.

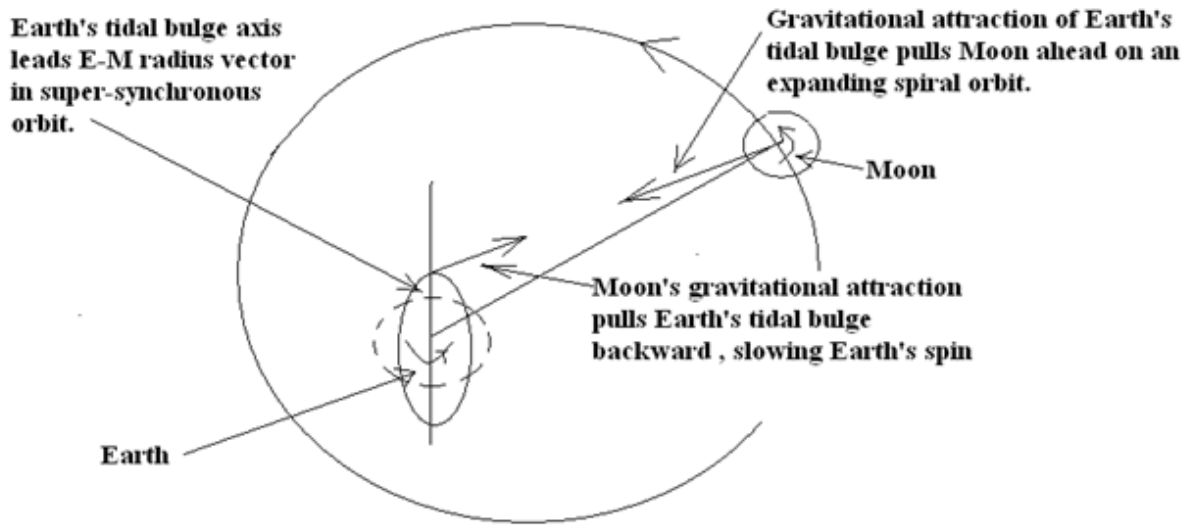

**Figure S3.1. In Earth-Moon System, Moon is in super-synchronous orbit. The off-setting of the line of bulge in Earth with respect to E-M radius vector creates a tidal drag and de-spinning of Earth leading to secular lengthening of the solar day. The de-spinning and consequent reduction in spin angular momentum of Earth leads to increase in the orbital**

angular momentum of E-M system in line with the conservation principle of total angular momentum of E-M system. Hence de-spinning of Earth is accompanied with expanding spiral orbit of Moon. During the conservative phase of E-M system, when Moon is just born at inner geo-synchronous orbit and the system is at near triple synchrony, by gravitational sling-shot impulsive torque Moon is catapulted on an expanding spiral path towards outer geo-synchronous orbit. After the conservative phase and during the dissipative phase there is no energy transfer from Earth spin to E-M expanding orbital system. Moon on its own, by virtue of the initial energy acquired during the conservative phase, coasts towards the outer geo-synchronous orbit where it gets finally locked-in. [Courtesy : Author]

### 3.2. Spin-up of the Primary in sub-synchronous orbits:

If the tidal bulge is lagging the radius vector of the secondary component then it is sub-synchronous orbit. In this configuration tidal secular acceleration or spin-up of the primary component occurs as in Mars-Phobos (M-P) system as shown in Figure 2.

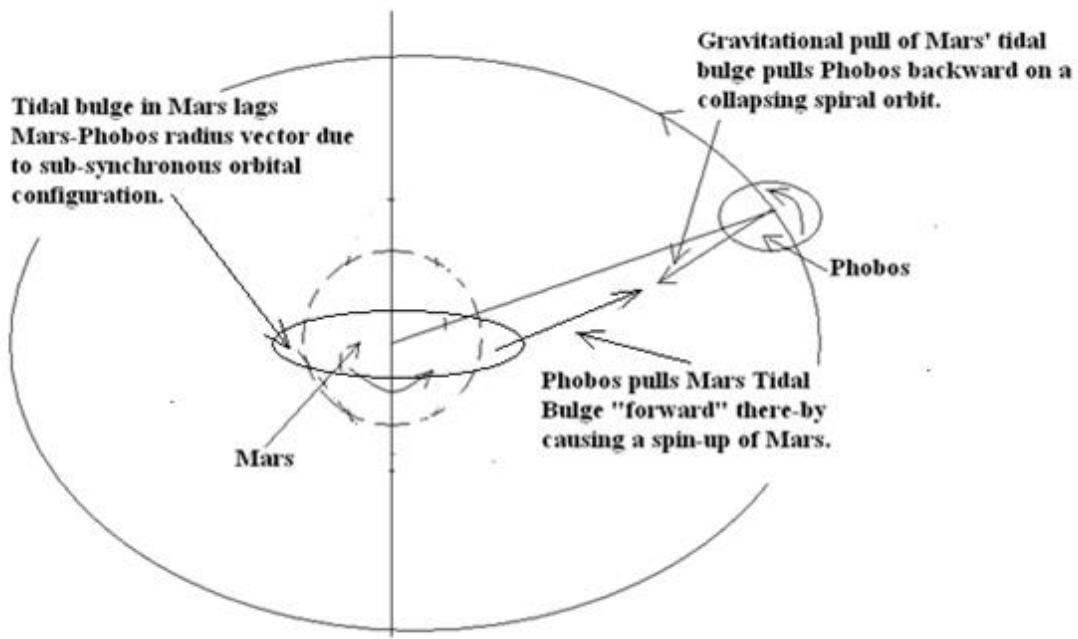

**Figure S3.2. In Mars-Phobos System, Phobos is in sub-synchronous orbit.The tidal bulge in Mars lags M-P radius vector hence Phobos is spinning up Mars. Conservation of angular momentum causes Phobos to be launched on gravitational runaway collapsing spiral orbit also known as death spiral.[Courtesy: Author].**

### 3.3. Lock-in at geo-synchronous orbits ($a_{G1}$ and $a_{G2}$ ) :

It is zero tidal interaction if the two bodies are tidally interlocked. When the primary and secondary are tidally interlocked, the lag angle/lead angle become zero and the system is a conservative system. This is Triple Synchrony state as defined by (2) and it is called geo-synchronous orbit in E-M system and it will be referred to as Clarke's orbits in all other binary pairs. As we will see all binary pairs have two Clarke's orbits, inner and outer Clarke's orbit. Inner Clarke's orbit is an equilibrium orbit but it is energy maxima hence unstable orbit. Outer Clarke's orbit is also an equilibrium orbit but energy minima and hence stable orbit (Sharma, Ishwar & Rangesh,2009; Sharma, 2011;Krasinsky, 2002). A binary pair originates at inner Clarke's Orbit. Any perturbation such as solar wind, cosmic particles, star dust or radiation pressure perturbs the secondary component to a new orbit within or beyond $a_{GI}$ .

Figure S3.1 illustrates lead angle in E-M system and Figure S3.2. illustrates the lag angle in M-P system. E-M system is in super-synchronous configuration that is Earth's tidal bulge leads Moon radius vector. Earth is experiencing a tidal brake by Moon and Moon is being pushed outward with a recession velocity of 3.82($\pm$0.07)cm/y as established by Apollo Mission 11 initiated Laser Ranging Experiment (Dickey et.al.1994). Figure S3.2 illustrates that M-P is in sub-synchronous configuration that is Mars' tidal bulge lags Phobos radius vector. Mars is being accelerated or spun-up by Phobos and Phobos is trapped in a death spiral i.e. it is traversing a collapsing spiral. In about 10My to 20My it is doomed to be tidally pulverized and spread like a ring around Mars just like Jupiter's or Saturn's ring (Black & Mittal 2015).

In perfect tidal lock-in position, the long axis of the tidal bulge of primary and secondary components are exactly aligned and both the components orbit the barycenter as one single body as is the case with Pluto-Charon. This lock-in pair was video photographed by New Horizon. in NH Lorri OPNAV Campaign.(Sharma and Ishwar 2004A). On 14th July 2015, Pluto-Charon were at the point of closest approach to New Horizon space probe and Pluto_Charon were seen making circular paths around the barycenter which lay outside the two globes. The tidal stretching and squeezing completely stops and hence tidal dissipation is zero. This perfect lock-in occurs when the two components are synchronized, the orbit of each component around the barycenter are circularized and the orbital planes of the two components are co-planer. This observation in reference to stellar binaries had been made by Zahn (1992,1975,1977a,1977b) :

"Eventually the Binary may settle in its state of minimum kinetic energy, in which the orbit is circular , rotation of both stars is synchronized with the orbital motion and the spin axis are perpendicular to the orbital plane. Whether the system actually reaches this state is determined by the strength of tidal interaction, thus by the separation of the two components, equivalently the orbital period. But it also depends on the efficiency of the physical process which are responsible for the dissipation of the kinetic energy."

Mars-Phobos is the example of sub-synchronous Satellite where Mars-Phobos Radius vector leads the tidal bulge in Mars (Sharma.& Ishwar2004A) , Phobos spins-up Mars and Phobos loses altitude. Mars spins-up because of transfer of angular momentum and orbital energy from Phobos to Mars. Earth-Moon is the example of super-synchronous Satellite where Earth-Moon Radius vector lags the tidal bulge in Earth, Moon is spinning down Earth (Earth's spin is slowing down at (2.3±0.1) ms per century(Stephenson& Morrison.1995). Earth was spinning at 5h per day, today it is spinning at 23.9344h per day and in its final lock-in orbit it will be spinning at 47d per spin period) and Moon is receding at 3.82(±0.07)cm/y presently(Dickey et.al. 1994). Here angular momentum is being transferred from the Earth to our Moon. Pluto-Charon is the example of tidally interlocked orbital configuration where the tidal bulge of both the components are aligned and both the components are orbiting the barycenter as one body in a perfect circle(Sharma.& Ishwar 2004A).

X. Shi, K. Willner and J. Oberst (2013) give the following tidal evolution equation:

$$a_t = \left[ a_0^{\frac{13}{2}} - \frac{13}{2} \times \frac{3k_2}{Q} \sqrt{\frac{G}{M}} \times mR^5 \Delta t \right]^{\frac{2}{13}} \qquad S3.2$$

$$where\ k_2 = Love\ Number\ of\ Mars, Q\ is\ the\ quality\ factor,$$

$$M\ and\ R\ are\ the\ mass\ and\ radius\ of\ of\ Mars,$$

$$m = mass\ of\ Phobos, a_0 = current\ semi-major\ axis.$$

This equation gives the orbital radius = $a_t$ at a time of $\Delta t$ seconds ago.

In Equation (S3.2), Love Number and Quality Factor depend upon density, rigidity, viscosity and rate of periodic forcing. These parameters are known with large uncertainties for different Planets and their Satellites and hence their Tidal Evolutionary History will be arrived at with equal uncertainty in Seismic Model based analysis.

## 4. Kinematic Model of E-M System

As already noted in the last section, analysis of Seismic model requires the knowledge of a large number of material properties of the celestial objects which are known with a lot of uncertainties. Hence Author developed Kinematic Model of E-M system which required only the globe-orbit parameters and the age of E–M system. These are known with a high confidence level..

Assumptions of Kinematic Model:

(1) E-M System is regarded as 3-body rotating system from its birth to its terminal point at the second geosynchronous orbit. The tidal drag of Sun for Lunar orbital radius greater than ten times Earth's radius is implicit in KM treatment. This is implicit in Advanced Kinematic Model(AKM) also (in the main text).

(2) Total angular momentum of E-M System has been assumed to be the scalar sum of the orbital angular momentum of E-M system, spin angular momentum of Earth and spin angular momentum of Moon. For mathematical tractability this approximation is made.

In the real world the obliquity angle($\phi$) or tilt angle of Earth's spin axis with respect to the Ecliptic plane normal, is 23.44º, the inclination angle($\alpha$) of Moon's orbital plane with respect to the ecliptic plane is $\alpha$ = 5.14º and axial tilt of Moon's spin axis with respect to Ecliptic Normal is $\beta$ (Moon's Obliquity angle) = 1.54º make the total angular momentum of E-M system ($J_{Total}$= $J_4$) the vector sum of orbital angular momentum ($J_0$), Moon's spin angular momentum ($J_1$) and Earth's spin angular momentum ($J_2$) .

By making the above two assumptions the calculation becomes tractable but due to scalar summation serious errors are introduced. In the present paper an exact analysis is done by including the obliquity of Earth , orbital inclination of Moon and obliquity of Moon and by including the tidal drag of Sun at Laplace Plane transition and beyond, At 23-24$R_E$ Earth had become rigid enough to retain the oblateness it had acquired at that point and then onward the moment of inertia has remained constant at its modern value namely C = $8.02 \times 10^{37}$Kg-m$^2$ . Hence evolving oblateness is not considered. Advanced Kinematic Model (AKM) considers the total angular momentum as the vectorial sum of Earth's spin angular momentum, Moon's spin angular momentum and orbital angular momentum of E-M system. After the semi-major axis reaches the Laplace Plane transition orbit namely a = 17$R_E$ then solar perturbation comes into picture and it has to be accounted and it is implicit in AKM treatment

Now we calculate the residual acceleration of Moon which according to Newton should be zero but is not zero and this precisely in the reason of instability outside geo-synchrony of E-M system:

$$Present\ centripetal\ acceleration = \frac{GM}{a_{present}^2} = 2.69756 \times 10^{-3} \frac{m}{s^2} \qquad S4.1.$$

$$Present\ centrifugal\ accleration = (\Omega_L)^2 \times a_{present} \times \frac{1}{1+\frac{m}{M}}$$

$$= 2.69026 \times 10^{-3} \frac{m}{s^2} \qquad S4.2.$$

Moon is effectively undergoing a radial deceleration of $0.0073\dot{x}10^{-3}$ m/sec$^2$. Its outward radial velocity is being decelerated until it becomes zero at the outer geo-synchronous orbit where it is in a triple synchrony state with as shown in Figure S4.1.:

$$\omega = \Omega = \Omega' = \frac{2\pi}{47d} \hspace{3cm} S4.3.$$

Either it will remain stay put in this outer geo-synchronous orbit or it will be deflected back on an inward collapsing spiral orbit due to Sun's perturbation. Sir James Jeans(1936) suggested that when our Moon will reach outer geo-synchronous orbit $a_{G2}$ then Moon will be orbiting Earth in 47 days, Earth will be spinning in 47 days but its orbital period around Sun will be more than 365.25d hence Earth will try to synchronize i.e. its spin will be try to equalize with orbital period hence Earth's spin period will further lengthen. As Earth's spin period becomes longer than 47 d , our Moon will fall in sub-synchronous orbit with respect to the Earth hence Moon will get trapped in a death spiral and eventually collapse into Earth. But much earlier Sun would have burnt all its fuel and become a Red Giant expanding to engulf Earth-Moon system (Schroder & Smith.2008.). Here a pertinent question arises.

Why does Moon recede from Earth when it is trapped in gravitational potential well created by the Earth? Where does the energy come for climbing up the potential well? The answer is 'Gravitational Sling Shot effect'.

**4.1. What is Gravitational Sling Shot effect ?**

Planet fly-by, gravity assist is routinely used to boost the mission spacecrafts to explore the far reaches of our solar system (Dukla et.al.2004;Jones,2005; Epstein, 2005.;Cook, 2005). Voyager I and II used the boost provided by Jupiter to reach Uranus and Neptune. Cassini has utilized 4 such gravity assists to reach Saturn. New Horizon space craft has used a similar Jupiter fly-by gravity assist in its interplanetary journey. By Jupiter flyby-gravity assist the journey to Pluto has been shortened by 3 years.

In E-M system, planet fly-by gravity assist maneuver a space-craft which passes " behind" the moon gets an increase in its velocity(and orbital energy) relative to the primary body. In effect the primary body launches the space craft on an outward spiral path. If the spacecraft flies "infront" of a moon, the speed and the orbital energy decreases. Traveling "above" and "below" a moon alters the direction modifying only the orientation (and angular momentum magnitude). Intermediate flyby orientation change both energy and angular momentum. Accompanying these actions there are reciprocal reactions in the corresponding moon.

The above slingshot effect is in a three body problem. In a three body problem, the heaviest body is the primary body. With respect to the primary body the secondary system of two bodies are analyzed.

In case of planet flyby, planet is the primary body and the moon- spacecraft constitute the secondary system.

**While analyzing the planetary satellites, Sun is the primary body and planet-satellite is the secondary system. But in our analysis, Sun has been neglected. This results in errors leading to erroneous LOD formalism which has a poor match with observed LOD curve given in Table 1 of the main text. In fact the general trend of evolution of our Moon has been misanalysed in** http://arXiv.org/abs/0805.0100.**.(Sharma 2005) In my personal communication Laplace Plane transition and Cassini State Transition has not been taken into account.**

Gravitational Sling Shot phenomena launched Moon on its Non-Keplerian Journey from (inner geo-synchronous orbit) $a_{G1}$ to (outer geo-synchronous orbit) $a_{G2}$ .

At inner and outer Geo-Synchronous Orbits, the Satellite is in Keplerian Orbit where centripetal and centrifugal forces are in equilibrium and radial acceleration and radial velocity are zero. But the Satellite is never allowed to stay in the inner Keplerian Orbits because it is energy maxima state as has been shown in subsequent section.

At the inner Geosynchronous Orbit slightest differential between $\omega$ and $\Omega$ due to solar wind, cosmic particles or radiation pressure perturbation causes the Satellite to tumble out of the Keplerian Orbit. If the Satellite is long of $a_{G1}$ it is launched on an outward expanding spiral path as our Moon is and if it is short of $a_{G1}$ it is injected into an inward collapsing spiral path as Phobos (Martian Satellite) is launched.

Initially at $a_{G1}$ both Energy Conservation and Angular Momentum Conservation are maintained. Hence as soon as the Satellite tumbles out of the inner Geo-Synchronous Orbit it enters a Gravitational Runaway Phase.

If it falls long of $a_{G1}$, the Satellite experiences a powerful sling-shot effect because of rapid transfer of Planet's Spin Rotational energy to Satellite's Orbital energy. This causes an outward radial acceleration peaking at $a_1$.(as shown in Figure 3.) The powerful sling-shot effect is like an impulsive torque.

The sling-shot phase or Gravitational Runaway phase is damped out as the differential between $\omega$ and $\Omega$ grows and tidal dissipation increases. This leads to the termination of the Gravitational Runaway Phase at $a_2$ where the Radial Acceleration becomes zero and Recession Velocity becomes maximum and where lom/lod = 2. This point is also referred to as Gravitational Resonance point or 2:1 Mean Motion Resonance(MMR) orbit (Rubicam.1975; Ward &

Canup2000). Thereafter the Satellite coasts along the outward spiral path on its own. Beyond $a_2$, Radial acceleration is negative and Radial outward Velocity is continuously decelerated until it becomes zero at $a_{G2}$ .

In Figure 3 and Figure 4, the radial velocity profile and radial acceleration profile have been illustrated.

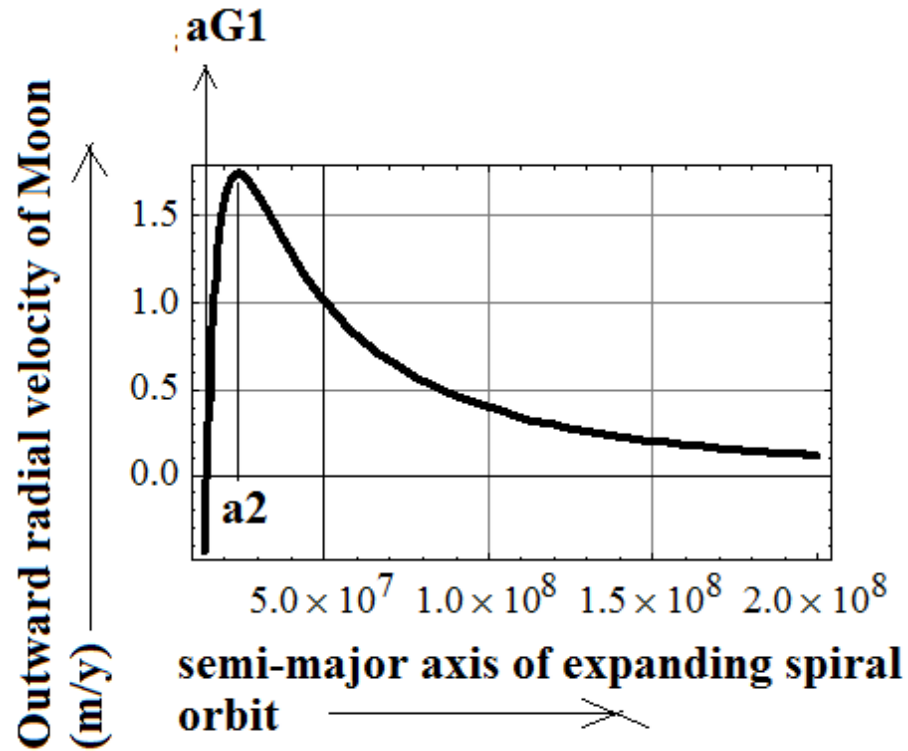

**Figure S4.1. Radial Velocity Profile of Moon from the birth to the final lock-in point at $a_{G2}$. X-axis is semi-major axis. Y-axis is velocity in m/y.[Courtesy : Author]**

The profile is drawn from $1 \times 10^7$ m to $1 \times 10^8$ m.

At $a_{G1} = 1.46 \times 10^7$ m and at $a_{G2} = 5.5335 \times 10^8$ m, the radial velocity = 0. These are the exact Keplerian equilibrium points. These are the triple synchrony orbits where the two components are tidally interlocked.

At $a_2$ (2:1 Mean Motion Gravity Resonance orbit) where Radial Velocity is maximum and Radial acceleration is zero as shown in Figure 4.

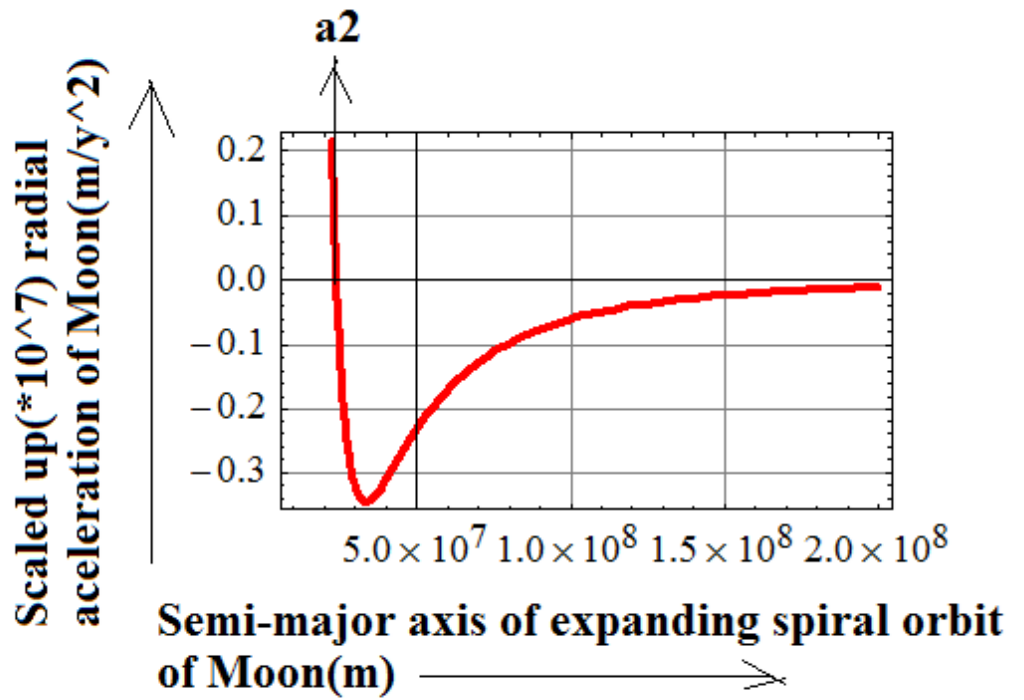

**Figure S4.2. Scaled up (×10⁷ times) Radial Acceleration Profile of Moon from birth to the final lock-in at aG2**

X-axis is semi-major axis (m). Y-axis is acceleration in m/y². [Courtesy: Author]

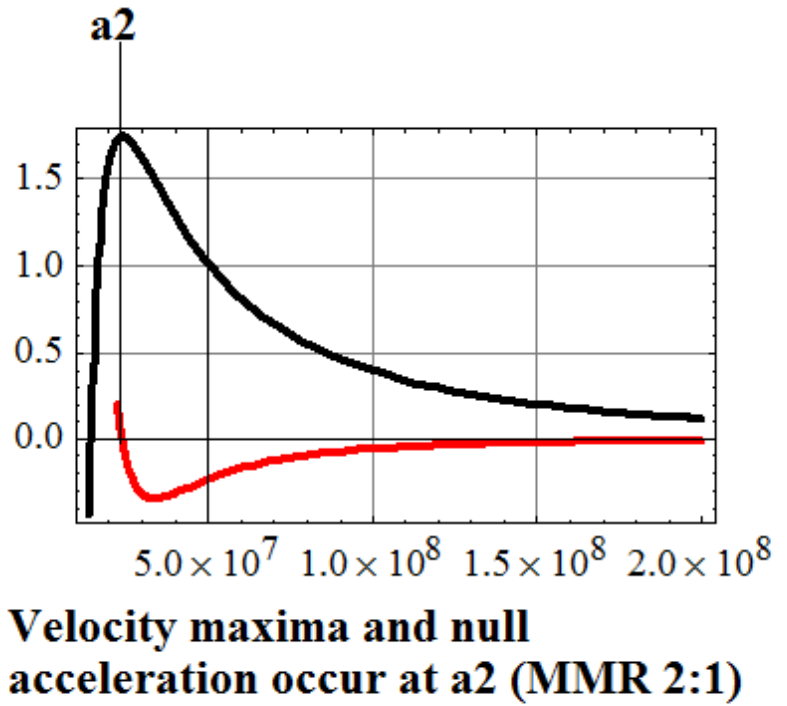

**Velocity maxima and null acceleration occur at a2 (MMR 2:1)**

**Figure S4.3. Superposition of Radial Velocity Profile and Scaled up (×10$^7$ times) Radial Acceleration Profile of Moon from birth to the final lock-in at a$_{G2}$ . Note the coincidence of Radial Velocity maxima and null radial acceleration at a$_2$ (MMR 2:1 orbital radius).[Courtesy: Author]**

Short of a$_{G1}$ the natural satellite is accelerated inward. Long of a$_{G1}$ from a$_{G1}$ to a$_2$ (2:1 MMR gravity resonance orbit) the Moon is rapidly accelerated outward under the influence of an impulsive gravitational torque due to rapid transfer of spin rotational energy from Earth to Moon. The maxima of the outward radial acceleration occurs at a$_1$. (This is the peak of the impulsive sling shot torque). From a$_1$ to a$_2$ the torque decays to zero because of the differential between $\omega$ and $\Omega$ leading to tidal dissipation. Hence up to a$_2$ Moon is accelerated to a maximum velocity V$_{max}$. Beyond a$_2$ , Moon climbs up the gravitational potential well and in the process gets decelerated . It reaches zero velocity at a$_{G2}$. At this terminal point, it will remain stay put as Charon is stay put with respect to Pluto or it will spiral in because of third body perturbation. Moon will spiral-in because of Sun's perturbation (Jeans 1936).

**S4.2. Deduction of outward radial velocity of Moon.**

In Kinematic Model, any binary system has two triple synchrony orbits which the Author refer to as inner and outer Clarke's Orbits and in Earth-Moon system they are referred to as inner geo-synchronous orbit (a$_{G1}$) and outer geo-synchronous orbit (a$_{G2}$).

Moon tidally evolves out of Inner Clarke's Orbit ($a_{G1}$). If it tumbles short of $a_{G1}$ , secondary rapidly spirals-in to its certain destruction and if it tumbles long of $a_{G1}$ then through Gravitational Sling Shot secondary is launched on an outward spiral path. But as the differential between orbital velocity and spin velocity of primary grows, tidal stretching and squeezing sets in the primary body which leads to tidal dissipation which causes a rapid exponential decay of the impulsive torque. In super synchronous orbit primary's tidal bulge leads the radius vector joining primary and secondary. This 'lead angle' causes secular deceleration of the primary and angular momentum transfer from primary to secondary for angular momentum conservation. From then onward the Moon coasts on its own until it locks into the outer Outer Clarke's Orbit ($a_{G2}$). But through out this tidal evolutionary history the Total Angular Momentum is conserved hence we have the following Conservation of Momentum equation:

$$J_T = C\omega + \left(m^*a_{present}^2 + I\right)\Omega = [C + (m^*a_{G1}^2 + I)]\Omega_{aG1}$$
$$= [C + (m^*a_{G2}^2 + I)]\Omega_{aG2} \qquad S4.4$$

In (S4.4):

C = Moment of Inertia of the Primary around its spin axis.

I = Moment of Inertia of the Secondary around its spin axis.

And m*= reduced mass of the secondary = m/(1+m/M) where m = the mass of the secondary and M= mass of the primary.

From Kepler's Third Law:

$$\Omega_{aG1} = \frac{B}{a_{G1}^{3/2}} \quad and \quad \Omega_{aG2} = \frac{B}{a_{G2}^{3/2}} \quad where \; B = \sqrt{G(M+m)} \qquad S4.5.$$

Substituting (S4.5.) in (S4.4.) we get:

$$J_T = C\omega + \left(m^*a_{present}^2 + I\right)\Omega = [C + (m^*a_{G1}^2 + I)]\frac{B}{a_{G1}^{3/2}}$$

$$= [C + (m^*a_{G2}^2 + I)]\frac{B}{a_{G2}^{\frac{3}{2}}} \qquad S4.6$$

Solving (S4.6.) we get the two roots of the Binary System namely $a_{G1}$ and $a_{G2}$. In classical Newtonian Mechanics two triple synchrony orbits do exist from total energy consideration as shown in S5. Hence Author calls this model as Kinematic Model (KM).

From Classical Mechanics the Synchronous Orbit is the same as the Inner Clarke's Orbit calculated in Kinematic Framework for vanishingly small mass ratio m/M. In Classical Mechanics, the synchronous orbit is defined as:

$$a_{sync}^{3/2} \Omega_{orb} = a_{sync}^{3/2} \omega_{primary} = B \qquad\qquad S4.7.$$

**Table S4.1. Comparative Study of Triple Synchrony Orbits of Earth-Moon, Mars-Phobos-Deimos , Pluto-Charon Systems, Sun-Jupiter and two stellar binaries (NN-Serpentis and RW Lac) from Classical Newtonian Mechanics and Kinematic Model.[The Globe-Orbit Parameters based on which the calculations have been made are given in S4(Appendix A in supplementary materials)**

| Planet-Sat | Mass-ratio (q) | a(present) (m) | B($m^{3/2}$/s) | $a_{G1}$ (m) | $a_{G2}$ (m) | $a_{sync}$ (m) from (S4.7.) |
|---|---|---|---|---|---|---|
| Earth-Moon | 1/81 | $3.84400 \times 10^8$ | $2.00811 \times 10^7$ | $1.46 \times 10^7$ | $5.53 \times 10^8$ | $4.234 \times 10^7$ |
| Mars-Phobos | $10^{-8}$ | $9.378 \times 10^6$ | $6.54 \times 10^6$ | $2.04 \times 10^7$ | $7.46 \times 10^{18}$ | $2.04 \times 10^7$ |
| Mars-Deimos | $10^{-9}$ | $23.459 \times 10^6$ | $6.54 \times 10^6$ | $2.04 \times 10^7$ | $1.69 \times 10^{20}$ | $2.04 \times 10^7$ |
| Pluto-Charon | 1/8 | $19.600 \times 10^6$ | $9.88 \times 10^5$ | $1.37672 \times 10^6$ | $1.95579 \times 10^7$ | $1.96133 \times 10^7$ |
| Sun-Jupiter | $9.55 \times 10^{-4}$ | $778.3 \times 10^9$ | $1.15256 \times 10^{10}$ | $1.06889 \times 10^9$ | $7.92465 \times 10^{11}$ | $2.53 \times 10^{10}$ |
| NN-Serpentis | 0.2074 | $6.49597 \times 10^8$ | $9.25989 \times 10^9$ | $4.44958 \times 10^7$ | $6.4986 \times 10^8$ | $6.49514 \times 10^8$ |
| RW-Lac | 0.9375 | $1.69267 \times 10^{10}$ | $1.54426 \times 10^{10}$ | $4.08908 \times 10^8$ | $1.69314 \times 10^{10}$ | $1.69252 \times 10^{10}$ |

In Table S4.1, all cases are consistent with Kinematic Formalism except Pluto-Charon (case no.4). This exception is due to large uncertainty in the Globe-Orbit parameters of Pluto-Charon.

Case 1: Moon is a significant fraction of Earth (1/81) hence our Moon has a definite Tidal Evolution History. It started its journey about 4.467Gya (The birth of the Solar System is the time when the condensation of the first solid took place from the Solar Nebula. This is taken as 4.567Gya. The last giant impact on Earth formed the Moon and initiated the final phase of core formation by melting the mantle of the Earth. The date of this last impact decides the birth date of Moon which was completed in a few hundred years by the accretion of the impact generated debris.Yin, et.al.2002,Jacobsen,2005 andTaylor, et.al. 2009 claim an age of Moon as 30My after the birth of Solar System.Toubol,et.al.2007, Allègre, et.al.2008.and Halliday .&Wood.2007claim an younger Moon formed after 50 to 100My after the first solid condensed. The concentration of highly siderophile elements (HSEs) in Earth's mantle constrains the mass of chondritic material added to Earth during Late Accretion(Chyba,1991; Bottke, et.al.2010). Using HSE abundance measurements (Becker, et.al.;2006; Walker, 2009), Jacobson et.al.(2014) determine a Moon-formation age of 95± 32 Myr after the condensation. This method is invariant of the geo-chemistry chronometer adopted by earlier researchers.So it will be realistic to take the age of Moon as 4.467Gya.since its birth just beyond Roche's Limit 15,000Km. By gravitational sling shot it was launched on an expanding spiral orbit from inner geo-synchronous orbit of 15,000Km orbital radius towards the outer geo-synchronous orbit of $5.53 \times 10^8$ m = 553,000Km. At the inner geo-synchronous orbit, the length of day = length of month = 5 hours and at the outer geo-synchronous orbit, the length of day = length of month = 47 days. Presently the lunar orbital radius is 384,400Km with sidereal length of day = 23.9344 hours and length of Sidereal Month = 27.32 Earth days. Earth-Moon started from geo-synchrony and will end in geo-synchrony. As predicted in Figure 1, for mass ratio = 1/81 the classical synchronous orbit is less than the outer geo-synchronous orbit.

Case 2 and 3: In case of Mars-Phobos-Deimos, since the mass ratio is insignificant hence Deimos launched on an orbit long of inner Clarke's Orbit has hardly evolved from its point of inception which is inner Clarke's Orbit. But Phobos is launched on an orbit short of inner Clarke's orbit hence it is on a gravitational runaway orbit, trapped in a death spiral. Deimos is stay-put in its orbit of inception which is 20,400Km but Phobos has lost altitude from its point of inception of 20,400Km to the present altitude of 9,378Km. Since the mass ratio is insignificant hence the classical synchronous orbit is the same for both Phobos and Deimos equal to 20,400Km same as the inner Clarke's Orbit. This is in exact correspondence with Figure 1.

Case 4. Pluto-Charon's classical synchronous orbit should be smaller than Outer Clarke's Orbit as required by Kinematic Analysis but the former is 0.28% larger. This is due to the uncertainty in Globe-Orbit parameters of Pluto-Charon.

Case 5. Mass ratio of Jupiter to Sun is $10^{-3}$ hence according to KM analysis Jupiter-Sun has a tidal evolutionary history with a rapid Time-constant of evolution of 4.275My. It has evolved from inner Clarke's Orbit $3.7859 \times 10^9$m to the present orbit of $778.3 \times 10^9$m where its evolution factor is 0.893 and eventually it will lock into second triple-synchrony state in the outer Clarke's Orbit of $871.161 \times 10^9$m. The classical synchronous orbit is at $25.3 \times 10^9$m, 97% smaller than outer Clarke's Orbit, as predicted by Figure S3.4 also.

In Paper No. B0.3-0011-12 Iapetus hypothetical sub-satellite re-visited and it reveals celestial body formation process in the KM Framework.presented at 39[th] COSPAR Scientific Assembly, Mysore, India from 14[th] July to 20[th] July 2012, the correspondence between Newtonian Formalism of Synchronous Orbit and Kinematic Case 6 and Case 7: These are stellar non-relativistic binaries. I call them non-relativistic because the mean apisidal motion is negligible. Here since the mass ratio is greater than 0.2, hence the original molecular cloud settles into a binary in Months-Years and gets locked-into outer Clarke's Orbit. In both cases the synchronous orbit is shorter than the Outer Clarke's Orbit by 0.05% and 0.04% respectively. This is consistent with Kinematic Analysis.

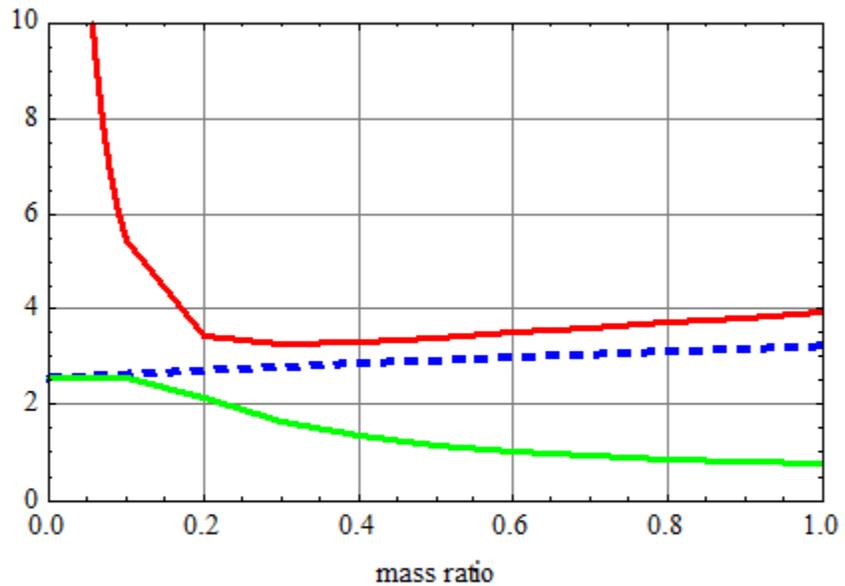

**Figure S4.4.[.tiff]. Plot of $a_{synSS}$ ($\times R_{Iap}$)[Dashed Blue], $a_{G1}$ ($\times R_{Iap}$)[Thick Green] and $a_{G2}$ ($\times R_{Iap}$)[Thick Red] as a function of 'q'=mass ratio. Y-axis is semi-major axis as a multiple of Iapetus Globe Radius. [Courtesy: Author]**

Inspection of Figure S4.4., tells us that at infinitesimal values of 'q' ,$a_{synSS}$ is the same as $a_{G1}$ and only one Clarke's Orbit is perceptible. But at larger mass ratios the two (classical and kinematic formalism for $a_{G1}$) rapidly diverge. Author's analysis till now has confirmed that $a_{G1}$ is the correct formalism for predicting the inner triple synchrony orbit in a binary system at q < 0.2..

At mass ratios greater than 0.2, $a_{G1}$ is physically untenable and only $a_{G2}$ is perceptible. Outer Triple Synchrony Orbit seems to converge but does not actually converge to the classical formalism but remains offsetted right till the limit of q =1. Here again only outer Clarke's Orbit is perceptible . The actual Star pairs satisfy the Kinematic formalism and not the classical formalism.

So Kinematic Formalism, though satisfies the correspondence principle at q ~ 0, is a theory in its own right. Till date there exists formalism for two triple synchrony orbits in Classical Newtonian Mechanics. In the mass ratio range 0.0001 to 0.2 through total energy analysis as shown in S5 the two triple synchrony orbis can be derived.

For mass ratio less than 0.0001, binaries remain in inner Clarke's Configuration stably which is predicted by Classical Newtonian Formalism also.

At mass ratios greater than 0.2 right up to unity, star pairs remain in outer Clarke's Configuration stably and its magnitude is more than Newtonian prediction.

For mass ratios $0.0001 < q < 0.2$, Outer Clarkes configuration is the only stable orbit and secondary is catapulted from $a_{G1}$ by Gravitational Sling Shot mechanism and it migrates out of that configuration. If it is at $a > a_{G1}$ the pair spirals out with a time constant of evolution and if a $< a_{G1}$ then the pair spirals-in on a collision course again with a characteristic time constant of evolution.

Time Constant of Evolution is in inverse proportion of some power of mass ratio (Sharma 2011).

For $q = 0.0001$, it is Gy and as q increases , time-constant decreases from Gy to My to kY to years. This is valid for mass scale encountered in Solar and Exo-Solar Systems. Between 0.2 to 1, a solar nebula falls into outer Clarke's Configuration by hydro-dynamic instability within months/years.

For q being vanishingly small, the calculation of the man-made Geo-synchronous Satellite's orbit of 36,000Km above the equator has been done by Kinematic Formalism. This calculation has been done by the Author in his personal communication:

http://arXiv.org/abs/0805.0100

**S4.2.1. Deductions of the kinematic parameter of E-M system**

Using Globe-Spin parameters of E-M system given in Appendix A of S4, we get the following kinematic parameters:

Reduced mass of Moon=$m/(1+m/M)$=$m^*$=$7.25284\times10^{22}$Kg.

B=$\sqrt{(G(M+m))}$=$2.00811\times10^7 m^{3/2}$/s, G=$6.67\times10^{-11}$ $m^3$/(kg-$s^2$),

C = moment of inertia of Earth around its spin axis=$0.33086MR^2$= $8.0209\times10^{37}$Kg-$m^2$,

I = moment of inertia of Moon around its spin axis=$0.394mR_{moon}^2$= $8.72791\times10^{34}$Kg-$m^2$,

Define $\theta_1$= I/C = 0.00108815 , $\theta_2$= $m^*$/C=$9.04243\times10^{-16}$ 1/$m^2$,

$J_T$= total angular momentum of E-M system= $3.43584\times10^{34}$Kg-$m^2$/s ,

Spin Period of Earth = 1 Solar Day =24 hours.

Sidereal Spin Period of Moon = 27.322 Solar Day ,

Sidereal Orbital period of E-M system =27.3217 Solar Day,

Moon is in synchronous orbit i.e. it is tidally locked and shows the same face to Earth. Generally this synchronism property is true for all compact binaries. Because of this synchrony of our Moon and because of the circular path, our Moon experiences no stretching and squeezing and hence no tidal heating and hence no volcanic effect. 'Io', a moon of Jupiter, is similarly placed as our Moon is with respect to Earth but still Io is the most volcanically active natural satellite because of tidal dissipation. In synchronous orbit there should have been no volcanic activity but its eccentric orbit makes it the most volcanically active natural satellite in our solar system. Io is in 2:1 resonance with Europa hence it has eccentric orbit and hence it is evolving and gradually circularizing. As its eccentricity is reduced in the process of circularization, the volcanic activity is on the wane.

Solving (S3.6) we obtain the two geo-synchronous orbits:

$$a_{G1} = 1.46402 \times 10^7 m, \qquad a_{G2} = 5.5247 \times 10^8 \qquad\qquad S4.8$$

According to Ida & Stewart (1997), tidal flexing does not allow the solid particles to coalesce within Roche's limit represented by the symbol $a_R$ .

$$a_R = 2.456(\frac{\rho_E}{\rho_M})^{1/3} R_E \sim 2.9 R_E = 18496.606 \; Km \; where \; \rho_E = \frac{5.5 gm}{cc} \; and \; \rho_M$$
$$= \frac{3.34 gm}{cc} \quad S4.9$$

and Roche Zone is defined within the range:

$$0.8 \; to \; 1.35 a_R \; or \; 2.32 \; to \; 3.915 R_E \; i.e. \; within \; 14,797 Km \; to \; 24,970 Km$$

This implies that impact generated debris will be prevented from accretion within $1.48 \times 10^7 m$ and those in $1.48 \times 10^7 m$ to $2.5 \times 10^7 m$ range also known as transitional zone will experience limited accretion growth whereas those lying beyond this zone will be unaffected by tidal forces. It is a happy coincidence that the Roche zone lies just beyond the inner Geo-Synchronous orbit of the Earth-Moon System. This implies that if accretional criteria of Canup & Esposito (1996)is satisfied along with the impact velocity condition that is the rebound velocity should be smaller than the mutual surface escape velocity then merged body formation of Moon starts within the Roche zone. The accreted Moon gradually migrates outward sweeping the remnant debris.

### S4.2.2. Deduction of LOM/LOD or ω/Ω equation.

Rewriting (S4.6) we obtain:

$$\frac{J_T}{C\Omega} = \left[\frac{\omega}{\Omega} + \left(\left(\frac{m^*}{C}\right)a_{present}^2 + \frac{I}{C}\right)\right] = \left[\frac{\omega}{\Omega} + \theta_2 a_{present}^2 + \theta_1\right] \qquad\qquad S4.10$$

Substituting Ω= B/a$^{3/2}$ in (S4.10) we obtain:

$$\left(\frac{J_T}{CB}\right)a^{\frac{3}{2}} = \left[\frac{\omega}{\Omega} + \theta_2 a_{present}^2 + \theta_1\right] \qquad S4.11$$

Rearranging the terms in (S4.11) we get:

$$\frac{\omega}{\Omega} = \frac{LOM}{LOD} = \left(\frac{J_T}{CB}\right)a^{\frac{3}{2}} - (\theta_2 a^2 + \theta_1) = Aa^{\frac{3}{2}} - Fa^2 \qquad S4.12$$

$$where\ A = \frac{J_T}{BC}\ and\ F = \left(\theta_2 + \frac{\theta_1}{a^2}\right)$$

Substituting the numerical values of the parameters we get:

A=$2.1331 \times 10^{-11}$ (1/m$^{3/2}$) and F=$9.0425 \times 10^{-16}$ (1/m$^2$).

Substituting these values in (S4.10) we get LOM/LOD =27.1479.

The actual value of LOM/LOD is 27.322. This error is due to uncertainty in Globe-Spin parameters. A and F are adjusted to obtain the exact value of 27.322.

The best fit values of LOM/LOD constants are:

A=$2.13853 \times 10^{-11}$ (1/m$^{3/2}$) and F=$9.05842 \times 10^{-16}$ (1/m$^2$).

The best fit parameters give the following geo-synchronous orbits:

$$a_{G1} = 1.46402 \times 10^7 m, \qquad a_{G2} = 5.5247 \times 10^8, a_2 = 2.40649 \times 10^7 m \qquad S4.13$$

Calculating LOM/LOD in (S4.12) using the best fit parameters of 'A' and 'F' we get :

LOM/LOD = 27.322 which is the present era Sidereal Lunar Month/Solar Day observed values.

### S4.2.3. The Tidal Torque formalism and Radial Velocity formalism..

For the calculation of the spiral trajectory we need the radial velocity of recession in case of super-synchronous configuration and velocity of approach in case of sub-synchronous configuration. The time integration of the reciprocal of radial velocity gives the non-Keplerian Transit time from its inception to the present orbit. This transit time should be equal to the age of the secondary or the natural satellites. The starting point of this time integral will be the tidal torque.

The Tidal Torque of Satellite on the Planet and of Planet on the Satellite = Rate of change of angular momentum hence

$$Tidal\ Torque = T\ = \frac{dJ_{orb}}{dt} \qquad S4.14$$

But Orbital Angular Momentum:

$$J_{orb} = m^*a^2 \times \frac{B}{a^{3/2}} = m^*B\sqrt{a} \qquad\qquad S4.15$$

Time Derivative of (S4.15) is:

$$T = \frac{dJ_{orb}}{dt} = \frac{m^*B}{2\sqrt{a}} \times \frac{da}{dt} \qquad\qquad S4.16$$

In super-synchronous orbit, the radius vector joining the satellite and the center of the planet is lagging planetary equatorial tidal bulge hence the satellite is retarding the planetary spin and the tidal torque is BRAKING TORQUE.

In sub-synchronous orbit, the radius vector joining the satellite and the center of the planet is leading planetary equatorial tidal bulge hence the satellite is spinning up the planet and the tidal torque is ACCELERATING TORQUE..

These two kinds of Torques are illustrated in Figure S3.1 and Figure S3.2.

I have assumed the empirical form of the Tidal Torque as follows:

$$T = \frac{K}{a^Q}\left[\frac{\omega}{\Omega} - 1\right] \qquad\qquad S4.17$$

(S4.17) implies that at Inner Clarke's Orbit and at Outer Clarke's Orbit, tidal torque is zero and (S4.16) implies that radial velocity at Inner Clarke's Orbit and at Outer Clarke's Orbit,is zero and there is no spiral-in or spiral-out.

At Triple Synchrony, Satellite-Planet Radius Vector is aligned with planetary tidal bulge and the system is in equilibrium. But there are two roots of $\omega/\Omega=1$: Inner Clarke's Orbit and Outer Clarke's Orbit. It has been shown in S5 that in Total Energy Profile, Inner Clarke's Orbit $a_{G1}$ is energy maxima and hence unstable equilibrium state and Outer Clarke's Orbit $a_{G2}$ is energy minima and hence stable equilibrium state. In any Binary System, secondary is conceived at $a_{G1}$. This is the CONJECTURE assumed in Kinematic Model. From this point of inception Secondary may either tumble short of $a_{G1}$ or tumble long of $a_{G1}$. If it tumbles short, satellite gets trapped in Death Spiral and it is doomed to its destruction. If it tumbles long, satellite gets launched on an expanding spiral orbit due to gravitational sling shot impulsive torque which quickly decays due to the growing differential of $\omega/\Omega$ and the resulting tidal heating. After the impulsive torque has decayed, the satellite coasts on it own toward final lock-in at $a_{G2}$.

Equating the magnitudes of the torque in (S4.16) and (S4.17) we get:

$$\frac{m^*B}{2\sqrt{a}} \times \frac{da}{dt} = \frac{K}{a^Q}\left[\frac{\omega}{\Omega} - 1\right] \qquad\qquad S4.18$$

Rearranging the terms in (S4.18) we get:

$$V(a) = Velocity\ of\ recession = \frac{2K}{m^*B} \times \frac{1}{a^Q}\left[Aa^2 - Fa^{2.5} - \sqrt{a}\right] m/s \qquad S4.19$$

The Velocity in (S4.19) is given in m/s but we want to work in m/y therefore (S4.19) R.H.S is multiplied by $31.5569088 \times 10^6$ s/(solar year).

$$V(a) = \frac{2K}{m^*B} \times \frac{1}{a^Q}\left[Aa^2 - Fa^{2.5} - \sqrt{a}\right] \times 31.5569088 \times 10^6 m/y \qquad S4.20$$

In (S4.20) 'a' refers to the semi-major axis of the evolving Satellite. There are two unknowns: exponent 'Q' and structure constant 'K'. Therefore two unequivocal boundary conditions are required for the complete determination of the Velocity of Recession.

First boundary condition is at a = $a_2$ which is a Gravitational Resonance Point where $\omega/\Omega$ = 2 (Rubicam 1975),

i.e. $(Aa^{3/2} - Fa^2) = 2$ has a root at $a_2$.

In E-M case, $a_2 = 2.40649 \times 10^7$ m.

At $a_2$ the velocity of recession maxima occurs. i.e. $V(a_2) = V_{max}$ .

Therefore at a = $a_2$, $(\delta V(a)/\delta a)(\delta a/\delta t)|_{a2} = 0$.

On carrying out the partial derivative of V(a) with respect to 'a' we get the following:

$$At\ a_2, \quad (2 - Q)A \times a^{1.5} - (2.5 - Q)F \times a^2 - (0.5 - Q) = 0 \qquad S4.21$$

Solving (S3.21) at 2:1 Mean Motion Resonance orbit '$a_2$' we obtain :

$$Q = 3.23771 \qquad S4.22$$

Now structure constant (K) has to be determined . This will be done by trial error so as to get the right age of Moon(Foot note 5) i.e. 4.467Gy. Rewriting (S4.20) and substituting the best fit values of the exponent and constants A and F we obtain the structure constant 'K'.

$$V(a) = \frac{2K}{m^*B} \times \frac{1}{a^Q}\left[A \times a^2 - F \times a^{2.5} - \sqrt{a}\right] \times 31.5569088 \times 10^6 m/y \quad S3.23$$

We will assume the age of Moon 4.46Gy as already mentioned in Foot Note 5. The Transit Time from $a_{G1}$ to the present 'a' is given as follows:

$$Transit\ Time = \int_{a_{G1}}^{a} \frac{1}{V(a)}\,da \qquad S4.24$$

Assuming a tentative value for $V_{max}$ and inserting it in (S3.23) at a = $a_2$ we deduce the value of 'K'. Using this 'K' in (S3.23) and inserting this trial expression in (S3.24) we carry out the time integral to get the transit time from $a_{G1}$ to present 'a' which should be the age of Moon as agreed upon 4.467Gy. Several iterations are carried by adjusting $V_{max}$. By this iteration method we obtain the best fit structure constant as:

$$K = 6.48548 \times 10^{42} (Newton - m^{Q+1})$$ S4.25

So the best fit velocity of recession formalism is:

$$V(a) = \frac{2 \times 6.48548 \times 10^{42}}{m^*B} \times \frac{1}{a^{3.23771}} \left[ A \times a^2 - F \times a^{2.5} - \sqrt{a} \right] \times 31.5569088$$
$$\times \frac{10^6 m}{y}$$ S4.26

Transit Time expression gives 4.4635Gy using (S4.24). This is the age of Moon as concluded previously.

Present recession velocity of Moon =2.3244cm/y            S4.27

LLR measurement of Moon's recession is anomalously higher than the calculated value.

This is the first indication that KM is oversimplified and not giving the correct result. The velocity of recession of Moon, the deceleration of Moon and the superposition of the two are shown in Figure S3.1., Figure S3.2., Figure S3.3.

## S4.3. The theoretical formalism of lengthening of day curve poorly validated by observed LOD curve.

Paleontologists have studied the length of day in the past geological epochs.John West Wells (1963,1964) through the study of daily and annual bands of Coral fossils and other marine creaturs in bygone era has obtained ten benchmark. Leschiuta& Tavella,2001,and Kaula& Harris,1975, have given the estimate of the synodic monthbased on the study of marine creature fossils. From the synodic month we can estimate the length of the Solar Day as given in S4.Appendix [B]. But the estimatin of LOD by synodic month is always an over-estimation hence the estimated LOD from synodic month have been rejected. In S4 Appendix [B] the estimation of LOD from synodic method and their over-estimation has been tabulated. One benchmark has been provided by Charles P. Sonnett et al,1998, through the study of tidaliesin ancient canals and estuaries. He gives an estimate of LOD=18.9h at about 900 million years B.P. in Proterozoic Eon, pre-Cambrian Age.Arbab(2008) has estimated LOD from the data of John West Wells and from cosmological consideration hence includingArbab data we have all together 24 data set which have been tabulated in Table S3.1. Using (S3.26) the orbital radius corresponding to the geologic epoch of observation has been calculated and tabulated in Column 3 of Table 3.1. From Table 3.2, the observed lengthening of day curve is generated as shown in Figure S3.7.

**Table S4.2, Observed Length of Day in different geological epochs as calculated in KM.**

| Data set # | Time B.P.(years) | Orbital radii(m)† | LOD(hrs)[1] | LOD[2](hrs) |
|---|---|---|---|---|
| 1 | Present | $3.844 \times 10^8$ | 24 | 24 |
| 2 | 65Ma | $3.8287 \times 10^8$ | 23.627 | 23.6 |
| 3 | 135Ma | $3.81213 \times 10^8$ | 23.25 | NA |
| 4 | 136Ma | $3.8118 \times 10^8$ | 23.2515 | 23.2 |
| 5 | 180Ma | $3.80129 \times 10^8$ | 23.0074 | 23 |
| 6 | 230Ma | $3.7891 \times 10^8$ | 22.7683 | 22.7 |
| 7 | 280Ma | $3.7768 \times 10^8$ | 22.4764 | 22.4 |
|  | 300Ma | $3.7718 \times 10^8$ | 22.3* |  |
| 8 | 345Ma | $3.76055 \times 10^8$ | 22.136 | 22.1 |
| 9 | 380Ma | $3.7517 \times 10^8$ | 21.9 | NA |
| 10 | 405Ma | $3.74535 \times 10^8$ | 21.8055 | 21.7 |
| 11 | 500Ma | $3.7208 \times 10^8$ | 21.276 | 21.3 |
| 12 | 600Ma | $3.6943 \times 10^8$ | 20.674 | 20.7 |
| 13 | 715Ma | $3.663 \times 10^8$ | NA | 20.1 |
| 14 | 850Ma | $3.6251 \times 10^8$ | NA | 19.5 |
| 15 | 900Ma | $3.61075 \times 10^8$ | 18.9 | 19.2 |
| 16 | 1200Ma | $3.5205 \times 10^8$ | NA | 17.7 |
| 17 | 2000Ma | $3.235 \times 10^8$ | NA | 14.2 |
| 18 | 2500Ma | $3.012 \times 10^8$ | NA | 12.3 |
| 19 | 3000Ma | $2.735 \times 10^8$ | NA | 10.5 |
| 20 | 3560Ma | $2.3143 \times 10^8$ | NA | 8.7 |
| 21 | 4500Ma | NA | NA | 6.1 |

[1]Length of Day according to John West Wells (Wells 1963,1965), and Charles P.Sonnet (1998).

[2]Length of Day according to Arbab (2009)

*Length of Day according to Leschiutta&Tavella (2001)

†Orbital Radius of Moon calculated from KM for classical Moon.

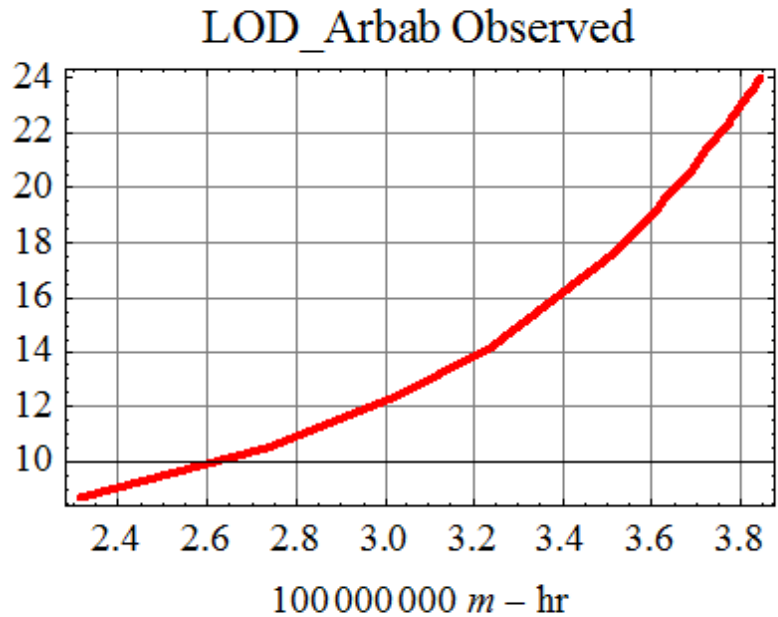

**Figure S4.5. Observed Lengthening of Day curve over a time span of 3.56Gy..[Courtesy: Author]**

.

**S4.3.1.Theoretical Formalism of Length of Earth Day in the entire life span of Moon.**

From Kepler's Third Law:

$$T_{orb} = \frac{2\pi \times a^{3/2}}{B} \qquad\qquad S3.27$$

From (S3.12):

$$\frac{T_{orb}}{T_E} = \left( A \times a^{\frac{3}{2}} - F \times a^2 \right) \qquad\qquad S\ 3.28$$

Substituting (S3.27) in (S3.28) we obtain LOD:

$$T_E = \frac{2\pi \times a^{3/2}}{B \left( A \times a^{\frac{3}{2}} - F \times a^2 \right)} \qquad\qquad S3.29$$

Substituting the best fit parameters in (S3.29) the theoretical lengthening of day curve is obtained in Figure S3.8.

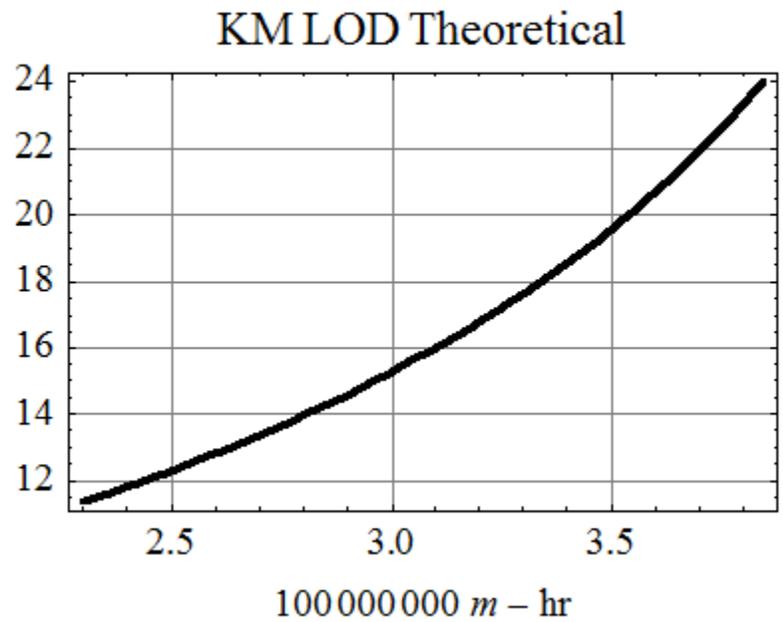

**Figure S4.6. Theoretical Lengthening of Day curve over a time span of 3.56Gy..**

Superposition of the theoretical curve and observed curve gives Figure S9.**[Curtesy: Author]**

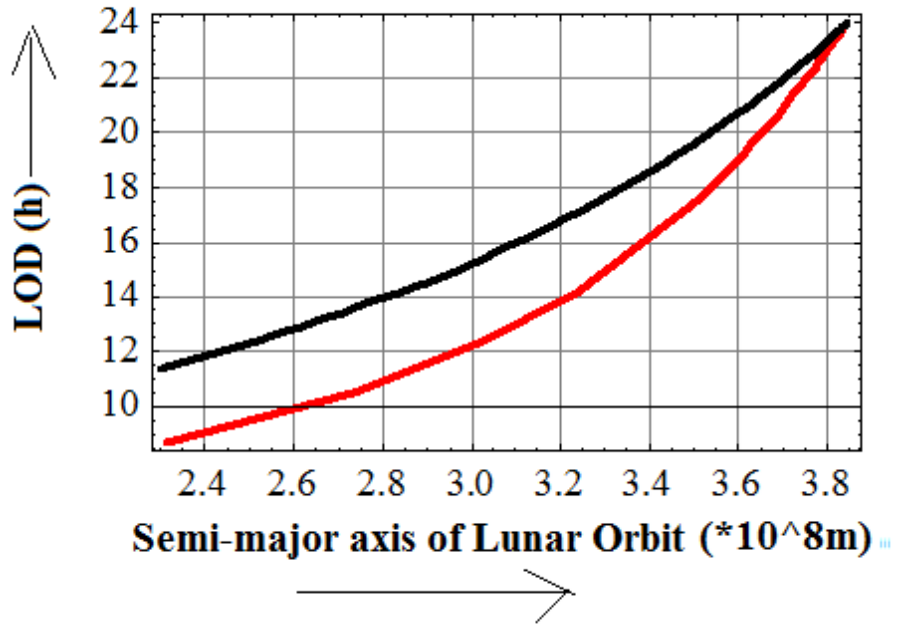

**Figure S3.7. Superposition of Theoretical and Observed Lengthening of Day curve over a time span of 3.56Gy. The worst case mismatch is 31%[Courtesy: Author]**

Clearly there is a mismatch over the entire time span with the worst case scenario occurring at 3.56Gy ago. The worst case mismatch is -31%.

Needless to say KM has failed to capture the true picture of tidally coupled Earth-Moon.

**S5. Appendix (A) and Appendix (B) of S3 and S4.**

**Appendix (A).Globe-Spin parameters of Earth-Moon System.**

**Fact Sheet of Earth-Moon :**
**http://nssdc.gsfc.nasa.gov/planetary/factsheet/moonfact.html**

| parameters | Earth | Moon |
|---|---|---|
| Mass(Kg) | $5.9726 \times 10^{24}$ | $0.07342 \times 10^{24}$ |
| GM(Km$^3$/s$^2$) | $0.3986 \times 10^6$ | $0.0049 \times 10^6$ |
| Volumetric Mean Radius | 6371 | 1737 |
| Or Median Radius($\times 10^3$ m) | | |
| Flattening(ellipticity) | 0.00335 | 0.0012 |
| Mean Density(Kg/m$^3$) | 5514 | 3344 |
| Moment of Inertia(I/(MR$^2$)) | 0.33086 | 0.394 |
| Sidereal Spin period | 23.9344h | 27.322d |
| Sidereal Orbital period(d) | - | 655.7208h (27.3217d) |
| a*(semi-major axis)($\times 10^8$m) | - | 3.84400 |
| Lunar Orbit eccentricity | - | 0.0549 |
| Lunar Orbital inclination w.r.t. Ecliptic | - | 5.145 degrees |
| B=$\sqrt{(G(M+m))}$   (m$^{3/2}$/s) | | 2.00873$\times 10^7$ |

*Mean Orbital Distance from the center of Earth.

## Appendix (B) Determination of LOD from the synodic month in a given geologic epoch.

Leschitua&Tavella(2001)[36] and Kaula& Harris (1975) [37] have estimated the synodic month in the past epochs. Their measurements are given in Table B1.

## Table S5.B1. Estimation of LOD from synodic month and from paleobotanical and paleotidal evidences

| T(B.P.) | T* | Synodic Month | Orbital radius | Estimated LOD | LOD by Wells |
|---|---|---|---|---|---|
| 900Ma | 3.56347Gy | 25d | 3.61075$\times 10^8$m | 19.8362h | 18.9h |
| 600Ma | 3.86347Gy | 26.2d | 3.6943$\times 10^8$m | 20.9669h | 20.674h |
| 300Ma | 4.16347Gy | 28.7d | 3.7718$\times 10^8$m | 23.0519h | 22.3h |
| 45Ma | 4.41847Gy | 29.1d | 3.8335$\times 10^8$m | 23.6h | NA |
| 2.8Ga | 1.66347Gy | 17d | 2.8537$\times 10^8$m | 13.5569h | NA |

The methodology of determination of LOD from Synodic Month.

$$T_{synodic} = apparent\ orbital\ period = \frac{T_{sidereal}}{1 - \frac{T_{sidereal}}{Z}}\ where\ T_{sidereal}$$

$$= actual\ orbital\ period\ and\ Z = \frac{365.242d/y}{T_E(d)}\quad where\ T_E$$

$$= Earth\ day; \hspace{4cm} S5.B1$$

From the geologic epoch the corresponding orbital radius is determined. Using the orbital radius of the given geologic epoch and (S3.12), LOM/LOD is determined.

$$\frac{LOM}{LOD} \text{ is determined from } (S3.12) \text{ at the geologic epoch 'a'}$$

Using (S4.B1) we get the expression:

$$T_{synodic} = \frac{\left(\dfrac{LOM}{LOD}\right) \times T_E(d)}{1 - \dfrac{\left(\dfrac{LOM}{LOD}\right) \times T_E(d)}{Z}} \qquad\qquad S5.B2$$

From (S4.B1) and (S4.B2) the LOD is obtained from Synodic months and tabulated in Table S4.B2.

**Table S5.B2. Comparative study of LOD from synodic method and those obtained from coral fossils and tidalites**

| Geologic epoch | a (orbital radii) | LOM/LOD | Synodic month | LOD1 | LOD2 |
|---|---|---|---|---|---|
| 900Ma | $3.61075 \times 10^8$m | 28.6282 | 25d | 19.8362h | 18.9h |
| 600Ma | $3.6943 \times 10^8$m | 28.2216 | 26.2d | 20.9669h | 20.674h |
| 300Ma | $3.7718 \times 10^8$m | 27.7835 | 28.7d | 23.0519h | 22.3h |
| 45Ma | $3.8335 \times 10^8$m | 27.3924 | 29.1d | 23.6h | NA |
| 2.8Ga | $2.8537 \times 10^8$m | 29.3245 | 17d | 13.5569h | NA |

LOD1 is the estimate from Synodic month.

LOD2 is the estimate from corals and tidalites

Clearly LOD obtained from synodic month are highly over estimated and hence rejected in this study.

## S5.1. Kinematic Model yields two geo-synchronous orbits of E-M system validated by Total Energy Analysis

The total energy analysis is utilized in case of E-M and the extremum points are obtained. The energy maxima happens to be the inner unstable geo-synchronous orbit ($a_{G1}$) and the energy minima happens to be outer stable geo-synchronous orbit($a_{G2}$) in case of E-M and M-P systems and their values from KM correspond to those obtained from Energy Analysis. This vindicates KM of binary pairs.

### S5.1.1. To determine the energy extremum points from Total Energy Profile of E-M binary system.

Total Energy of Earth-Moon System = Rotational Kinetic Energy + Potential Energy + Translational Kinetic Energy.
Translational Kinetic Energy of the order of $1 \times 10^8$Joules due to recession of Moon for all practical purposes is negligible as compared to Rotational Kinetic Energy of the order of $1\times10^{30}$ Joules . Hence Translational Kinetic Energy is neglected in future analysis.

Moon is trapped in potential well created by the Earth.

Moon's potential energy = - $GM_{Earth}M_{Moon}/a$
G =Gravitational Constant=$6.673\times10^{-11}$ N-m$^2$/Kg$^2$;

$M_{Earth}$ = mass of the Earth = $5.9742\times10^{24}$ Kg;

$M_{Moon}$ = mass of the Moon = E/81 = $7.348\times10^{22}$ Kg;

a= semi-major axis of Moon's orbit around the Earth= $3.844\times10^8$ m;

Rotational Kinetic Energy of Earth-Moon System = Spin Energy of the Earth + Orbital Energy of the Earth-Moon System + Spin Energy of the Moon =

$$\frac{1}{2}C\omega^2 + \frac{1}{2}\left(\frac{M_{Moon}}{1+\frac{M_{Moon}}{M_{Earth}}}\right)a^2 \times \Omega^2_{orbital} + \frac{1}{2}\times(0.4M_{Moon}R^2_{Moon})\Omega^2_{spin} \qquad S5.1$$

Where C = moment of inertia around polar axis = $0.3308M_{Earth}R_{Earth}^2$=$8.02\times10^{37}$Kg-m$^2$;

Equatorial Radius of Earth=$6.37814\times10^6$ m;

Equatorial Radius of Moon=$1.738\times10^6$ m;

Earth angular spin velocity = $\omega = 2\pi/T_E = [2\pi/(86400)]$radians/sec;

In this analysis we will consider all rates of rotation to be in Solar Days. We will consider one solar day as the present spin-period of Earth. Similarly while calculating Earth-Moon orbital angular momentum we will use present sidereal month expressed in 27.3 solar days.

Earth-Moon Orbital Angular Velocity = $\Omega = [2\pi/(27.3\times86400)]$radians/sec where sidereal month =27.3 d;

Since Moon is in synchronous orbit i.e. it is tidally locked with the Earth hence we see the same face of Moon and Moon's Orbital Angular Velocity = Moon's Spin Angular Velocity = $\Omega$;

Therefore total rotational Kinetic Energy Equation 2 reduces to:

$$\frac{1}{2}C\omega^2 + \frac{1}{2}\left(\frac{M_{Moon}}{1+\frac{M_{Moon}}{M_{Earth}}}\right)a^2 \times \Omega^2 + \frac{1}{2}\times(0.4M_{Moon}R^2_{Moon})\Omega^2 \qquad S5.2$$

Similarly total angular momentum of Earth-Moon System is as follows:

$$J_T = C\omega + \left(\frac{M_{Moon}}{1 + \frac{M_{Moon}}{M_{Earth}}}\right) a^2 \times \Omega + (0.4 M_{Moon} R_{Moon}^2)\Omega \qquad S5.3$$

Substituting the numerical values in Eq.S5.3 we obtain:

$J_T$ = 3.44026×10$^{34}$ Kg-m$^2$/sec ;

From Sharma (2011) (8), we have the following relation between Length of Sidereal Month and Length of Sidereal Day:

$$\frac{LOM}{LOD} = \frac{\omega}{\Omega} = A \times a^{1.5} - F \times a^2 \qquad S5.4$$

$$Where \; A = \frac{J_T}{BC} \; and \; F = \left(\frac{M_{Moon}}{C(1 + \frac{M_{Moon}}{M_{Earth}})}\right)$$

$$Here \; B = \sqrt{G(E + m)}$$

Substituting the numerical values we get:

B=20.08884482×10$^6$ m$^{3/2}$/s;

A=2.13531×10$^{-11}$ m$^{-3/2}$ ;

F = 9.05036×10$^{-16}$ m$^{-2}$ ;

If the numerical values of A and F are substituted in Eq.S5.4.and the present value of 'a' is substituted we get LOM/LOD = 27.2 whereas we should get 27.3. This is because of margin of error in Globe-Orbit parameters of E-M system. By tweaking A and F we get the observed value of LOM/LOD=27.322.

$$\frac{LOM}{LOD} = \frac{\omega}{\Omega} = A \times a^{1.5} - F \times a^2 = 1 \qquad S5.5$$

The roots of S5.5 yield inner geo-synchronous orbit $a_{G1}$ and $a_{G2}$ . We get exactly the same values from the extremum points of Total Energy analysis.

Rewriting total rotational Kinetic Energy expression from Eq.S5.5 we get:

$$KE = \frac{1}{2} C \omega^2 + \frac{1}{2} \left( \frac{M_{Moon}}{1 + \frac{M_{Moon}}{M_{Earth}}} \right) a^2 \times \Omega^2 + \frac{1}{2} \times (0.4 M_{Moon} R_{Moon}^2) \Omega^2$$

Reshuffling the angular velocity terms we get:

$$KE = \frac{1}{2} \Omega^2 \left[ C \left( \frac{\omega}{\Omega} \right)^2 + \left( \frac{M_{Moon}}{1 + \frac{M_{Moon}}{M_{Earth}}} \right) a^2 + (0.4 M_{Moon} R_{Moon}^2) \right]; \qquad S5.6$$

Substituting Eq.S5.4 in Eq.S5.6 we get:

$$KE = \frac{1}{2} \Omega^2 \left[ C(A \times a^{1.5} - F \times a^2)^2 + \left( \frac{M_{Moon}}{1 + \frac{M_{Moon}}{M_{Earth}}} \right) a^2 \right.$$
$$\left. + (0.4 M_{Moon} R_{Moon}^2) \right]; \qquad S5.7$$

According to Kepler's 3$^{rd}$ Law:

$$a^3 \Omega^2 = G(M_{Earth} + M_{Moon}) \qquad S5.8$$

Substituting Eq.S5.8 in Eq.S5.7 we obtain:

$$KE = \frac{1}{2} \times \frac{G(M_{Earth} + M_{Moon})}{a^3} \left[ C(E \times a^{1.5} - F \times a^2)^2 + \left( \frac{M_{Moon}}{1 + \frac{M_{Moon}}{M_{Earth}}} \right) a^2 \right.$$
$$\left. + (0.4 M_{Moon} R_{Moon}^2) \right]; \qquad S5.9$$

Therefore total energy of the E-M System is:

$$TE = KE + PE$$

$$TE = \frac{1}{2} \times \frac{G(M_{Earth} + M_{Moon})}{a^3} \left[ C(E \times a^{1.5} - F \times a^2)^2 + \left( \frac{M_{Moon}}{1 + \frac{M_{Moon}}{M_{Earth}}} \right) a^2 \right.$$

$$\left. + (0.4 M_{Moon} R_{Moon}^2) \right] - \frac{GM_{Earth}M_{Moon}}{a} \qquad S5.10$$

To determine the stable and unstable equilibrium points in non-keplerian journey of Moon we must examine the Plot of Eq.S5.10 from 'a' = $8\times10^6$ m to 'a' = $6\times10^8$m;

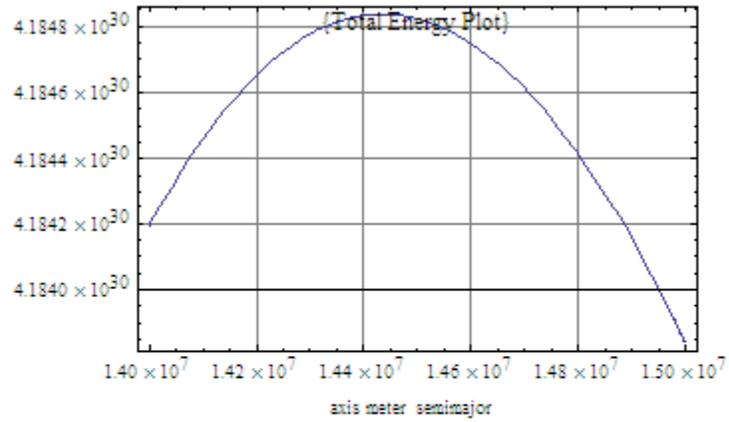

**Total Energy Plot from a=1.4\*10$^7$m to a=1.5\*10$^7$m**
**Inner Geosynchronous Orbit= aG1=1.46177\*10$^7$m**

**Figure S5.1. Plot of total energy in the range 1.4×10$^7$m and 1.5×10$^7$m around the inner geo-synchronous orbit of a = 1.46×10$^7$m.[Courtesy: Author]**

We find an energy Maxima at inner geo-synchronous orbit ($a_{G1}$=1.46×10$^7$m) hence it is unstable equilibrium point. When Moon is at inner-geosynchronous orbit, any perturbation launches Moon on either a sub-synchronous orbit or on extra-synchronous (or super-synchronous orbit). If it is launched on sub-synchronous orbit then it rapidly spirals in towards the primary body and if it is launched on extra-synchronous orbit then it spirals out from inner to outer geosynchronous orbit. In our case, Moon is fully formed beyond Roches' Limit which is 18,000Km [31] just beyond inner Geo-synchronous Orbit hence Moon is launched on expanding spiral orbit towards outer Clarke's Orbit or outer Geo-synchronous Orbit.

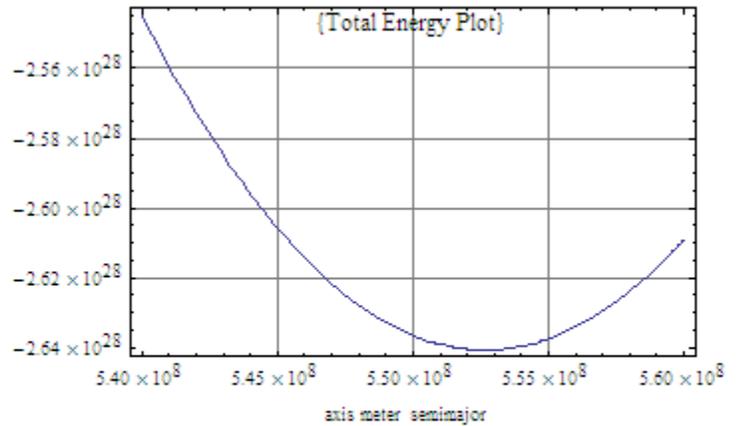

**Total Energy Plot from a=5.4*10^8 m to a= 5.6*10^8 m.**
**Outer geo-synchronous orbir = aG2=5.52656*10^8 m.**

**Figure S5.2. Plot of total energy in the range $5.4{\times}10^8$m and $5.6{\times}10^8$m around the outer geo-synchronous orbit of $a_{G2} = 5.527{\times}10^8$m.[Courtesy: Author]**

As seen in Figure S5.2 at outer geosynchronous orbit (**$a_{G2} = 5.527{\times}10^8$m.**)there is energy minima hence it is stable equilibrium point. Secondary body can never move beyond this orbit. Either it is stay-put in that orbit or it gets deflected back into a contracting spiral orbit.

**The outer Geosynchronous Orbit defines the sphere of gravitational influence of Earth in much the same way as Hill Radius does for Earth in presence of Sun.**

$$\boldsymbol{Hill\ Radius\ =\ R_H = R \times (\frac{M_+}{3M_\odot})^{\frac{1}{3}}} \qquad\qquad \boldsymbol{S5.11}$$

R = 1AU=$1.49598{\times}10^{11}$m.

Substituting the mass of Earth and Sun, Hill Radius is $1.49{\times}10^9$m whereas $a_{G2}=5.527{\times}10^8$m.

The results of KM are validated by an alternate method namely total energy profile analysis method

**S6. Theoretical Formalism of observed LOD within AKM framework. (The methodology of LOD theoretical formalism is given in Protocol Exchange URL http://doi.org/10.1038/protex.2019.017**

**S6.1. Data Set of Observed LODs known with high confidence level.**

John West Wells (1963,1966) using coral fossils and geo-chronometry has given 10 data points: Length of Day(LOD) in hours extending from the present time to pre-Cambrian era. Kaula and Harris(1975) have given two data points. Charles P.Sonnet and Chan(1998)(38) have given one data point at 900Ma.Leschiuta and Tavella(2001)have given 3 data point. Arbab (2009)(39) has given 18 data points.

For analysis purposes Leschiuta and Tavella (2001)(36) and Kaula and Harris (1975) data set have been rejected because LOD in those cases are based on synodic months and synodic months give an over-estimation of LOD as shown in S4 Appendix B.

LODs data set known with high level of confidence have been tabulated in Table S6.1.

**Table S6.1. List of LODs and their corresponding geological epochs.**

| Data set # | Time B.P.(years)[1]. | Orbital radii $(\times 10^8 m)^2$ | Orbital radii $(\times 10^8 m)^3$ | V(a)[4] (cm/y) | LOD(h) | Comment |
|---|---|---|---|---|---|---|
| 1 | Present | $3.844=60.336 R_E$ | $3.844=60.336 R_E$ | 3.7 | 24 | |
| | 45Ma | | $3.8275=60.07 R_E$ | | 23.566(Ref.1) | Ref 1 |
| 2 | 65Ma | $3.8287=59.63566 R_E$ | $3.8198=59.965 R_E$ | 3.755 | 23.627(Ref.2) | Ref 2 |
| 3 | 135Ma | $3.81213=59.83566 R_E$ | $3.79315=59.547 R_E$ | 3.814 | 23.25(Ref.2) | Ref 2 |
| 4 | 136Ma | $3.8118=59.83 R_E$ | $3.7927=59.54 R_E$ | 3.815 | 23.2(Ref.5) | Ref 5 |
| 5 | 180Ma | $3.80129=59.6655 R_E$ | $3.7756=59.27 R_E$ | 3.8526 | 23(Ref.2) | Ref 2 |
| 6 | 230Ma | $3.7891=59.474 R_E$ | $3.7558=58,96 R_E$ | 3.897 | 22.7684(Ref.2) | Ref 2 |
| 7 | 280Ma | $3.7768=59.28 R_E$ | $3.7357=58.645 R_E$ | 3.94 | 22.4765(Ref.2) | Ref 2 |
| | 300Ma | | $3.7275=58.516 R_E$ | | 22.3(Ref.4) | Ref 4 |
| 8 | 345Ma | $3.76055=59.28 R_E$ | $3.7068=58,19 R_E$ | 4 | 22.136(Ref.2) | Ref 2 |
| 9 | 380Ma | $3.7517=58.887 R_E$ | $3.69418=57.99 R_E$ | 4.034 | 21.9(Ref.2) | Ref 2 |

| 10 | 405Ma | 3.74535=58.787R$_E$ | 3.6825=57.8257 R$_E$ | 4.05815 | 21.8(Ref.2) | Ref 2 |
|---|---|---|---|---|---|---|
| 11 | 500Ma | 3.7208=58.4R$_E$ | 3.6422=57.177 R$_E$ | 4.15 | 21.27(Ref.2) | Ref 2 |
| 12 | 600Ma | 3.6943=57.986R$_E$ | 3.5968=56.4646 R$_E$ | 4.25357 | 20.674(Ref.2) 20.7(Ref.4) | Ref 2 Ref 4 |
| 13 | 715Ma | 3.663=57.495R$_E$ | 3.5423=55.609 R$_E$ | 4.377 | 20.1(Ref.5) | Ref 5 |
| 14 | 850Ma | 3.6251=56.9R$_E$ | 3.4748=54.549 R$_E$ | 4.531 | 19.5(Ref.5) | Ref 5 |
| 15 | 900Ma | 3.61075=56.67R$_E$ | 3.4485=54.1365 R$_E$ | 4.59 | 18.9(Ref.3) 19.2(Ref.4) (Ref.5) | Ref 3 Ref 4 Ref 5 |
| 16 | 1200Ma | 3.5205=55.26R$_E$ | 3.2775=51.452 R$_E$ | 4.98 | 17.7(Ref.5) | Ref 5 |
| 17 | 2000Ma | 3.235=50.8R$_E$ | 2.6=40.8R$_E$ (not permissible)† | Oscillatory In AKM | 14.2(Ref.5) | Ref 5 |
|  | 2450Ma (3.28×10$^8$ m) |  | 1.96=30.7R$_E$ (not Permissible)† | Oscillatory In AKM |  |  |
| 18 | 2500Ma | 3.012=47R$_E$ |  |  | 12.3(Ref.5) | Ref 5 |
| 19 | 3000Ma | 2.735=43R$_E$ |  |  |  |  |
| 20 | 3560Ma | 2.3143 |  |  |  |  |
| 21 | 4500Ma | 0.18 |  |  |  |  |

[1]Based on annual bands in coral fossils.

[2]Orbital radii based on monotonically evolving Moon with Moon's age 4.467Gy(KM).

[3]Orbital radii based on accelerated Moon.(AKM)

[4]Velocity of recession of Moon based on presently accelerated Moon(AKM)

Reference 1:Kaula& Harris (1975)(37)

Reference 2:John West Wells (1965,1966) (34,35)

Reference 3:Charles P. Sonnett& Chan(1998) (38)

Reference 4:Leschiuta&Tavella (2001) (36)

Reference 5:A.J.Arbab (2009) (39)

† 33R$_E$ orbital radius is Cassini State Transition orbit where E-M system becomes unstable due to the transition in Lunar Spin axis from State 1 to State2.

Table S6.1 clearly brings out the analogy of our classical Moon and the real Moon with tortoise and hare parable. Our classical Moon was slow and steady but our real Moon first moved in fits and interruptions and in the last 1.2Gy real Moon has bounded to its goal. Real Moon has raced to its present orbit and hence we see 3.82±0.07cm/y lunar orbital recession rate which is anomalously high recession rate indicative of anomalously high tidal dissipation rate in Earth..

Advanced Kinematic Model of Earth-Moon system was introduced in CELE-D-17-00144 and is described in Sharma(2019).. It is being used here to predict the theoretical LOD curve .

Earth's obliquity angle is not defined at 30$R_E$ ,35$R_E$ and 40$R_E$ hence in AKM the lunar orbital radius from 45$R_E$ to 60.336$R_E$ is the permissible range of analysis (Sharma 2019).

### S6.2. Methodology of theoretical formalism of LOD curve in AKM (in the main text).

Rewriting its basic equation from the main text containing LOM/LOD we get:

$$(N)^2 \times a^3 = X^2 + (F \times \sqrt{1-e^2})^2 \times (a^2)^2 + G^2 +$$

$$2\left(F \times \sqrt{1-e^2} \times a^2\right)(G)\left\{\sqrt{1-D^2}\sqrt{1-A^2} - AD\right\} + 2 \times X$$

$$\times \sqrt{\left(F \times \sqrt{1-e^2} \times a^2\right)^2 + (G)^2 + 2\left(F \times \sqrt{1-e^2} \times a^2\right)(G)\left\{\sqrt{1-D^2}\sqrt{1-A^2} - AD\right\}}$$

$$\times \left\{\sqrt{1-A^2}\sqrt{1-B^2} - A.B\right\} \qquad S6.20$$

Where the different symbols are defined as follows:

$$\frac{J_4}{C \times B} = N$$

$$here\ C = 0.3308 \times M_{Earth} \times R_{Earth}^2$$

$$= present\ day\ spin\ moment\ of\ inertia\ of\ Earth\ ,$$

$$B = \sqrt{G(M+m)} = 2.00873 \times 10^7 \frac{m^{3/2}}{s},$$

$$J_4 = total\ vector\ sum\ of\ the\ angular\ momentums\ of\ E - M\ system \quad S9.21$$

$$\frac{F^*}{C} = F\ and\ \frac{I}{C} = G\ and\ X = \frac{LOM}{LOD}$$

$$here\ I = 0.394 \times m \times R_{Moon}^2 = spin\ moment\ of\ Inertia\ of\ Moon \qquad S6.22$$

Sin[β] = D and Cos[β] =√(1-D$^2$),

Sin[α] = A and Cos[α] =√(1-A$^2$),

Sin[Φ] = B and Cos[Φ] =√(1-B$^2$),

$$Cos[\alpha]Cos[\Phi]-Sin[\alpha]Sin[\Phi]=\sqrt{(1-A^2)}\sqrt{(1-B^2)} - A.B. \qquad\qquad S6.23$$

The empirical relation describing the evolution of Moon's orbital plane inclination with respect to the ecliptic is(Sharma 2019):.

$$Inclination\ angle\ \alpha = \frac{1.18751 \times 10^{25}}{a^3} - \frac{7.1812 \times 10^{16}}{a^2} + \frac{1.44103 \times 10^{8}}{a}$$
$$-8.250567342 \times 10^{-3} \qquad\qquad S6.31$$

The empirical relation describing the evolution of Moon's obliquity angle (β) is given as below(Sharma 2019):

$$Moon's\ Obliquity\ angle\ \beta$$
$$= 3.36402 - 1.37638 \times 10^{-8}a + 1.32216 \times 10^{-17}a^2 \qquad S6.32$$

The empirical relation describing the evolution of Moon's orbit eccentricity is(Sharma 2019):.

$$e = 0.210252 + 8.38285 \times 10^{-10}a - 3.23212 \times 10^{-18}a^2 \qquad\qquad S6.33$$

$$\frac{LOM}{LOD} = \frac{\omega}{\Omega} = -12.0501 + 2.6677 \times 10^{-7} \times a - 4.27538 \times 10^{-16} \times a^2 \qquad S6.16$$
$$\varphi = -0.732299 + 2.97166 \times 10^{-9} \times a \qquad\qquad S6.17$$

(S6.31), (S6.32) , (S6.33) , (S6.16) and  (S6.17) will be used to solve the quadratic equation given in (S6.20). Two roots of (S6.20) are obtained ,out of which positive root is retained and will be used for analysis purpose.

We have altogether 5 spatial function (S6.31), (S6.32) , (S6.33) , (S6.16) and  (S6.17)  describing the evolution of inclination angle (α), Moon's obliquity (β), eccentricity(e) of lunar orbit, LOM/LOD and Earth's obliquity (Φ) respectively through different geologic epochs. Table S6.2 gives the evolution of these parameters through past geologic epochs.

**Table S6.2. Evolutionary history of  ω/Ω (LOM/lOD) , α (Inclination angle) , β (lunar obliquity), e (eccentricity) and Φ (terrestrial obliquity).**

| a (×R_E) | a (×10⁸m) | ω/Ω | α radians | β | e | Φ(rad) | Sin[Φ] |
|---|---|---|---|---|---|---|---|
| 30 | 1.9113 | 23.3752 | 0.480685 (27.4°) | 1.21635 (69.69°) | 0.2524 | unstable | -0.464076 |
| 35 | 2.22985 | 26.1194 | 0.26478 (15.17°) | 0.952317 (54.56°) | 0.236 | unstable | -0.216896 |
| 40 | 2.5484 | 28.1147 | 0.168969 (9.68°) | 0.71512 (40.97°) | 0.214 | 0.0213773 | 0.0213757 |
| 45 | 2.86695 | 29.2938 | 0.124631 (7.1408°) | 0.504756 (28.92°) | 0.1849 | 0.113792 (6.51°) | 0.113547 |

| 50 | 3.1855 | 29.5965 | 0.103801 (5.04736°) | 0.321225 (18.4°) | 0.1493 | 0.220227 (12.6°) | 0.218451 |
|---|---|---|---|---|---|---|---|
| 55 | 3.50405 | 28.9877 | 0.0941394 (5.39379°) | 0.164527 (9.4267°) | 0.10714 | 0.314929 (18°) | 0.309749 |
| 60 | 3.8226 | 27.4 | 0.0898729 (5.149°) | 0.03466 (1.986°) | 0.0584 | 0.398676 (22.84°) | 0.388198 |
| 60.336 | 3.844 | 27.32 | 0.08971 (5.14°) | 0.0268 (1.54°) | 0.0549 | 0.409105 (23.44°) | 0.397788 |

### S6.2.1. The formalism of the velocity of recession of Moon in AKM.

The radial velocity of lunar recession is the same as in the classical case:

$$V(a) = \frac{2K}{m^*B} \times \frac{1}{a^Q} \left[ A \times a^2 - F \times a^{2.5} - \sqrt{a} \right] \times 31.5569088 \times 10^6 m/y \qquad S6.23$$

Except in AKM it is rewritten as:

$$V(a) = \frac{2K}{m^*B} \times \frac{\sqrt{a}}{a^Q} [X - 1] \times 31.5569088 \times 10^6 m/y \qquad S6.24$$

In (S6.24) 'a' refers to the semi-major axis of the evolving Satellite. There are two unknowns: exponent 'Q' and structure constant 'K'. Therefore two unequivocal boundary conditions are required for the complete determination of the Velocity of Recession.

Equation (S6.20) gives the expression of the permissible X in advanced Kinematic Model. That permissible X is substituted in (S6.24) for analysis purpose.

By classical E-M model Q is calculated to be Q = 3.22684.

K = 5.5×10$^{42}$Newton-m$^{Q+1}$ ,Transit Time (from 3.012×10$^8$m to 3.844×10$^8$m) =2.38Gy.

This gives present epoch velocity of recession of Moon as =2.4cm/y.        S6.A.

In 'Fits and Bound model of E-M system:

K = 8.33269×10$^{42}$Newton-m$^{Q}$,

Transit Time (from 3.012×10$^8$m to 3.844×10$^8$m) = 1.57732 Gy.

This gives present epoch velocity of recession of Moon as = 3.7cm/y.        S6.B

So for our calculations we will retain the structure constant in (S6.B). This helps achieve correspondence with LLR result = 3.7cm/y.

Now this can be justified.

From 3R$_E$ to  45R$_E$ Moon does not have a smooth transit. Infact it is bumpy. It is chaotic, gets stuck in resonances and comes out of the resonances and gets stalled and resumes its tidal evolution. In fact Moon takes 3.267Gy to spirally expand from  3R$_E$ to  45R$_E$ in fits and stalled manner. From  45R$_E$ to  60.336R$_E$, Moon is literally accelerated though smoothly coasts in 1.2Gy to its present orbit.This accelerated spiral expansion results in present day velocity of recession of 3.7cm/y.

Since Adv.KM is well defined  from 45R$_E$ (Cassini State2)  to 60.336R$_E$ so data set within this range  only is considered,

**S6.2.2. Theoretical formalism of LOD.**

Our basic assumption has been:

$$X = \frac{LOM}{LOD} \ \ where \ LOM = P_{ORB}(E - Msystem)$$

$$= \frac{2\pi}{B} \times a^{3/2} \ from \ Kepler's \ third \ law. \qquad S6.25$$

From S6.25, reshuffling the terms we get:

$$LOD = \frac{2\pi a^{3/2}}{B} \times \frac{1}{X} \qquad\qquad S6.26$$

Equation (S6.20) gives the expression of the permissible X in advanced Kinematic Model. That permissible X is substituted in (S6.26) for generating Theoretical LOD curve.

**S6.2.3. New features of Lunar Tidal evolution in AKM.**

In Figure S6.1, tidal evolution of Moon's orbital radius is given based on the new study of MatijaCuk et.al.(2016)

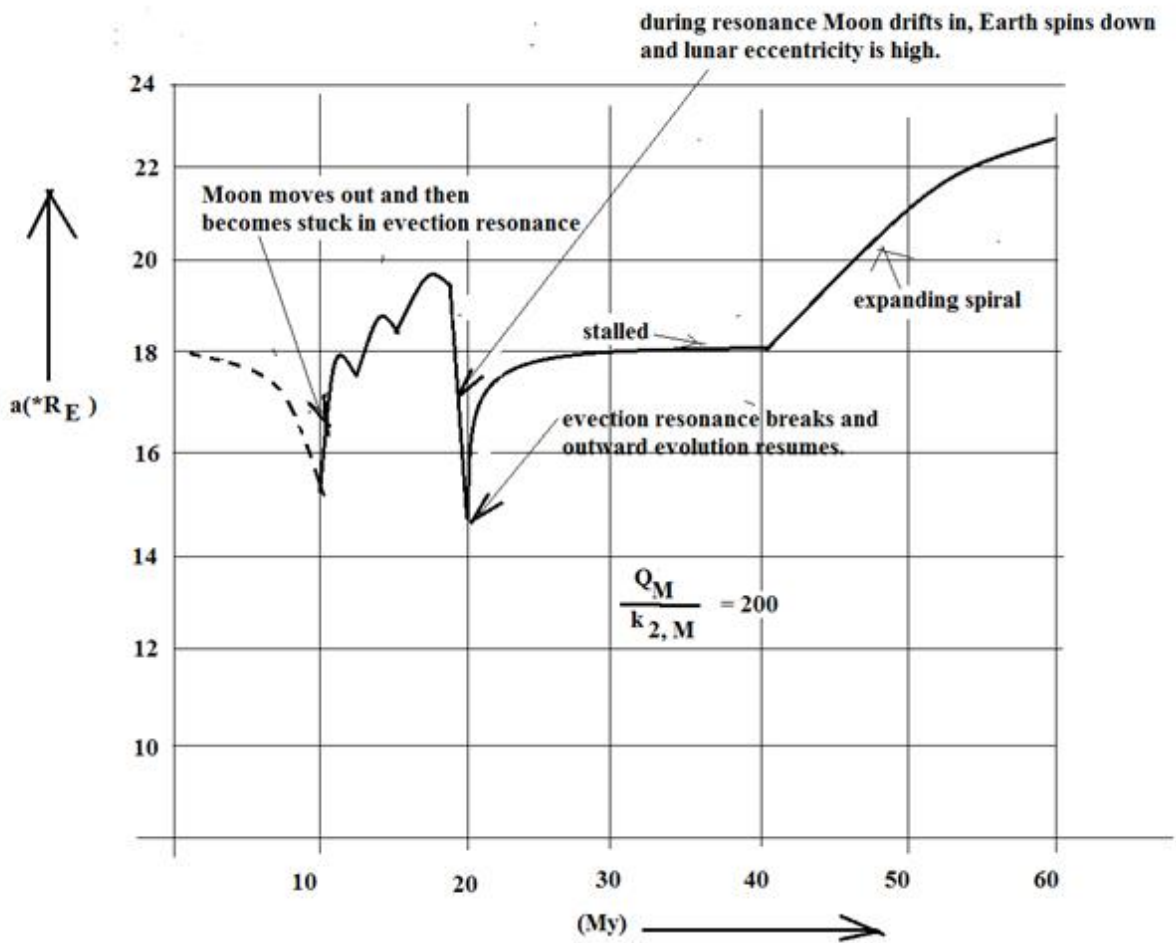

**Figure S6.1. Evolution of semi-major axis of Moon for first 60My when Laplace plane transition is encountered at 20My and a= 17R$_E$ .[ Courtesy:** modified form of Extended Data Figure 5. Early tidal evolution of the Moon with Q$_E$/k$_{2,E}$ = 200 through out the simulation in MatijaCuk,; Douglas P. Hamilton,; Simon J. Lock,; Sarah T. Stewart (2016)."Tidal evolution of of the Moon from high obliquity, high angular momentum Earth", *Nature,* **539,** 402-406, (2016) doi: 10/1038/nature19846]

Moon must have been formed at giant impact stage of terrestrial planets formation. In this final stage of terrestrial planet formation proto-planets collide with one another to form planets. All collisions may not lead to accretion, specifically for high velocity and/or large impact parameters. Accretion condition is derived for protoplanet collisions in terms of impact velocity, angles and masses of colliding bodies. EiichiroKokubo and HidenariGenda,(2010) have adopted, realistic accretion condition in N-body simulation of terrestrial planet formation from proto-planets. In realistic accretion models about half of collisions do not lead to accretion, spin angular velocity obeys Gaussian Distribution and obliquity obeys isotropic distributions independent of accretion condition.

In real life, Laplace Plane Transition plays a significant role in E-M tidal evolution due to high obliquity Earth due to giant impact. Laplace Plane shifts from Earth's equatorial plane to Solar System's ecliptic plane as Moon recedes from Earth. For any perturbed orbit, there exists a Laplace Plane around the normal of which the normal of the orbital plane of the perturbed orbit precesses. The Laplace Plane undergoes a transition during lunar tidal evolution when the Moon recedes from inner region dominated by perturbation of the Earth's equatorial bulge to the region dominated by solar perturbation. At 17R$_E$ the shift occurs. This referred to as r$_L$ .

$$r_L(laplace\ plane\ transition\ orbit\ radius) = (2J_2 \frac{M_E}{M_S} R_E^2 a_E^3 \times (1 - e^2)^{3/2})^{1/5} \qquad S6.1$$

J$_2$ = oblateness moment of Earth. As Earth spin slows down, oblateness decreases leading to r$_L$ moving inward.

$$r_L = 16 \sim 22 R_E \qquad\qquad S6.2$$

In Figure S6.1 we clearly see the interrupted and stalled tidal evolution of Moon upto 45R$_E$ . For Earth's obliquity Φ less than 68.9° , Laplace Plane transition is smooth and inclination and eccentricity remain zero. But for Earth's obliquity Φ greater than 68.9° , Laplace Plane transition causes  orbital instability, acquires substantial eccentricity and substantial inclination driven  by solar secular perturbation that operate at high inclination as seen in Kozai resonance (Adobe & Ida 2007).Tidal evolution of the Moon from high obliquity Earth is followed by inclination damping  at the Cassini state transition due to lunar obliquity tides which (that is lunar obliquity) becomes as high as 77° periodically for a brief period of time.

Tides raised on the Earth by the Moon have caused an expansion of the spiral orbit. Tides raised on Moon by Earth have de-spun the Moon to synchronous rotation and driven the spin axis of Moon to a Cassini State. Cassini State is a co-processing configuration. In  co-processing configuration ,  Lunar's spin axis becomes coplanar with the lunar orbit normal  and with the normal of the Laplace plane (which at present is coincident with the normal of the ecliptic).Moon is pushed-out due to Earth's tides but is pushed-in due to Moon's tide..

After Laplace Plane transition, Moon continues to recede and lunar spin axis passes through Cassini state transition. Regardless of the nature of the lunar rotation state, Moon's obliquity is very high during Cassini state transition and immediately following it, leading to  the damping of lunar inclination. Lunar inclination is damped from 30° ( obtained during Laplace plane transition)  to the present value of 5.334° if we assume the long term average  tidal properties for Earth and non-dissipative Moon in synchronous rotation states.

At first the Moon's orbit normal and Moon's spin axis are on the same side of the normal to the Ecliptic, indicating that the Moon is in Cassini State 1.Once Cassini state is de-stabilized after some wobbling, the Moon settles in non-synchronous state some what similar to Cassini State 2 (with orbit normal and spin axis being on the opposite sides of the normal to the ecliptic).During this time both the inclination and obliquity (which is forced by inclination) are being damped by lunar obliquity tides. At a = 35.1$R_E$ the Moon becomes synchronous again and enters Cassini State 2, where it stays for the rest of the simulation (this event is visible as 5° jump in obliquity)(Cuk et.al.2016).

In the "Animation of relative orientation of earth's spin and Moon's orbit during the Laplace Plane Transition" given in (Cuk et.al.2016), the simulation covers a time span of 60My and Moon's spiral orbit expands in fits from 10$R_E$ to 18$R_E$ . Subsequently there is monotonic expansion from 18$R_E$ to 23$R_E$ . As animation progresses, Laplace Plane shifts from Earth's equatorial plane to Solar System's ecliptic plane. At 17$R_E$ the shift occurs.

**Table S6.3. The history of Earth's obliquity (Φ) and Moon's orbital plane inclination (α) during Laplace Plane transition.**

| Time (My) | Earth's Spin(h) | Ecliptic Component Of J(%) | Laplace Plane | Earth's obliquity (Φ) | Moon's orbital plane inclination (α) | comment |
|---|---|---|---|---|---|---|
| 0.25 | 2.86 | 153.1 | Equatorial plane of Earth | 70° | 0° | |
| 0.99 | 2.95 | 151.8 | Equatorial plane of Earth | 60° | 0° | |
| 3.4 | 3.21 | 131.8 | Equatorial plane of Earth | 55° | oscillatory | Falls into secular resonance |
| 9.5 | 3.73 | 127.2 | Equatorial plane of Earth | 50° | oscillatory | Moon moves out and gets stuck in evection resonance |
| 12 | 3.98 | 139 | Transition | 40° | oscillatory | unstable |
| 20.5 | 5.66 | 112.1 | Ecliptic | 30° | oscillatory | Comes out of resonance |
| 25 | | | | | | stalled |
| 28.6 | 6.4 | 102.8 | Ecliptic | 20° | oscillatory | stalled |
| 40 | | | | | | stalled |
| 44.8 | 7.04 | 103.6 | Ecliptic | 15° | 20° | Expansion resumed |
| 59.5 | 7.35 | 103.1 | ecliptic | 10° | 25° | expanding |

**S6.3. Theoretical Validation of Observed LOD for accelerated MOON**

MatijaCuk et.al.(2016) have proposed a radically different model where Moon tidally evolves in fits and bound. Fits are due to stalling of Moon tidal evolution due to strong lunar obliquity tides created in Laplace Plane transition. Bound is due to accelerated transit time of 1.2Gy in spiraling out from $3.274\times 10^8$m to the present lunar orbit of $3.844\times10^8$m as compared to 1.9Gy for the classical Moon for an identical orbital radius expansion. Application of Advanced Kinematic Model to fits and bound model of Moon at one stroke removes the tension between Lunar Laser Ranging measurement of 3.7cm/y (Dickey et.al.1994) and theoretically predicted Lunar recession of 2.3cm/y assuming 4.467Gy for the classical Moon on one hand and gives a perfect match between observed LOD curve and theoretically predicted LOD curve for last 900My and near perfect match over last 1.2Gy with a mismatch of -1.3%.

### 6.3.1.Theoretical Formalism of observed LOD curve over 900My time span.

10 Data points given byJohn West Wells (1965,1966),and Charles P. Sonnettand M.A.Chan.(1998)are considered and tabulated in Table S9.4.

Data set given by Kaula& Harris (1975) and Leschiuta&Tavella (2001) have been rejected because of over-estimation errors.

**Table S6.4. Tabulation of LOD in past geologic epochs for accelerated Moon.(Structure constant K= 8.333269 N-m[Q], Q= 3.22684, present velocity of recession = 3.7cm/y)**

| Data set # | Time B.P.(years)[1] . | Orbital radii ($\times10^8$m)[3] | LOD(h) |
|---|---|---|---|
| 1 | Present | 3.844 | 24 |
| 2 | 45Ma | 3.8275 | 23.566(Ref.1) |
| 3 | 65Ma | 3.8198 | 23.627(Ref.2) |
| 4 | 135Ma | 3.79315 | 23.25(Ref.2) |
| 5 | 136Ma | 3.7927 | 23.2(Ref.5) |
| 6 | 180Ma | 3.7756 | 23(Ref.2) |
| 7 | 230Ma | 3.7558 | 22.7684(Ref.2) |
| 8 | 280Ma | 3.7357 | 22.4765(Ref.2) |
| 9 | 300Ma | 3.7275 | 22.3(Ref.4) |
| 10 | 345Ma | 3.7068 | 22.136(Ref.2) |
| 11 | 380Ma | 3.69418 | 21.9(Ref.2) |
| 12 | 405Ma | 3.6835 | 21.8(Ref.2) |
| 13 | 500Ma | 3.6422 | 21.27(Ref.2) |
| 14 | 600Ma | 3.5968 | 20.674(Ref.2) 20.7(Ref.4) |
| 15 | 715Ma | 3.5423 | 20.1(Ref.5) |
| 16 | 850Ma | 3.4748 | 19.5(Ref.5) |
| 17 | 900Ma | 3.4485 | 18.9(Ref.3) 19.2(Ref.4) (Ref.5) |

| 18 | 1200Ma | 3.2775 | 17.7(Ref.5) |
|----|--------|--------|-------------|
| 19 | 2000Ma | 2.6(Not Permisible) | 14.2(Ref.5) |
|    | 2450Ma ($3.28\times10^8$m) | 1.96(Not Permisible) | |
| 20 | 2500Ma | | 12.3(Ref.5) |
| 21 | 3000Ma | | |
| 22 | 3560Ma | | |
| 23 | 4500Ma | | |

[1]Based on annual bands in coral fossils.

[3]Orbital radii based on accelerated Moon.

**6.3.2.ListPlot of LOD for accelerated Moon for the time span of 900My.**

Figure S6.2 gives the observed LOD curve for fits and bound Moon covering a time span of 900My.

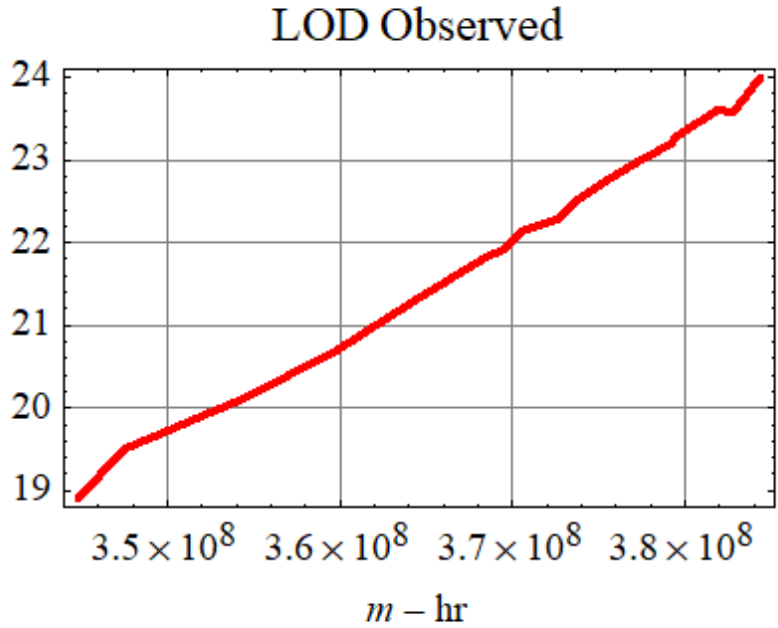

*Figure S6.2. Observed LOD curve for fits and bound Moon covering a time span of 900My.[Courtesy:Author]*

### 6.3.3. Generation of the theoretical LOD curve from Equation S6.26.

Equation S6.26 is as follows:

$$LOD = \frac{2\pi a^{3/2}}{B} \times \frac{1}{X} \qquad\qquad S6.26$$

Using (S6.26) theoretical curve is generated and is displayed in Figure S6.3.

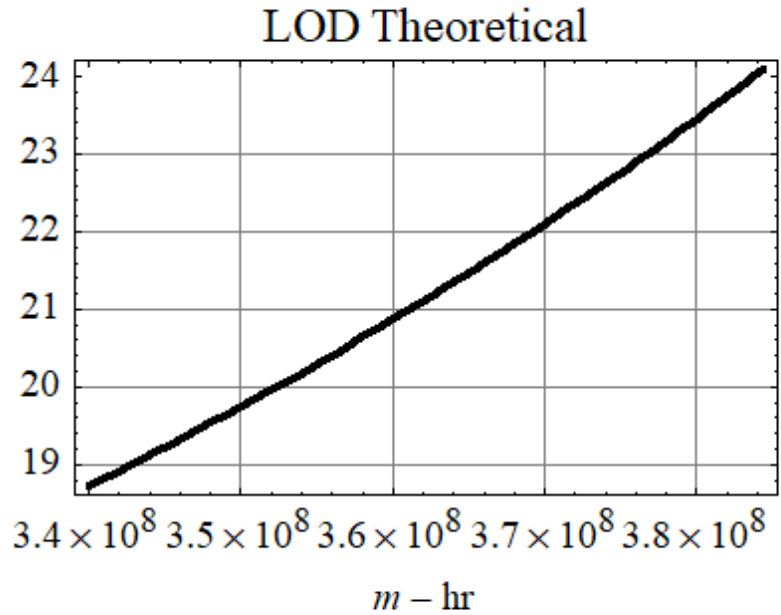

**Figure S6.3.Theoretical Plot for LOD curve for accelerated Moon over 900My from the present.[Courtesy:Author]**

### 6.3.4. Superposition of the Observed curve and the theoretical curve.

Superposing the two curves we get the match between theory and observation as shown in Figure S6.4.

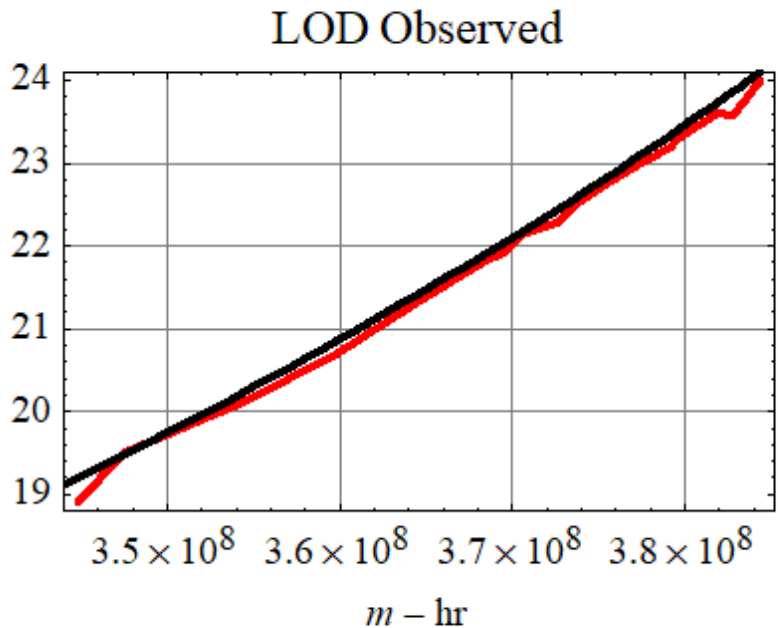

**Figure S6.4. Precise match between Theoretical and Observed LOD Plot over last 900My.[Courtsey:Author]**

As we see the two curves in Figure S6.4. have 100% match. Hence formalism given in S9.26 is an exact match between Observed LOD curve and theoretical LOD curve.

### 6.3.2.Theoretical Formalism of observed LOD curve over 1.2Gy time span ,the permissible range in AKM.

**Validation of Observed LOD foraccelerated MOON(from 3.274×10⁸m to the present orbit) using 17 Data points given by** John West Wells (1965,1966),Kaula& Harris (1975),Charles P. Sonnett& Chan.(1998) and Leschiuta&Tavella (2001) and Arbab(2009).

As seen in Table S6.4, Arbab (2009) provides LOD as far back as 2500My in remote past.. But his data points fall beyond the permissible range of 45$R_E$ to 60.336$R_E$ .Below 45$R_E$ the parameters particularly obliquity is not uniquely defined. Numerical simulation (Cuk et.al.2016) has shown that Cassini State has instability and oscillations of lunar orbital plane, hence theoretical analysis below 45$R_E$ has been kept out of the permissible range of AKM analysis. Therefore two data points of Arbab (2009) have been rejected namely {2500Ma-12.3h} and

{2000Ma-14.2h} have been rejected. Data points up to {1200My ago -17.7h} have been included in the analysis. Figure S6.5, gives the observed LOD curve over a time span of 1.2Gy. Figure S6.6 gives the theoretical plot of LOD curve over a time span of 1.2Gy.Figure S6.7 gives the precise match between observed and theoretical LOD curves over 1.2Gy.

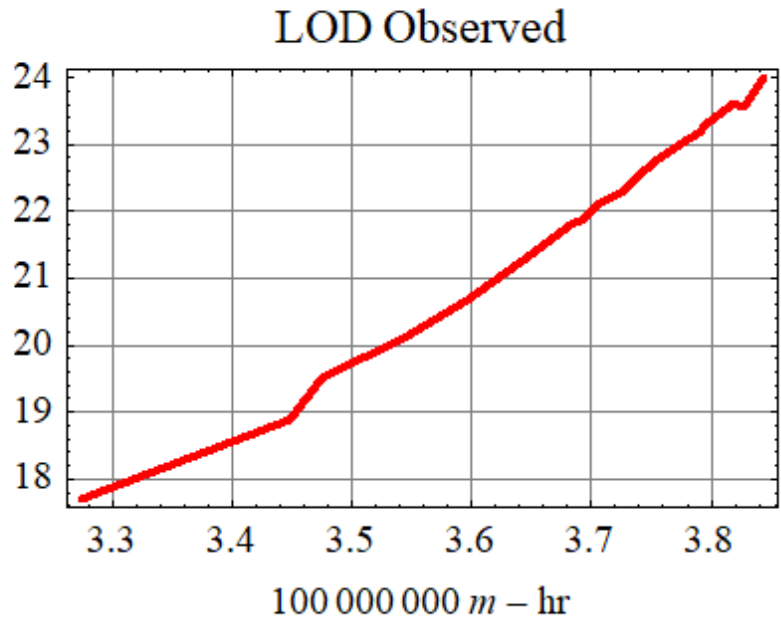

**Figure S6.5..Observed LOD curve for accelerated Moon using 17 data points covering a time span of 1.2Gy.[Courtesy:Author]**

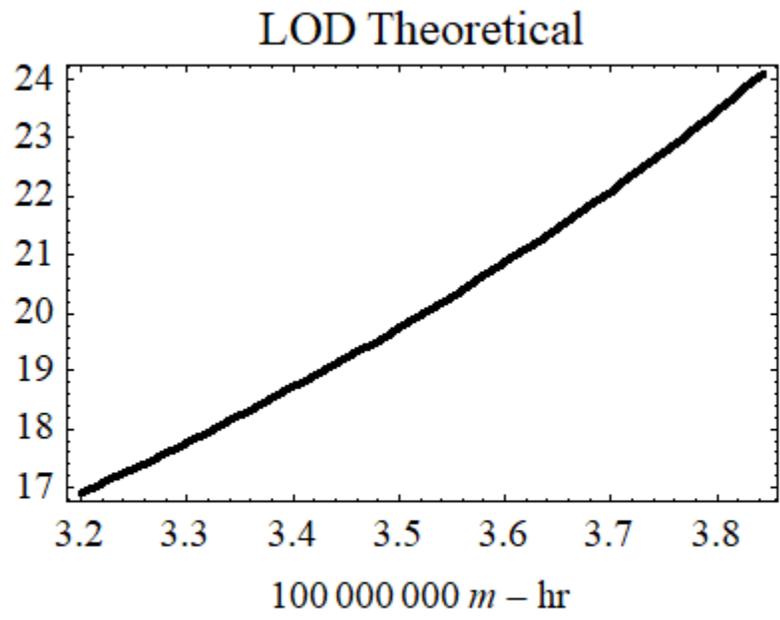

**Figure S6.6.Theoretical Plot for LOD curve for accelerated Moon covering a time span of 1.2Gy.[Courtesy:Author]**

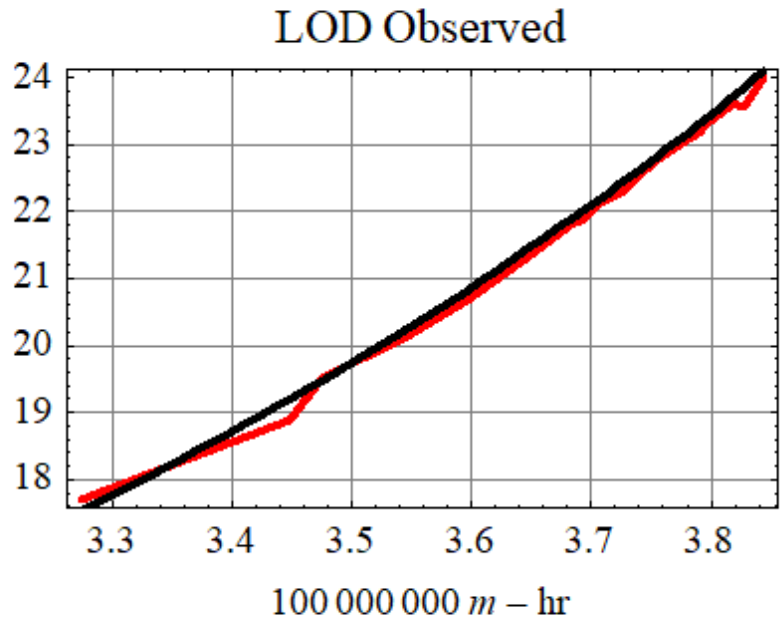

**Figure S6.7. Repeat precise match between Theoretical and Observed LOD Plot over the permissible time span 1.2Gy and spatial range of 45R$_E$ to 60.336R$_E$ of the Advanced kinematic Model of E-M system.[Courtesy:Author]**

As seen in superposition graph Figure S6.7 there is precise match between observed and theoretical LOD curve in the time slot of 1.2Gy.

For classical Moon , from $3.274 \times 10^8$m to the present orbit the transit time is 1.9Gy.

For fits and bound Moon  , from $3.274 \times 10^8$m to the present orbit the transit time is 1.2Gy.

S6.4. **Discussion;**

The Advanced Kinematic Model (in the main text) has been developed by including Earth's obliquity (Φ), Moon's orbital plane inclination with respect to ecliptic(α) as well as lunar obliquity(β) with respect to the lunar orbital normal and by including the vectorial summation of angular momentum vectors. The Laplace Plane transition and Cassini State transition occurring at a =(17 to 19)R$_E$ and at 33R$_E$ respectively have  been kept out of the permissible range of Advanced KM. Advanced KM covers the range of Moon's tidal evolution from 45R$_E$ to 60.335R$_E$ (the present orbital radius). Because of instability and unpredictability of  Laplace Plane transition and Cassini State transition, the range from 3R$_E$ to 45R$_E$ has been kept out of range for Advanced KM. In classical Model of E-M system to satisfy the Age of Moon a lunar recession rate of 2.3cm/y was being adopted which was completely distorting the time scale of tidal evolution of Moon for last 1 Gy. If lunar recession rate of 3.7cm/y was being adopted then too short a Moon's age of 2.7Gy was being obtained which was contrary to the observed Moon's age as obtained from Moon's rocks analysis obtained in Apollo Mission . This conundrum got resolved only after the publication of. land mark paper by Matija Cuk,; Douglas P.Hamilton,; Simon J.Lock,; Sarah T.Stewart (2017)on Moon's tidal evolution in high obliquity, high angular momentum Earth. By adopting the Lunar Laser Ranging data of lunar recession , the AKM in one stroke became self consistent in all respects namely we obtained present Earth day of 24h, present LOM/LOD= 27.322, present lunar recession rate of 3.7cm/y and most of all we got a precise match between observed LOD curve and theoretical LOD curve. This high obliquity, high angular momentum Earth as the origin of Moon also resolved isotopic conundrum and gave a robust mechanism for arriving at climate friendly low obliquity Earth.

## S6.5. Conclusion .

Using Advanced Kinematic Model to obtain a perfect match between observed LOD curve and theoretical LOD curve has been a crowning achievement as well as the ultimate vindication of Advanced KM. This paper has laid to rest all the nagging doubts which have been there for the empirical nature of the tidal torque developed in Kinematic Model. This paper has proved with 95% confidence level that Advanced KM is a valid model and well tested model which can be used with accuracy and reliably for analyzing two body tidally interacting systems. The application range of Advanced KM can be planet-satellite, planet hosting star and planet, star binary, Neutron star binary, neutron star-black hole binary or black hole binary. In subsequent papers the validity of advanced KM will be proved in this wide range of binary pairs.

Most of all the precise theoretical formalism of observed LOD curve has paved the path for searching the necessary precursors for early warning and forecasting methods (EWFM) for earthquakes and sudden volcanic eruptions.

## S7.Gutenberg-Richter Law of Chaos.

The actual sidereal day length does not secularly increase. It has periodic variations as well as non-periodic variations. If it is plotted over a secularly lengthening of day curve there will be deviations. The irregular deviations follow Gutenburg-Richter Law of chaos and hence it is chaotic and not random. This chaos implies that it has underlying causative factors.The various causative factors are plate-tectonics, ice-caps expansion and thawing, El-Nino Ocean

Currents,electromagnetic coupling between core and mantle and global wind pattern interactions with various mountain ranges. If the various signatures are identified then we may isolate the precursors of Earth-quake and Sudden Volcanic Eruptions in l.o.d curve fluctutations.

It is clear that long range correlation of deviation implies chaos and chaos implies causative factors. Subduction of oceanic plates deep under the continental plates forces the folding of the Earth's crust into mountain ranges. It also causes molten magma / lava to rise through weak portions of Earth or through mid-oceanic ridges which is rupturing apart as sudden volcanic eruptions.

Apart from head-on collisions of tectonic plates there are sideway sticking and slipping. These lateral movement give rise to rupturing faults or premonitory creeps.

Rupturing faults behave like sand-piles during stasis. Premonitory creeps are non-critical. Graduallu it develops into an universal critical class. Avalanche rupture occurs. The length of rupturing fault determines the magnitude of Earth-quake .

$N(E) = 1/(E)^P$

where N(E) = number of Earthquake event of size E in a given observation period.

E = the energy released in a given Earthquake.

P = power law = the signature of the chaos.

Plotting of the real-time sidereal day length on the theoretical curve will give a chaotic scatter and chaotic scatter should be related to geo-tectonic movements which is the underlying causative factor of Avalanche Break-down in form of Earth-Quakes or sudden volcanic eruption of dormant volcanoes. This Avalanche Breakdown is followed by long periods of stasis. The chaotic nature becomes evident if the deviation follow Gutenburg-Richter Law. Gutenburg-Richter law comes from Sand-Pile Theory(Buchanan,1997;Coontz,1998).

**The legends of the Figures in supplementary materials.**

**Figure S2.1[.tiff].In Earth-Moon System, Moon is in super-synchronous orbit. The off-setting of the line of bulge in Earth with respect to E-M radius vector creates a tidal drag and de-spinning of Earth leading to secular lengthening of the solar day.The de-spinning and**

consequent reduction in spin angular momentum of Earth leads to increase in the orbital angular momentum of E-M system in line with the conservation principle of total angular momentum of E-M system. Hence de-spinning of Earth is accompanied with expanding spiral orbit of Moon. During the conservative phase of E-M system, when Moon is just born at inner geo-synchronous orbit and the system is at near triple synchrony, by gravitational sling-shot impulsive torque Moon is catapulted on an expanding spiral path towards outer geo-synchronous orbit. After the conservative phase and during the dissipative phase there is no energy transfer from Earth spin to E-M expanding ornital system. Moon on its own, by virtue of the initial energy acquired during the conservative phase, coasts towards the outer geo-synchronous orbit where it gets finally locked-in.[Courtesy:Author]

Figure S2.2. In Mars-Phobos System, Phobos is in sub-synchronous orbit.The tidal bulge in Mars lags M-P radius vector hence Phobos is spinning up Mars. Conservation of angular momentum causes Phobos to be launched on gravitational runaway collapsing spiral orbit also known as death spiral[Courtesy: Author].

Figure S3.1[.tiff]. Radial Velocity Profile of Moon from the birth to the final lock-in point at $a_{G2}$.
   X-axis is semi-major axis. Y-axis is velocity in m/y.[Courtesy:Author]

Figure S3.2[.tiff]. Radial Acceleration Profile of Moon from birth to the final lock-in at $a_{G2}$
   X-axis is semi-major axis. Y-axis is acceleration in $m/s^2$.[Courtesy:Author]

Figure S3.3[.tiff]. Superposition of radial Velocity profile and radial acceleration profile of Moon.[Courtesy:Author]

Figure S3.4[.tiff].Plot of $a_{synSS}$ ($\times R_{Iap}$)[Dashed Blue], $a_{G1}$ ($\times R_{Iap}$)[Thick Green] and $a_{G2}$ ($\times R_{Iap}$)[Thick Red] as a function of 'q'=mass ratio. Y-axis is semi-major axis as a multiple of Iapetus Globe Radius.[Courtesy:Author]

Figure S3.5[.tiff].Observed Lengthening of Day curve over a time span of 3.56Gy.[Courtesy:Author]

Figure S3.6[.tiff]. Theoretical Lengthening of Day curve over a time span of 3.56Gy.[Courtesy:Author]

Figure S3.7[.tiff]. Superposition of Theoreticaland Observed Lengthening of Day curve over a time span of 3.56Gy.Worst case mismatch is -31%.[Courtesy:Author]

Figure S5.1[.tiff]. Plot of total energy of E-M system in the range $1.4\times10^7$m and $1.5\times10^7$m around the inner geo-synchronous orbit of $a = 1.46\times10^7$m.[Courtesy:Author]

Figure S5.2[.tiff]. Plot of total energy of E-M system in the range $5.4\times10^8$m and $5.6\times10^8$m around the outer geo-synchronous orbit of $a_{G2} = 5.527\times10^8$m.[Courtesy:Author]

**Figure S6.1[.tiff]. Evolution of semi-major axis of Moon for first 60My when Laplace plane transition is encountered at 20My and a= 17R$_E$[ Courtesy:** modified form of Extended Data Figure 5.Early tidal evolution of the Moon with $Q_E/k_{2,E} = 200$ through out the simulation in MatijaCuk,; Douglas P.Hamilton,; Simon J.Lock,; Sarah T.Stewart (2016)."Tidal evolution of of the Moon from high obliquity, high angular momentum Earth", *Nature,* **539,** 402-406, (2016) doi: 10/1038/nature19846]

**Figure S6.2[.tiff]. Observed LOD curve for fits and bound Moon covering a time span of 900My.[Courtesy: Author]**

**Figure S6.3[.tiff].Theoretical Plot for LOD curve for accelerated Moon over 900My from the present.[Courtesy: Author]**

**Figure S6.4[.tiff].Exact match between Theoretical and Observed LOD Plot over last 900My.[Courtesy: Author]**

**Figure S6.5[.tiff]..Observed LOD curve for accelerated Moon using 17 data points covering a time span of 1.2Gy.[Courtesy: Author]**

**Figure S6.6[.tiff].Theoretical Plot for LOD curve for accelerated Moon using 17 Data points covering a time span of 1.2Gy.[Courtesy: Author]**

**Figure S6.7[.tiff].Repeat precise match between Theoretical and Observed LOD Plot over the permissible time span 1.2Gy and spatial range of 45R$_E$ to 60.336R$_E$ of the Advanced kinematic Model of E-M system.[Courtesy: Author]**